\documentclass[]{aa}

\usepackage{graphicx}
\usepackage{txfonts}
\usepackage{natbib}
\usepackage{longtable}
\usepackage{lscape}

\providecommand{\sorthelp}[1]{}

\begin{document}

\title{Galactic cold cores
V. Dust opacity
\thanks{
{\it Planck} \emph{(http://www.esa.int/Planck)} is a project of the European Space
Agency -- ESA -- with instruments provided by two scientific consortia funded
by ESA member states (in particular the lead countries: France and Italy) with
contributions from NASA (USA), and telescope reflectors provided in a
collaboration between ESA and a scientific Consortium led and funded by
Denmark.
}
\thanks{{\it Herschel} is an ESA space observatory with science instruments provided
by European-led Principal Investigator consortia and with important
participation from NASA.}
\thanks{Tables 1 and E.1 are only available in electronic form
at the CDS via anonymous ftp to cdsarc.u-strasbg.fr (130.79.128.5)
or via http://cdsweb.u-strasbg.fr/cgi-bin/qcat?J/A+A/}
}

\author{M.     Juvela\inst{1},
I.     Ristorcelli\inst{2,3},
D.J.   Marshall\inst{4},
J.     Montillaud\inst{1,5},
V.-M.  Pelkonen\inst{1, 6},
N.     Ysard\inst{7},
P.     McGehee\inst{8},
R.     Paladini\inst{8},
L.     Pagani\inst{9},        
J.     Malinen\inst{1},
A.     Rivera-Ingraham\inst{2},
C.     Lef\`evre\inst{9},
L.V.   T\'oth\inst{10},                
L.A.   Montier\inst{2,3},
J.-P.  Bernard\inst{2,3},        
P.     Martin\inst{11}
}

\institute{
Department of Physics, P.O.Box 64, FI-00014, University of Helsinki,
Finland, {\em mika.juvela@helsinki.fi}                                
\and
Universit\'e de Toulouse, UPS-OMP, IRAP, F-31028 Toulouse cedex 4, France   
\and
CNRS, IRAP, 9 Av. colonel Roche, BP 44346, F-31028 Toulouse cedex 4, France  
\and
Laboratoire AIM, IRFU/Service d’Astrophysique - CEA/DSM - CNRS -
Université Paris Diderot, Bât. 709, CEA-Saclay, F-91191,
Gif-sur-Yvette Cedex, France
\and
Institut UTINAM, CNRS UMR 6213, OSU THETA, Universit\'e de Franche-Comt\'e, 
41 bis avenue de l'Observatoire, 25000 Besan\c{c}on, France
\and
Finnish Centre for Astronomy with ESO (FINCA), University of Turku,
V\"ais\"al\"antie 20, FI-21500 Piikki\"o, Finland
\and
IAS, Universit\'e Paris-Sud, 91405 Orsay cedex, France                
\and
IPAC, Caltech, Pasadena, USA                                          
\and
LERMA, CNRS UMR8112, Observatoire de Paris, 61 avenue de l'observatoire 75014 Paris, France
\and
Lor\'and E\"otv\"os University, Department of Astronomy,
P\'azm\'any P.s. 1/a, H-1117 Budapest, Hungary (OTKA K62304)          
\and
CITA, University of Toronto, 60 St. George St., Toronto, ON M5S 3H8,
Canada
}

\authorrunning{M. Juvela et al.}

\date{Received September 15, 1996; accepted March 16, 1997}

\abstract { 
The project $Galactic$ $Cold$ $Cores$ has carried out {\it Herschel} photometric
observations of interstellar clouds where the $Planck$ satellite survey has located cold
and compact clumps. The sources represent different stages of cloud evolution from
starless clumps to protostellar cores and are located in different Galactic
environments.
} 
{
We examine this sample of 116 {\it Herschel} fields to estimate the submillimetre dust
opacity and to search for variations that might be attributed to the evolutionary stage
of the sources or to environmental factors, including the location within the Galaxy.
}
{
The submillimetre dust opacity was derived from {\it Herschel} data, and near-infrared
observations of the reddening of background stars are converted into near-infrared
optical depth. 
We investigated the systematic errors affecting these parameters and used modelling
to correct for the expected biases.
The ratio of 250\,$\mu$m and J band opacities is correlated with the
Galactic location and the star formation activity. We searched
for local variations in the ratio
$\tau(250\mu$m)/$\tau(J)$  using the correlation plots and
opacity ratio maps.
}
{
We find a median ratio of $\tau(250\mu{\rm m})/\tau(J)= (1.6\pm 0.2) \times
10^{-3}$, which is more than three times the mean value reported for the diffuse
medium. Assuming an opacity spectral index $\beta=1.8$ instead of $\beta=2.0$, the
value would be lower by $\sim30\%$. No significant systematic variation is detected
with Galactocentric distance or with Galactic height.  Examination of the
$\tau(250\mu{\rm m})/\tau(J)$ maps reveals six fields with clear indications of a
local increase of submillimetre opacity of up to $\tau(250\mu{\rm m})/\tau(J) \sim 4
\times 10^{-3}$ towards the densest clumps. These are all nearby fields with
spatially resolved clumps of high column density.
}
{We interpret the increase in the far-infrared opacity as a sign of grain growth in the
densest and coldest regions of interstellar clouds.}

\keywords{
ISM: clouds -- Infrared: ISM -- 
Submillimeter: ISM -- dust, extinction -- Stars: formation -- 
Stars: protostars
}

\maketitle

\section{Introduction}

The all-sky survey of the {\it Planck} satellite \citep{Tauber2010} has enabled a new
approach to studying the earliest stages of star formation. The sub-millimetre
measurements, with high sensitivity and an angular resolution down to $\sim4.5\arcmin$,
have enabled detecting and classifying of a large number of cold and compact
Galactic sources. These probably represent different phases in the evolution of
dense interstellar clouds that leads to the formation of stars. Careful analysis of the
{\it Planck} data has led to a list of more than 10000 objects that form the Cold Clump
Catalogue of Planck Objects \citep[C3PO, see][]{planck2011-7.7b}. At {\it Planck}
resolution, it is not possible to resolve gravitationally bound cores even in the
nearest molecular clouds. The low colour temperature of most of the sources ($T<14$\,K)
strongly suggests that the {\it Planck} clumps must have high column densities,
possibly at scales not resolved by {\it Planck}, and they probably contain even
dense cores. A significant fraction of the clumps may be transient structures
produced by turbulent flows, however.

Within the {\it Herschel} Open Time Key Programme {\em Galactic Cold Cores}, we have
carried out dust continuum emission observations of 116 fields that were selected based
on {\it Planck} detections listed in C3PO. The fields, which are typically
$\sim$40$\arcmin$ in size, were mapped with {\it Herschel} PACS and SPIRE instruments
\citep{Pilbratt2010, Poglitsch2010, Griffin2010} at wavelengths 100--500\,$\mu$m. The
higher angular resolution of {\it Herschel} \citep{Poglitsch2010, Griffin2010} enables studying the internal structure of the {\it Planck} clumps, detecting
individual cores, and, in conjunction with mid-infrared data, studying
protostellar sources \citep{Montillaud2014}. The inclusion of far-infrared wavelengths
helps to determine the physical characteristics of the regions and, in particular, to
study the properties of dust emission. First results have been presented in
\citet{planck2011-7.7b, planck2011-7.7a} and in \citet{Juvela2010, Juvela2011, GCC-III}
(papers I, II, and III, respectively). \citet{Montillaud2014} presented the analysis of all
clumps and cores found in our {\it Herschel} fields, including a comparison with the
population of young stellar objects (YSOs). Further studies concentrated especially on
high latitude clouds (\citet{Malinen2014}, Rivera-Ingraham et al. and Ristorcelli et
al., in preparation).

In this paper we concentrate on dust properties and especially on the submillimetre
dust opacity. 
Variations of dust emission properties have been investigated with far-infrared (FIR)
and submillimetre observations of diffuse and molecular clouds, using data from IRAS,
COBE, ISOPHOT, the PRONAOS balloon-borne experiment, and ground-based telescopes.  It
was shown that the low temperatures found in a sample of molecular clouds
\citep{Laureijs1991, Abergel1994, Abergel1996} and the translucent Polaris Flare cirrus
cloud \citep{Bernard1999} cannot be explained by the extinction of the radiation
field. An increase of the dust emissivity by a factor of 3 compared to the standard
diffuse value was needed to reproduce the cold temperatures observed in the Taurus
filament L1506 \citep{Stepnik2003}. In dense regions, several studies have shown an
opacity increase by a factor between 2 to 4 \citep{Cambresy2001, Kramer2003,
Bianchi2003, delBurgo2003, Kiss2006b, Ridderstad2006, Lehtinen2007}. More recently,
similar results have been obtained with {\it Herschel} and {\it Planck} in molecular
clouds and cold cores \citep{Juvela2011, planck2011-7.13, Martin2012, Roy2013}. In
their detailed modelling of {\it Herschel} observations of the L1506 filament,
\citet{Ysard2013} characterised the dust evolution toward the dense part of the
filament. The dust emissivity in the outer layers of the filament was found to be
consistent with standard grains from the diffuse medium, whereas the emissivity
increases by a factor of $\sim$2 above gas densities of a few times $10^{3} {\rm
cm}^{-3}$. This change has been attributed to the formation of fluffy aggregates in the
dense medium, resulting from grain coagulation, as suggested by previous studies
\citep{Cambresy2001, Stepnik2003, Bernard1999, delBurgo2003, Kiss2006b,
Ridderstad2006}.  The average size of aggregates required to fit the FIR,
submillimetre, and extinction profiles in the L1506 filament is about 0.4 $\mu$m
\citep{Ysard2013}. This value is close to the smallest grain size needed to scatter
light efficiently in the mid-IR, which produces the 'coreshine' observed toward a
number of dense cores, which has also been interpreted as a result of grain growth
\citep{Pagani2010, Steinacker2010}.

On the theoretical side, various dust optical property calculations have predicted
a significant increase in the emissivity of aggregates at long wavelengths compared to
compact spherical grains \citep{Wright1987, Bazell1990, Ossenkopf1993, Ossenkopf1994,
Stognienko1995, Kohler2011, Kohler2012}. This variation is shown to be mainly due to
the increase in the porosity fraction with aggregate growth, but the shape, structure,
material composition, and accretion of mantles can also all contribute.

Moreover, \citet{Malinen2011, JuvelaYsard2012, Ysard2012} have investigated the impact
of radiative transfer on the results derived from observations under the assumption of
a single average colour temperature. They showed that the mixing of different
temperatures along the line of sight produces a tendency that is opposite to the
observed one. They concluded that in the absence of internal heating sources, the
observed emissivity increase toward dense clouds cannot be explained by radiative
transfer effects. It must originate in intrinsic variations of the optical
properties of the grains.

It is, however, important to note that the dust emissivity increase is not
systematically observed in the interstellar medium (ISM, see \citet{Nutter2008, Juvela2009, Paradis2009}) .
These intriguing results call for a broader investigation, making use of the large
observations statistics provided by {\it Herschel} and {\it Planck}, probing different
Galactic environments. The key questions are still open today: when, where, and how dust
evolves between diffuse and dense regions, what the physical conditions enhancing
(or preventing) the efficiency of the coagulation process are, what the time scales are,
and whether the process is directly related to specific stages in the cloud or core evolution.
Understanding these questions is critical since knowing the dust
opacity has a direct impact on many key parameters derived from dust emission, such as
the column densities, masses, and volume densities of the clouds. For this reason, it
is also necessary to investigate the possible systematic effects on the emission
and extinction measurements that could cause errors in the opacity estimates. 

The structure of the paper is as follows: The observations are described in
Sect.~\ref{sect:obs}. The main results are presented in Sect.~\ref{sect:results},
including the estimates of submillimetre and near-infrared (NIR) optical depths, the
correlations between these variables, and the correlations with environmental factors.
The results are discussed in Sect.~\ref{sect:discussion} before we list the
final conclusions in Sect.~\ref{sect:conclusions}.

\section{Observations and basic data analysis}  \label{sect:obs}

\subsection{Target selection}

The selection of the {\it Herschel} fields is described in \citet{GCC-III} and an
overview of all the maps is given in \citet{Montillaud2014}. We only repeat the
main points here. 

{\it Planck} sub-millimetre observations, together with IRAS 100\,$\mu$m data, enabled
the detection of over 10000 compact sources in which the dust is significantly colder than
in the surrounding regions \citep{planck2011-7.7b}. The detection procedure is based on
this colour temperature difference and, furthermore, limits the size of the detected
clumps to values below $\sim$12$\arcmin$ \citep{Montier2010}. The sources are believed
to be Galactic cold clumps or, at larger distance, entire clouds
\citep{planck2011-7.7b}.

The fields for {\it Herschel} follow-up observations were selected using a binning of
{\it Planck} cold clumps with respect to the Galactic longitude and latitude, the
estimated dust colour temperature, and the clump mass. At the time of source selection,
distance estimates existed for approximately one third of the sources in C3PO and,
therefore, some sources of unknown mass were also included. The binning ensured
full coverage of the clump parameter space, especially of the high Galactic latitudes
and of the outer Galaxy. Galactic latitudes $|b|<1^{\degr}$ were excluded because that
area is covered by the Hi-GAL programme \citep{Molinari2010}. Similarly, regions
included in other {\it Herschel} key programmes such as the Gould Belt survey
\citep{Andre2010} and HOBYS \citep{Motte2010} were avoided.

A total of 116 separate fields were observed. The SPIRE maps are on average over
40$\arcmin$ in linear size, with an average area of $\sim1800$\,(arcmin)$^2$. The PACS
maps are smaller, with an average area of $\sim 660$\,(arcmin)$^2$. Most fields contain
more than one {\it Planck} clump, the maps altogether covering $\sim$350 individual
{\it Planck} detections. The range of probed column densities extends from diffuse
fields with ${N({\rm H}_2)} \sim 10^{21}$\,cm$^{-2}$ to cores with $N({\rm
H}_2)>10^{23}$\,cm$^{-2}$. The fields are listed in Table~\ref{table:fields} and the
{\it Herschel} observation numbers are included in Table~\ref{table:nicer}.

\subsection{{\it Herschel} data} \label{sect:Herschel_data}

\subsubsection{{\it Herschel} data reduction} \label{sect:reduction}

The fields were mapped with the SPIRE instrument at wavelengths 250, 350, and
500\,$\mu$m and with the PACS instrument at wavelengths 100 and 160\,$\mu$m. One field
was observed with SPIRE alone (G206.33-25.94, part of the Witch Head Nebula, IC~2118). The
{\it Herschel} observations are discussed in detail in \citep{GCC-III} and
\citep{Montillaud2014}. The SPIRE observations at 250\,$\mu$m, 350\,$\mu$m, and
500\,$\mu$m were reduced with the {\it Herschel} Interactive Processing Environment
HIPE v.10.0.0, using the official pipeline with the iterative destriper and the
extended emission calibration options. The maps were produced with the naive map-making
routine. The PACS data at 100\,$\mu$m and 160\,$\mu$m were processed with HIPE v.
10.0.0 up to level 1, and the maps were then produced with Scanamorphos v20
\citep{Roussel2013}. In the order of increasing wavelength, the resolution of the maps
is 7$\arcsec$, 12$\arcsec$, 18$\arcsec$, 25$\arcsec$, and 36$\arcsec$ for the five
bands. 
The raw and pipeline-reduced data are available via the {\it Herschel} Science
Archive, the user-reduced maps are available via ESA site\footnote{\em
http://herschel.esac.esa.int/UserReducedData.shtml}. 

The accuracy of the absolute calibration of the SPIRE observations is expected to be
better than 7\%\footnote{SPIRE Observer's manual, \\ {\em
http://herschel.esac.esa.int/Documentation.shtml}}. For PACS we assume a calibration
uncertainty of 15\%. This is a conservative estimate that is compatible with the
differences of PACS and Spitzer MIPS measurements of extended emission
\footnote{http://herschel.esac.esa.int/twiki/bin/view/Public/PacsCalibrationWeb}.

\subsubsection{Estimating intensity zero points}
\label{sect:zeropoint}

To determine the zero point of the intensity scale, we compared the {\it Herschel} maps
with {\it Planck} data complemented with the IRIS version of the IRAS 100\,$\mu$m data
\citep{MAMD2005}. The {\it Planck} and IRIS measurements were interpolated to {\it
Herschel} wavelengths using fitted modified blackbody curves, $B_{\nu}(T_{\rm dust})
\nu^{\beta}$, with a fixed value of the spectral index, $\beta=2.0$. The linear
correlations between {\it Herschel} and the reference data were extrapolated to zero
{\it Planck} (+IRIS) surface brightness to determine offsets for the {\it Herschel}
maps. For SPIRE the uncertainties of these fits are typically $\sim$1\,MJy\,sr$^{-1}$
at 250\,$\mu$m and at longer wavelengths smaller in absolute value. The derived
intensity zero points are independent of {\it Planck} calibration and of any
multiplicative errors in the comparison. For PACS the correlations are often less well
defined, and the zero points were set directly based on the comparison of the average
values of the {\it Herschel} maps and the corresponding interpolated {\it Planck} and
IRIS data. 
The zero points were calculated iteratively, including colour corrections calculated
using colour temperatures that were estimated from SPIRE data with a fixed spectral
index of $\beta$=2.0.

\subsubsection{Calculating submillimetre optical depth} \label{sect:tau250}

The {\it Herschel} maps were converted into estimates of dust optical depth at
250\,$\mu$m. The surface brightness maps were convolved to a common resolution of
40$\arcsec$ , and colour temperatures were calculated by fitting the spectral energy
distributions (SEDs) with modified blackbody curves with a constant opacity spectral
index of $\beta=$2.0. The 250\,$\mu$m optical depth was obtained from 
\begin{equation}
\tau(250\mu{\rm m}) = \frac{I_{\nu}(250\mu{\rm m})}{B_{\nu}(T)},
\end{equation}
using the fitted 250\,$\mu$m intensity $I_{\nu}(250\mu{\rm m})$ and the colour
temperature $T$. The calculations were made with 250--500\,$\mu$m data and
160--500\,$\mu$m data. The fits were weighted according to 15\% and 7\% error estimates
for PACS and SPIRE surface brightness measurements, respectively  (see
Sect.~\ref{sect:reduction}).

The assumed value of $\beta$=2.0 may be appropriate for dense clumps, although at lower
column densities the average value is lower, $\beta\sim 1.8$ \citep{Boulanger1996,
planck2011-7.13}, and the value of $\beta$ may further depend on the Galactic location
and the wavelength range \citep[e.g.][]{planck2013-XIV}. If the true value of $\beta$
were 1.8 instead of 2.0, our colour temperature estimates would be higher by $\sim$1\,K
and the $\tau(250\mu$m) values lower by $\sim$30\%. Furthermore, if the values of
$\beta$ were correlated with column density, the slope of $\tau(250\mu$m) vs.
$\tau_{J}$ would be similarly affected. 
We return to these effects in Sects.~\ref{sect:biasbias} and \ref{sect:discussion}.

To estimate the statistical uncertainty of $\tau(250\mu$m) values, we used Markov chain
Monte Carlo (MCMC) runs. The prior distribution of temperature values is flat but
limited between 5.0\,K and 35\,K. In addition to the relative errors quoted above, we
included the uncertainty of the intensity zero points. These are typically much smaller
than the assumed relative errors, but may be important at low column densities,
especially at 160\,$\mu$m. The zero-point errors are systematic but are included simply
as another component of statistical noise. Their effect is thus reflected in the error
estimates of individual pixels. The error distribution of $\tau(250\mu$m) is nearly
Gaussian, and we used the standard deviation of the MCMC $\tau(250\mu$m) samples as the
error estimates. These estimates were calculated separately for each pixel of the
$\tau(250\mu$m) maps.

Because of line-of-sight temperature variations, the derived $\tau(250\mu {\rm m})$
estimates probably systematically underestimate the true values \citep{Shetty2009a,
Malinen2011}. We cannot directly determine the magnitude of these errors but, with some
assumptions, we can use radiative transfer modelling to estimate the magnitude of the
bias. The simulations, described in detail in Appendix~\ref{sect:Herschel_simu}, were
used to derive bias maps that are taken into account when the data were correlated with
$\tau_{J}$ values.

\subsection{Near-infrared data} \label{sect:NIR_data}

We used the Two Micron All Sky Survey \cite[2MASS,][]{Skrutskie2006} to derive estimates
of dust column density that are independent of dust emission. We used the method NICER
\citep{Lombardi2001} and the standard extinction curve \citep{Cardelli1989} to convert
the reddening of the background stars to estimates of J-band optical depth, $\tau_{J}$.
Because the calculations involve only near-infrared bands, the results are expected to be
insensitive to the value of the ratio of total to selective extinction, $R_{\rm V}$
\citep{Cardelli1989}. 
The shape of the NIR extinction curve is believed to be relatively stable, even at
high extinctions \citep[e.g.][]{Draine2003ARAA, Indebetouw2005, Lombardi2006_Pipe,
RomanZuniga2007, Ascenso2013, Wang2014}. Some variations are observed with Galactic
location and/or density, but generally only at a level of 5\% of the NIR power-law index
\citep[e.g.][]{Stead2009, Fritz2011}.
This question is discussed in more detail in Sect.~\ref{sect:comparison}. The
$\tau(J)$ values are derived using both the J-H and H-K colours but, with the
extinction curve used, we have the correspondence of $E_{J-K}=0.65\,\tau(J)$. Flags
in the 2MASS catalogue were used to avoid galaxies ({\em ext\_key} not null or {\em
gal\_contam} not zero) and sources with uncertain photometry ({\em ph\_qual} worse
than {\em C}).

Five of our fields are fully covered in the VISTA Hemisphere Survey, VHS
\citep{McMahon2013}, which has more than ten times the sensitivity of 2MASS (in $H$ band
VHS has a 5-$\sigma$ detection threshold of 19.0, compared to a 2MASS point source
catalogue completeness limit of $\sim$16\ mag).
One of these fields is too distant to obtain a reliable extinction map, but the data for
the four other fields were analysed and the results compared with those obtained with
2MASS data. The fields are G4.18+35.79 (LDN~134), G21.26+12.11, G24.40+4.68, and
G358.96+36.75. For the fields G21.26+12.11 and G24.40+4.68, only J- and Ks-band data
exist. The data are available in VISTA Survey
Archive\footnote{http://horus.roe.ac.uk/vsa/index.html} , and the VISTA Data Flow System
pipeline processing and science archive are described in \citet{Irwin2004} and
\citet{Hambly2008}. 

Extinction maps are produced by averaging extinction estimates of individual stars with
a Gaussian weighting function with FWHM=180$\arcsec$. We also tested a
higher resolution of FWHM=120$\arcsec$. For distant sources, the extinction of the
target clouds cannot be reliably reproduced because of the poor resolution and the
increasing number of foreground stars. This is the main factor that limits the number
of fields where the ratio $\tau(250\mu {\rm m})/\tau(J)$ can be reliably estimated. The
extinction measurements can be significantly biased even in nearby fields if these
contain steep column density variations. No special steps were taken to eliminate
the contamination by foreground stars \citep[see, e.g.,][]{Schneider2011}, apart from
the sigma clipping procedure that is part of the NICER method and was performed at
3$\sigma$ level. The reliability of the extinction maps and the bias caused by sampling
problems and the presence of foreground stars was examined with simulations (see
Sect.~\ref{sect:NIR_simu}). The results of these simulations are used to derive maps of the expected uncertainty and the bias of the ${\tau(J)}$ values for each field.

\subsection{Correlations between sub-millimetre and NIR opacity}
\label{sect:correlations}

The ratio $k$ of sub-millimetre opacity ${\rm \tau(250\mu m)}$ and the NIR opacity
$\tau_{J}$ was estimated for all 116 fields. The $\tau(250\mu$m) maps were convolved to
the 3$\arcmin$ resolution of the $\tau_{J}$ maps. The $\tau(250\mu$m) and the
${\tau(J)}$ data were read at 90$\arcsec$ steps (half-beam sampling), excluding the map
borders where the result of the convolution to 3$\arcmin$ resolution is poorly
defined. For local background subtraction, only areas where the signal was more
than 2$\sigma$ above the average value of the reference area were used
(see~\ref{sect:zeropoint}). Here $\sigma$ is the standard deviation of the values in
the reference region.  This is a conservative limit because part of the fluctuations is
caused by real surface brightness variations and not by noise alone.

For $\tau_{J}$ the error estimates were provided by the NICER routine. For ${\rm
\tau(250\mu m)}$ these were obtained from MCMC calculations (see
Sect.~\ref{sect:tau250}). The comparison between the different cases (for example,
regarding the use of 160\,$\mu$m data, background subtraction, or gradient corrections)
provides information on the uncertainty caused by some sources of systematic errors.

The $\tau(250\mu$m) vs. ${\tau(J)}$ points of individual fields and samples of fields
were fitted with a linear model to derive the ratio $\tau(250\mu {\rm m})/\tau(J)$.
These total least-squares fits take into account the uncertainties in both variables.
The fits were made using either all data points or only data below or above a given
$\tau(J)$ limit. To reduce the bias caused by these cuts, the data were divided with
the help of a preliminary linear fit to all data points (see
Sect.~\ref{sect:apparent} for
details). The limiting value of $\tau(J)$ thus corresponds to a position on this line,
and the cut was performed using a line that is perpendicular in a coordinate system
where the average uncertainties of the two variables are equal. 

The $\tau(250\mu {\rm m})/\tau(J)$ ratios were also calculated for alternative versions
of the $\tau(250\mu{\rm m})$ data, using local background subtraction or using ancillary
data in an attempt to correct for possible large-scale errors in the surface brightness
data. These alternative data are discussed in Appendix~\ref{sect:alternative}.

\section{Results} \label{sect:results}

\subsection{Apparent ${\rm \tau(250\mu m)}$/$\tau(J)$ values} \label{sect:apparent}

We calculated $\tau(J)$ and ${\rm \tau(250\mu m)}$ maps of the 116 fields as described in
Sects.~\ref{sect:Herschel_data} and~\ref{sect:NIR_data}. The correlations between
$\tau(J)$ and ${\rm \tau(250\mu m)}$ were calculated at a resolution of 180$\arcsec$. In
addition to the full range of column densities, the relationships were examined
separately below and above the limit of $\tau(J)$=0.6 (see
Sect.~\ref{sect:correlations}). This corresponds to visual extinctions $A_{\rm V} \sim
2.3$\,mag and $A_{\rm V} \sim 2.0$\,mag for the $R_{\rm V}$ values of 3.1 and 5.0,
respectively \citep{Cardelli1989}. Instead of a higher limit, we selected the relatively
low number of ${\tau(J)}=0.6$ to maximise the number of fields where a linear fit could
also be made above the $\tau(J)$ threshold. The $\tau(250\mu {\rm m})$ values were
derived from {\it Herschel} data with either 250--500\,$\mu$m or 160--500\,$\mu$m (see
Appendix~\ref{sect:alternative} for analysis with additional alternative data sets).

In a given field, the number of points either below or above the $\tau(J)$ limit is often
insufficient to determine any reliable value for the slope $k={\rm \Delta \tau(250\mu
m)}$/${\Delta \tau(J)}$. In a few fields no reliable value of $k$ can be determined at
all, mainly because of the low quality of the $\tau(J)$ data. This especially affects the
most distant fields because of the contamination by foreground stars and because the
structures are too small to be resolved with the 3$\arcmin$ beam. The formal errors of
the $k$ parameter were used to exclude the clearly unreliable fits. The criterion
$\delta k/k<0.1$ leaves in the default case 106 fits to all data in a field,
103 fits below $\tau(J)$=0.6, and 38 fits above ${\tau(J)}$=0.6. These fits appear
relatively reliable also based on visual inspection.

\begin{figure}
\includegraphics[width=8.8cm]{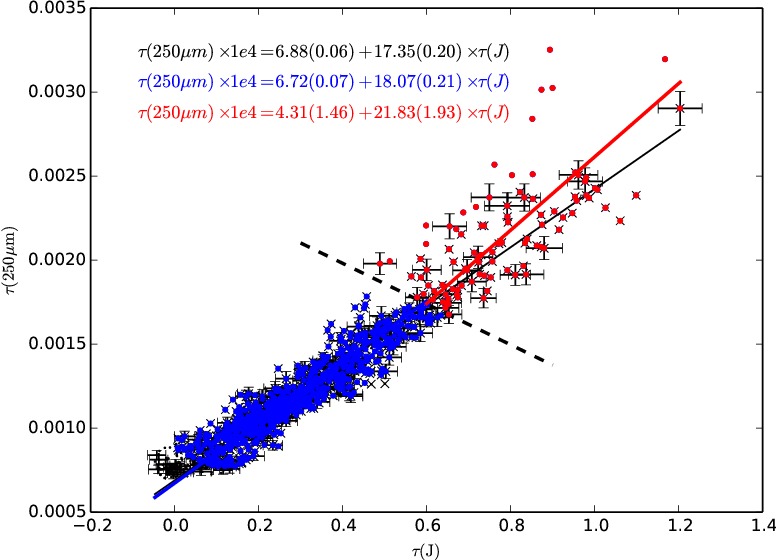} 
\caption{
Relation between ${\rm \tau(250\mu m)}$ and $\tau(J)$ in the field G95.76+8.17. The black
solid line is a linear weighted total least-squares fit to all data points. The blue and
red points and lines of the corresponding colour show the data and the fits below and
above the threshold of $\tau(J)=$0.6, the dashed line indicates the division. The values
of the slopes $k$ are given in the plot. Error bars are shown for a set of random data
points. 
}
\label{fig:corr_sample}
\end{figure}

\begin{figure*}
\includegraphics[width=18.5cm]{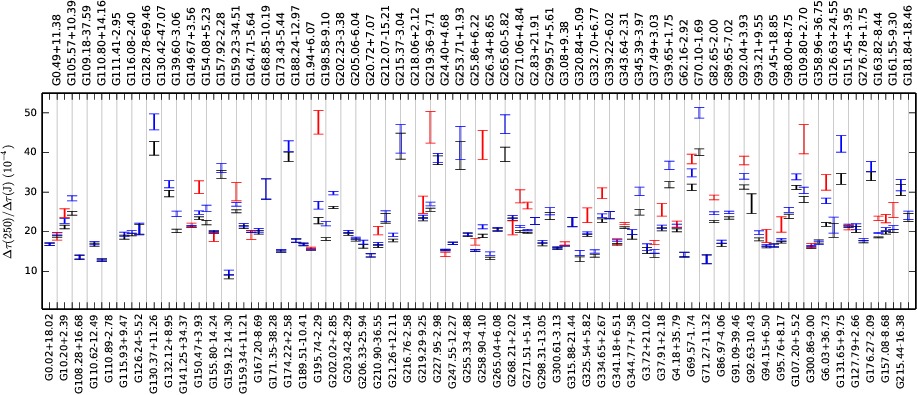} 
\caption{
Slopes $k= \Delta \tau(250\mu {\rm m})$/$\Delta \tau(J)$ for all cases with 
uncertainties $\delta k/k<0.1$. The black, blue, and red symbols correspond to values
derived for the full $\tau(J)$ range and for data below and above the limit of
$\tau(J)$=0.6. The values of ${\rm \tau(250\mu m)}$ have been derived from SPIRE data
without the subtraction of the local background. Neither $\tau(250\mu$m) nor $\tau(J)$
has been corrected for the expected bias.
}
\label{fig:corr_all}
\end{figure*}

Figure~\ref{fig:corr_sample} shows an example of the recovered dependence between
$\tau(J)$ and ${\rm \tau(250\mu m)}$ values, including linear fits to the three $\tau(J)$
ranges. In this example, the slope appears to become steeper as $\tau(J)$ increases. This
might be an indication of an increase in the dust submillimetre opacity,
which in turn might be attributed to grain growth
\citep[e.g.,][]{Ossenkopf1994,Stepnik2003,Ormel2011,Ysard2013}. However, before drawing
any such conclusions, we must consider the systematic effects that affect the two
parameters. Figure~\ref{fig:corr_all} shows a summary of all the ${\rm \Delta \tau(250\mu
m)}$/${\Delta \tau(J)}$ values where ${\rm \tau(250\mu m)}$ values are based on {\it
Herschel} 250-500\,$\mu$m data. Before any bias corrections (see below), the values are
seen to cluster around $\sim 2.0 \times 10^{-3}$, with some tendency for higher values in
the higher $\tau(J)$ range.

\begin{figure}
\includegraphics[width=8.5cm]{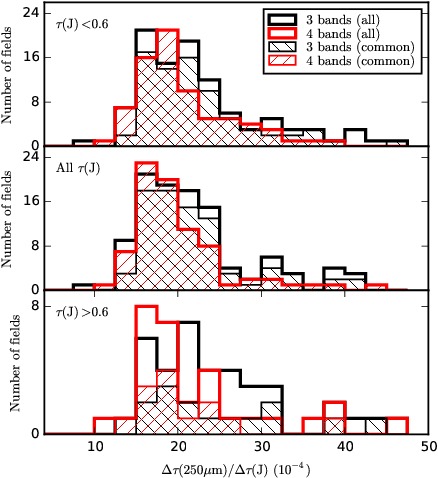} 
\caption{
Comparison of ${\rm \Delta \tau(250\mu m)}$/${\Delta \tau(J)}$ values in three $\tau(J)$
intervals (three frames), obtained using either three of four {\em Herschel} bands in
deriving the ${\rm \tau(250\mu m)}$ values. The two histograms without hatching
(thick outlines) show all fields where the estimated uncertainty is below 10\%. The two
hatched histograms contain only the intersection with better than 10\% accuracy with both
three and four bands (79, 83, and 14 fields for the three panels, respectively).
}
\label{fig:hist}
\end{figure}

Figure~\ref{fig:hist} summarises the statistics of the dust opacity measurements as
histograms, including all fits where the formal error of the slope of the least-squares
fit ${\rm \tau(250\mu m)}$ vs. $\tau(J)$ is below 10\%. 

We need to observe a sufficient number of background stars for each resolution
element, even at high column densities. This means that $\tau(J)$ estimates and the
comparison with submillimetre emission can be made only at a low resolution 
(2--3$\arcmin$). Averaged over such large areas, the statistical uncertainty of {\it
Herschel} data is very small. In Sect.~\ref{sect:bias_tauJ} we show that the bias is probably also dominated by the errors in $\tau(J)$.
In the following, we rely mainly on the {\em Herschel} data set that consists of
observations 250--500\,$\mu$m (the ``default'' data set).
There are three reasons. First, in theory, the inclusion of the 160\,$\mu$m data reduces
statistical uncertainty of the colour temperature estimates but increases systematic
errors caused by line-of-sight temperature variations \citep{Shetty2009a, Malinen2011}.
Second, because of the smaller (and, for parallel mode, different) area covered by the
PACS observations, the use of the 160\,$\mu$m band significantly reduces the area where
correlations with ${\tau(J)}$ can be calculated. 
Third, 160\,$\mu$m data may be affected by additional systematic effects related to the
relative calibration of the two instruments, uncertainties in the zero-point
determination (interpolation between IRAS and {\it Planck} channels and the contribution
of stochastically heated grains in the IRAS 100${\rm \mu m}$ band) and to imperfections
in the map making that could be increased by the smaller size of the PACS maps (see
Sect.~\ref{sect:grad}). We are particularly interested in the coldest regions where
250--500\,${\rm \mu m}$ data provide adequate constraints on the dust temperature.

\subsection{Bias in $\tau(J)$ values} \label{sect:bias_tauJ}

Bias in $\tau(J)$ values is very likely a significant problem, especially for distant fields in which all high column density structures are not resolved and the
results begin to be affected by foreground stars. Both effects decrease the
$\tau(J)$ estimates, especially towards column density peaks. We estimated the
extent of the problem with simulations using the stellar statistics in
low-extinction areas near each field. The contamination by foreground stars was
evaluated with the help of the Besan\c{c}on model of the Galactic stellar
distribution \citep{Robin2003}. We used the {\it Herschel} column density maps as a
model of the column density structure, simulated the distribution of foreground and
background stars, analysed the simulated observations with NICER routine, and
compared the results with the known input $\tau(J)$ map. The procedure is described
in detail in Appendix~\ref{sect:NIR_simu}. We obtained for each field a map of the
expected systematic relative error in $\tau(J)$  that gives a multiplicative
correction factor for the $\tau(J)$. The simulations do not consider the effect of
cloud structures at scales below 18$\arcsec$ but the procedure probably provides a
reasonable estimate of the magnitude of the effect.

\begin{figure}
\includegraphics[width=8.5cm]{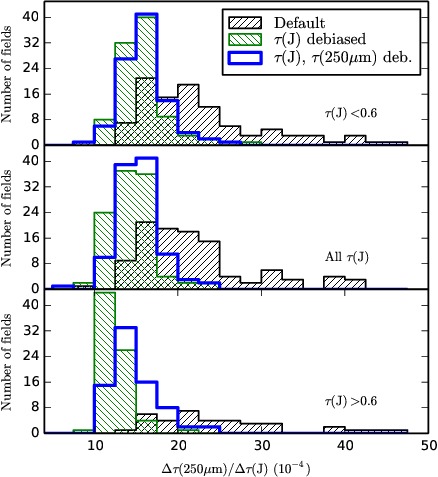} 
\caption{
Comparison of ${\rm \Delta \tau(250\mu m)}$/${\Delta \tau(J)}$ distributions without bias
corrections (``Default'') and with bias corrections applied either to $\tau(J)$ or to
both $\tau(J)$ and ${\rm \tau(250\mu m)}$. The three frames correspond to different
ranges of $\tau(J)$ values.
}
\label{fig:hist_bias}
\end{figure}

We repeated the analysis of the previous section and replaced the original ${\tau(J)}$ maps
with bias-corrected estimates. Figure~\ref{fig:hist_bias} compares the ${\rm \Delta
\tau(250\mu m)}$/${\Delta \tau(J)}$ distributions for the default case with and without
bias correction. The statistics include all fits for which the formal error of the slope is
below 10\%. This corresponds to Fig.~\ref{fig:hist}, but the number of points is
different. Because the bias corrections depend on the cloud distance, fields without
distance estimates had to be dropped. However, in the $\tau(J)>0.6$ interval the number
of fields fulfilling the $\delta k/k<0.1$ criterion has doubled.

\begin{figure*}
\begin{center}
\includegraphics[width=17.0cm]{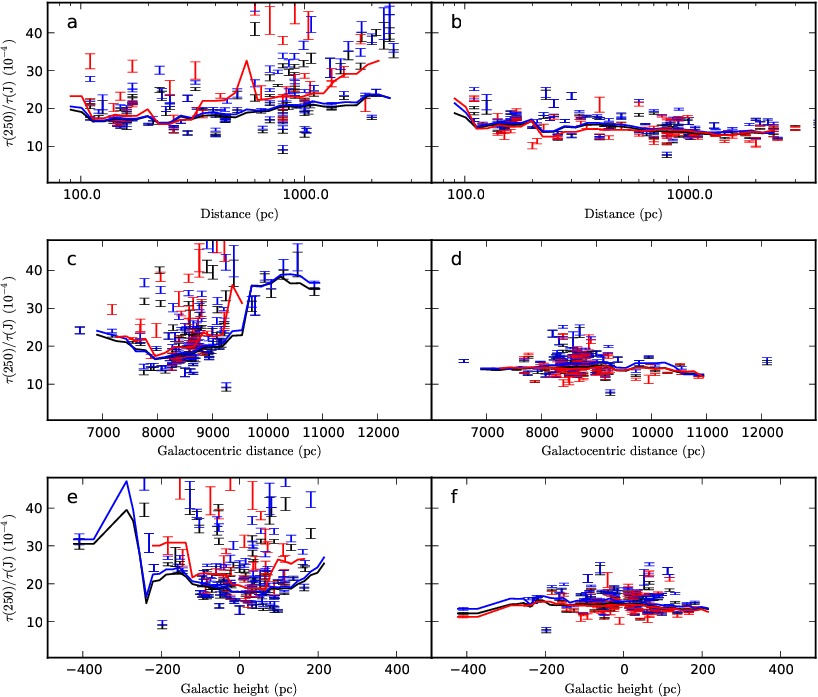}
\end{center}
\caption{
Slope values $k={\rm \tau(250\mu m)}$/$\tau(J)$ of Fig.~\ref{fig:corr_all} as a
function of estimated distance ({\em frames a, b}), galactocentric distance ({\em frames
c, d}), and galactic height ({\em frames e, f}).  The left frames show the original
slopes, the right frames the slopes after the bias corrections of $\tau(250\mu {\rm
m})$ and $\tau(J)$. The black, blue, and red colours correspond to the
full $\tau(J)$ range and to data below and above $\tau(J)$=0.6,
respectively. The solid curves with the
same colours are the weighted moving averages (window sizes 30\% in distance, 800\,pc in
Galactocentric distance, and 100\,pc in Galactic height). All $\tau(250\mu {\rm m})$
values are calculated with SPIRE bands alone.
}
\label{fig:vs_distance}
\end{figure*}

\subsection{Bias in ${\rm \tau(250\mu m)}$ values} \label{sect:bias_tau250}

The systematic errors in ${\rm \tau(250\mu m)}$ values were estimated with radiative
transfer modelling. The line-of-sight temperature variations are expected to be the
main source of error that, in standard analysis, leads to overestimation of the
mass-averaged dust temperature and, subsequently, to underestimation of ${\rm
\tau(250\mu m)}$ \citep[e.g.][]{Ysard2012}.

We constructed for each field a three-dimensional radiative transfer model that covered
a projected area of 30$\arcmin \times 30\arcmin$ with a 10$\arcsec$ pixel size. The
modelling assumed spatially constant dust properties, the dust model \citep{Draine2003}
corresponding to $R_{\rm V}$=5.5 (see Appendix~\ref{sect:Herschel_simu} for details).
The density distribution and external heating were adjusted until the model exactly
reproduced the observed 350\,$\mu$m surface brightness and, for the area above median
column density, the average 250$\mu$m/500\,$\mu$m ratio. The model-predicted surface
brightness maps were analysed as in the case of the actual observations, to produce maps
of ${\rm \tau(250\mu m)}$. To estimate the bias, these values were compared to the
actual ${\rm \tau(250\mu m)}$ values of the model to derive multiplicative
correction factors. 

The results depend on the assumed cloud structure in the line-of-sight direction
\citep{Juvela2013_colden}. In our models, the line-of-sight density distribution only has
one peak. This enhances temperature contrasts and increases our bias estimates. On the
other hand, the densest observed cores are probably even more compact, and in their
case we may be underestimating the bias. If the clouds contain embedded sources, the
actual bias may again locally be very different and often lower than predicted by our
models. Although the bias estimation is more difficult than for $\tau(J)$, the
models should again provide a reasonable estimate of the magnitude of the effect.  The
relative systematic errors are smaller in ${\rm \tau(250\mu m) }$ than in $\tau(J)$ so
that their effect on the final result is less strong.

The ${k = \rm \Delta \tau(250\mu m)}$/${\Delta \tau(J)}$ values were re-calculated
including bias corrections in both variables. The resulting histograms are included in
Fig.~\ref{fig:hist_bias}. The bias correction makes the distributions significantly 
narrower. Figure~\ref{fig:hist_bias} also shows that the corrections are much stronger for
$\tau(J)$ than for $\tau(250\mu{\rm m})$. As a result, the average value of ${\rm \Delta
\tau(250\mu m)}$/${\Delta \tau(J)}$ now decreases with increasing $\tau(J)$. The
number of fields fulfilling the $\delta k/k<0.1$ criterion has doubled to 76 fields (using
SPIRE bands). Compared to the original data, the median values of $k\times 10^4$ have
decreased from 20.2, 21.1, and 23.4 to 15.3, 16.0, and 12.2, the numbers corresponding to the full $\tau(J)$ range, data below $\tau(J)=0.6$, and data above
$\tau(J)=0.6, $ respectively. The strong change in the $k$ values suggests that the uncertainty of $k$ is
probably often several tens of per cent, especially in the ${\tau(J)>0.6}$ interval.
Therefore, Fig.~\ref{fig:hist_bias} does not exclude a systematic
increase of $k$ as the function of $\tau(J)$, if that becomes visible only at high
column densities.

Figure~\ref{fig:vs_distance} displays the slopes $k={\rm \Delta \tau(250\mu m)}$/${\Delta
\tau(J)}$ as function of distance and Galactic location. The left frames show the
relations without and the right frames with the bias corrections applied to $\tau(J)$ and ${\rm \tau(250\mu m)}$. Only fits with $\delta k/k<0.1$ are included. The
original data showed some trends, including an increase in $k$ as a function of
distance and galactocentric distance. The first is visible especially in the high
$\tau(J)$ interval, but was expected because $\tau(J)$ values of distant sources can be
severely underestimated. After bias corrections, the scatter of the $k$ values is
significantly reduced. This suggests that the corrections are of correct magnitude. The
distance dependence has changed so that in the corrected data there is a slight decrease
in $k$ values as a function of distance. This could point to some over-correction of the
$\tau(J)$ estimates, although the bias correction should not only depend on distance, but
even mainly on the cloud structure. However, the decrease of $k$ can be an indication of
selection effects or direct resolution effects. For example, higher $k$ values might be
found in individual dense clumps that are only resolved at short distances.

There is little difference between the $k$ values found in the three $\tau(J)$ intervals.
In the next section we examine in more detail the global $\tau(J)$ dependence of the
$\tau(250\mu {\rm m}) / \tau(J)$ ratios, especially regarding the highest observed column
densities.

\subsection{Global relation ${\rm \tau(250\mu m)}$ vs. $\tau(J)$}
\label{sect:global}

To further test the hypothesis that ${\tau(250\mu{\rm m}) / \tau(J)}$ ratios may change
systematically as a function of column density, we carried out non-linear fits
${\tau(250\mu m) = A + B \times \tau(J) + C \times \tau(J)^2}$.  The fits were first
performed using the combined data of all fields. To reduce the mismatch in the zero
levels of individual fields, we subtracted from each $\tau(J)$ and ${\rm \tau(250\mu m)}$
map the local background using the off regions listed in Table~\ref{table:fields}. The
off regions are not completely void of emission but provide a common reference point for
the quantities.
Thus, the relation is expected to develop via the origin for each field separately, the parameter $A$
being close to zero for the combined data as well.
The sign of the fitted parameter $C$ indicates the possible increase or decrease of
$k={\Delta \tau(250\mu {\rm m})/ \Delta \tau(J)}$ as the function of column density.

\begin{center}
\begin{figure}
\includegraphics[width=8.5cm]{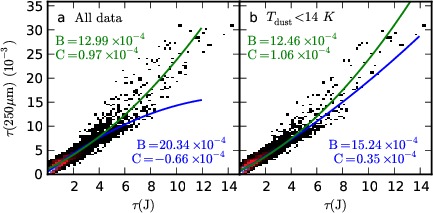} 
\caption{
Fit of ${\tau(250\mu {\rm m}) = A + B \times \tau(J) + C \times \tau(J)^2}$ to the
combined data of all fields in which individual linear fits showed a strong correlation with
$\delta k/k<0.2$. The blue and the green lines correspond to fits to the
full column density range and to data points $\tau(J)>1.0$ alone,
respectively. In the second frame,
only data with colour temperatures below 14\,K are used.
}
\label{fig:MCMC_TLS}
\end{figure}
\end{center}

Figure~\ref{fig:MCMC_TLS} shows the results obtained with the bias-corrected data.  We
included data from all fields in which individual linear fits had $\delta k/k<0.2$, thus
relaxing the previous constraint of $\delta k/k<0.1$. The second-order polynomial was
fitted to all data and separately to data points with $\tau(J)>1.0$. In the previous
section a threshold value of $\tau(J)=0.6$ was used. However, some 70\% of all points are
below $\tau(J)=0.6$ and, when included, they dominate the fits that systematically
underestimate the data above $\tau(J)\sim 5$. With the combined data set, there are
enough high column density data points so that the lower limit can be moved upwards. The
use of the $\tau(J)=1.0$ threshold enables an adequate fit to all data with higher
$\tau(J)$ values.
The $\tau(J)$ calculations employ a different off region for each field. These may
contain different amounts of extinction, which leads to small relative shifts along the
$\tau(J)$ axis. Based on dust emission, the extinction in the off regions is typically
$\tau(J)$=0.2--0.4. The uncertainty of the relative zero points contributes to the
scatter in Fig.~\ref{fig:MCMC_TLS}, but the effect is weaker
than the total
dispersion and the non-linearity seen at high extinctions. 

To prevent the $\tau(J)$ cut itself from biasing the fits, the data were selected using
lines perpendicular to a linear least-squares line fitted to all data (cf. 
Sect.~\ref{sect:correlations}). Thus, the quoted $\tau(J)$ limits correspond to a point
on the fitted line, and the cut itself is perpendicular to the fitted line. All fits
take into account the uncertainties in both variables, which are assumed to be
uncorrelated. The error distributions of the parameters $A$-$C$ were calculated with an
MCMC.

The sign of the parameter $C$ depends on the range of $\tau(J)$ values but is less
dependent on the field selection, for example regarding the $\delta k/k$ limit that was
used to select the fields. Most fields with high $\tau(J)$ values (and thus with a
wide dynamical range) have low values of $\delta k/k$. When all points are included,
the value of the parameter $C$ is negative, but the fit is very poor at high $\tau(J)$.
When pixels $\tau(J)<1.0$ are excluded, $C$ becomes positive (see
Fig.~\ref{fig:MCMC_TLS}a), which points to an increase in the submillimetre opacity above
$\tau(J) \sim 1$. Systematic additive errors in either parameter might also explain the
different behaviour at very low $\tau(J)$. When the fit is made using all data
$\tau(J)>0.6$ (not shown), the parameter $C$ is marginally positive, but beyond
$\tau(J)=10$ the fitted line is below all the data points. The second frame of
Fig.~\ref{fig:MCMC_TLS} shows the fits when pixels with colour temperatures above 14\,K
are excluded. The values of $C$ are now higher and positive even when data
$\tau(J)<1.0$ are not excluded. 
The best fit to the high $\tau(J)$ end of the relation is still obtained by excluding data
with $\tau(J)<1.0$, this results in the relation
\begin{equation}
\tau(250\mu {\rm m}) = 0.73\times 10^{-3} + 1.25\times 10^{-3}\,\tau(J) + 0.11\times
10^{-3}\, \tau(J)^2    
.\end{equation}
The formal error estimates of the parameters $A$-$C$ are of the order of 5\%, probably
lower than the systematic uncertainties. All data beyond $\tau(J)\sim 5$ are affected
by large bias corrections and, consequently, the value of $C$ also depends on the
accuracy of these corrections. Thus, Fig.~\ref{fig:MCMC_TLS} strongly suggests but does
not yet provide a final proof of the variations of the ratio $\tau(250\mu {\rm m})/\tau(J)$ . The
positive offset $A$=0.7$\times 10^{-3}$ results from the facts that at low column
densities the relation is linear, the curvature increases only beyond ${\tau(J)\sim 5}$,
and the lowest data points ${\tau(J)<1.0}$ are not part of the fit.

\begin{figure}
\includegraphics[width=8.5cm]{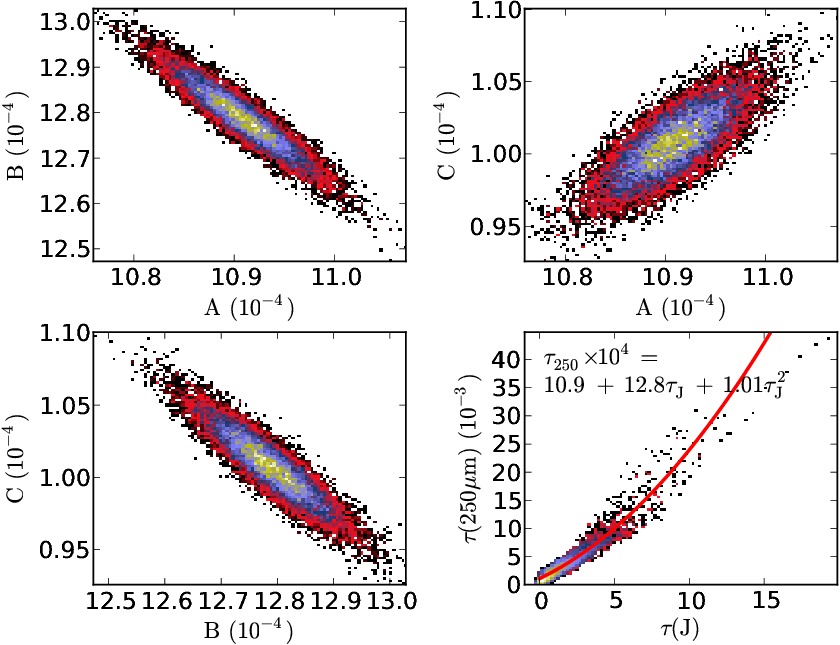} 
\caption{
Distributions of the parameters of the fit $\tau(250\mu {\rm m}) = A + B \times \tau(J)
+ C \times \tau(J)^2$. The fit is limited to data with $\tau(J)>1$.
}
\label{fig:MCMC_TLS_samples}
\end{figure}

Figure~\ref{fig:MCMC_TLS_samples} shows the error distributions of the parameters A--C.
The fit was made to data $\tau(J)>1.0$ for all fields with a
linear fit accuracy $\delta
k/k<0.2$. In most fields, the formal error estimates of $\delta \tau(J)$ and $\delta
\tau(250\mu {\rm m})$ are smaller than the actual scatter of points. Therefore we used the residuals of the linear fits before the MCMC calculation to determine a
scaling factor, typically 2.0--3.0, that makes the error estimates in each field
consistent with the actual scatter. Even after this increase of uncertainties, MCMC gives a
100\% probability for a positive value of C. In reality, the result is not that strong
because the uncertainty may be dominated by systematic errors. The sign of $C$ was already
seen to change depending on the range of $\tau(J)$ values fitted. The result also depends
on a relatively small number of fields with data above $\tau(J) > 5$. Therefore, we must
consider  the $\tau(250\mu {\rm m})$ vs. ${\tau(J)}$ relation in individual
fields in more detail.

\begin{center}
\begin{figure*}
\begin{center}
\includegraphics[width=6.0cm]{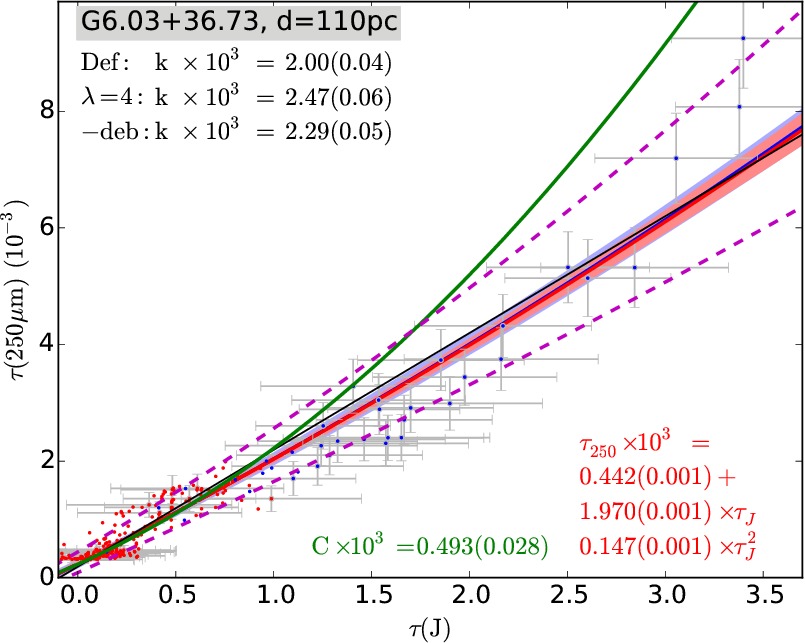} 
\includegraphics[width=6.0cm]{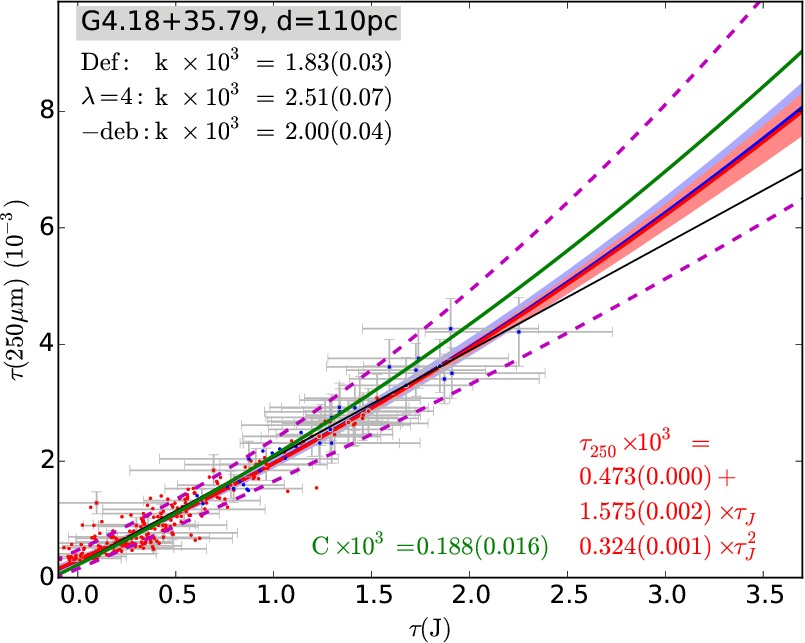} 
\includegraphics[width=6.0cm]{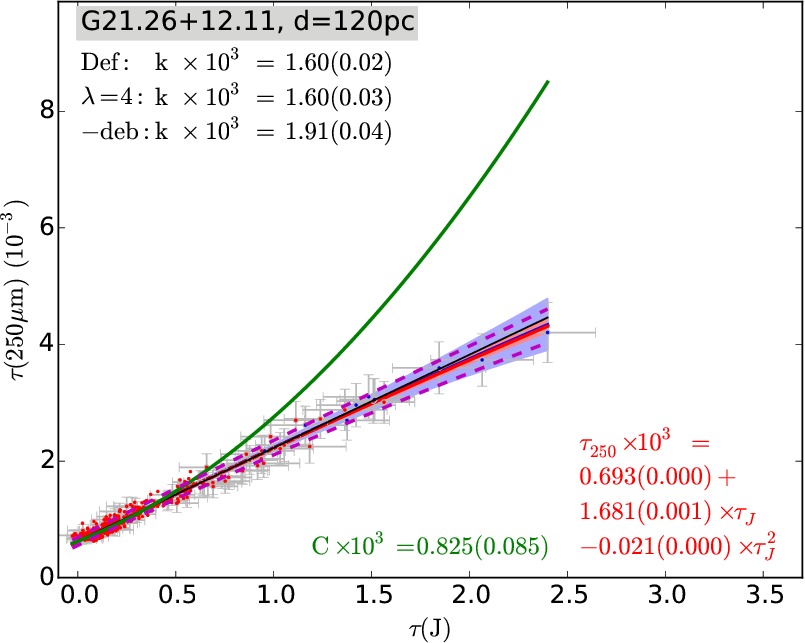} 
\includegraphics[width=6.0cm]{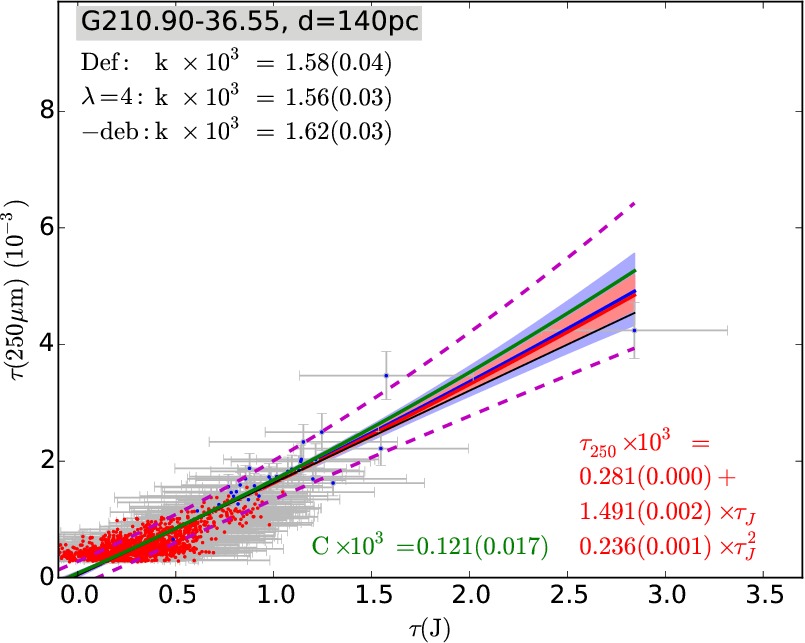} 
\includegraphics[width=6.0cm]{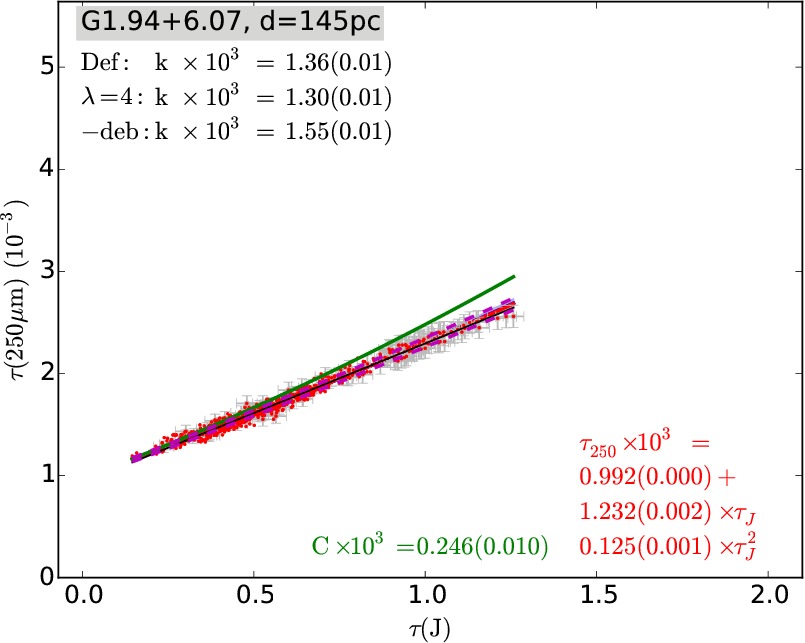} 
\includegraphics[width=6.0cm]{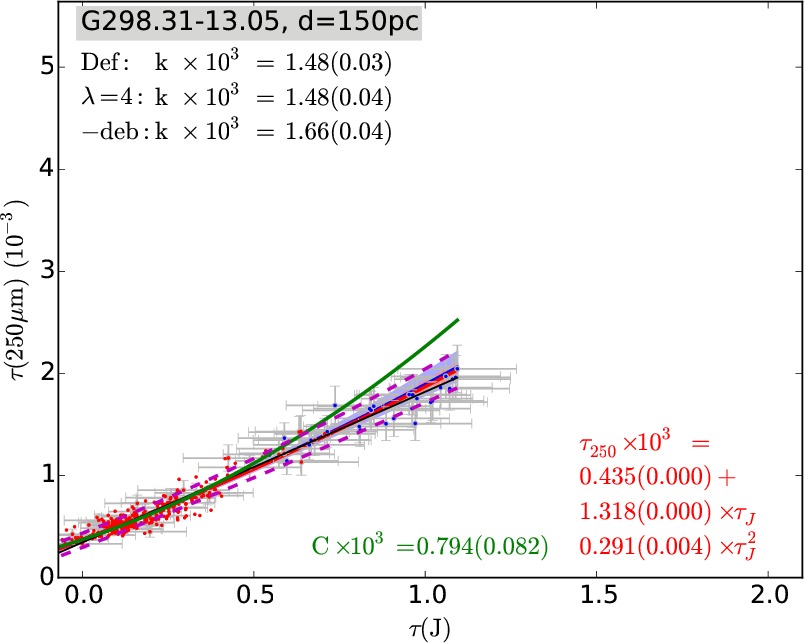} 
\includegraphics[width=6.0cm]{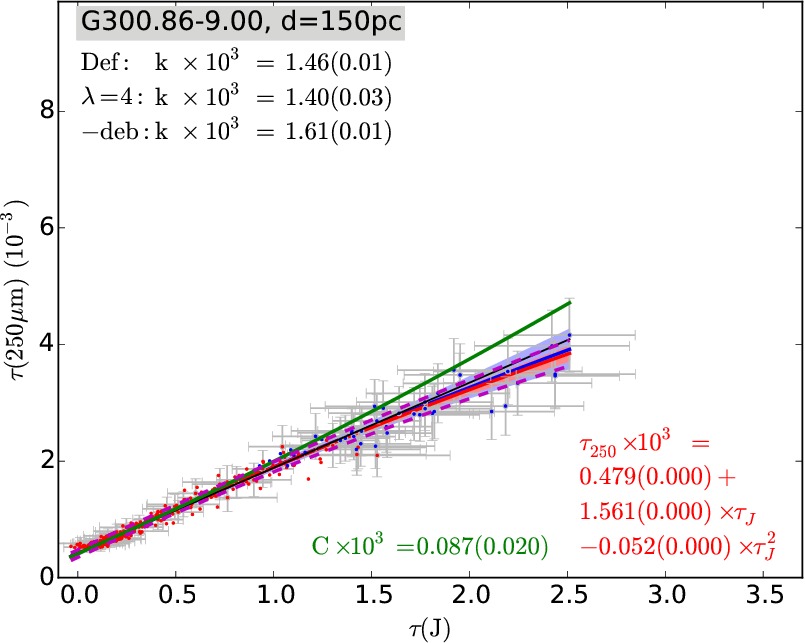} 
\includegraphics[width=6.0cm]{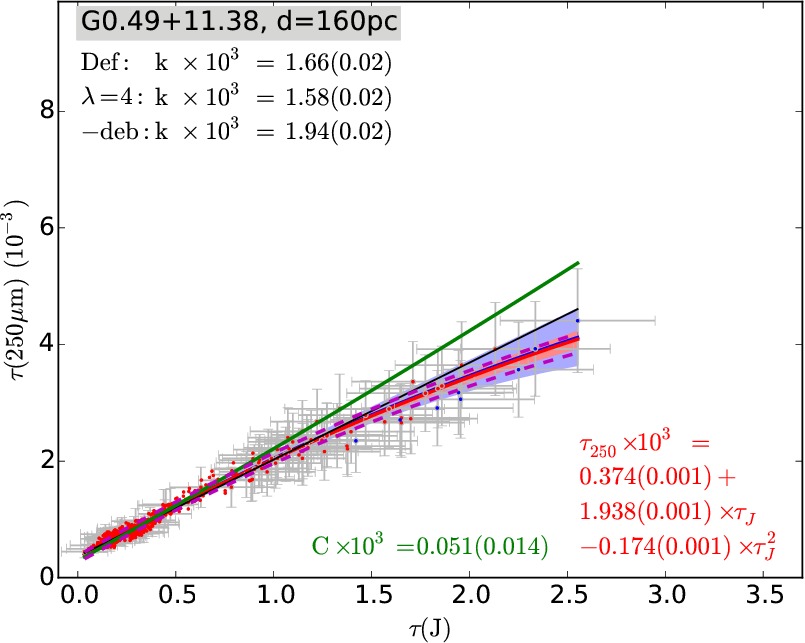}
\includegraphics[width=6.0cm]{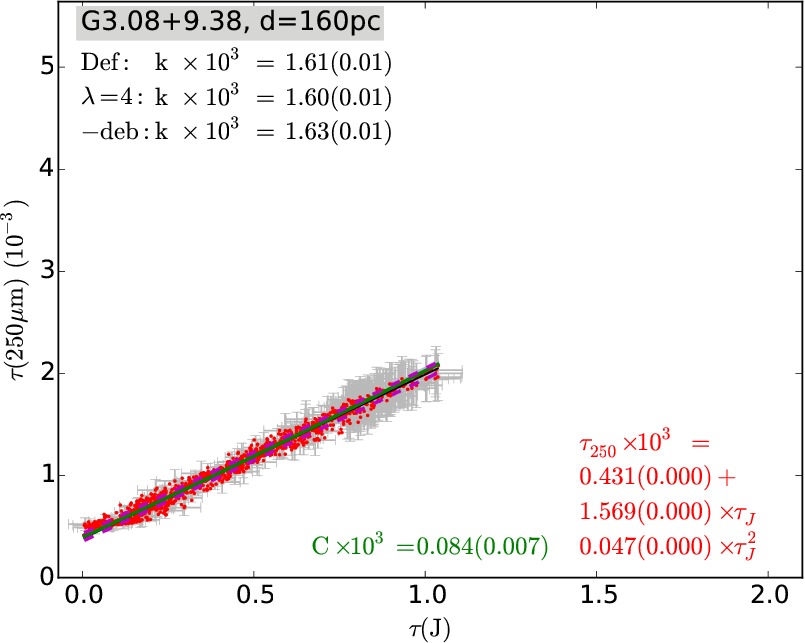}
\includegraphics[width=6.0cm]{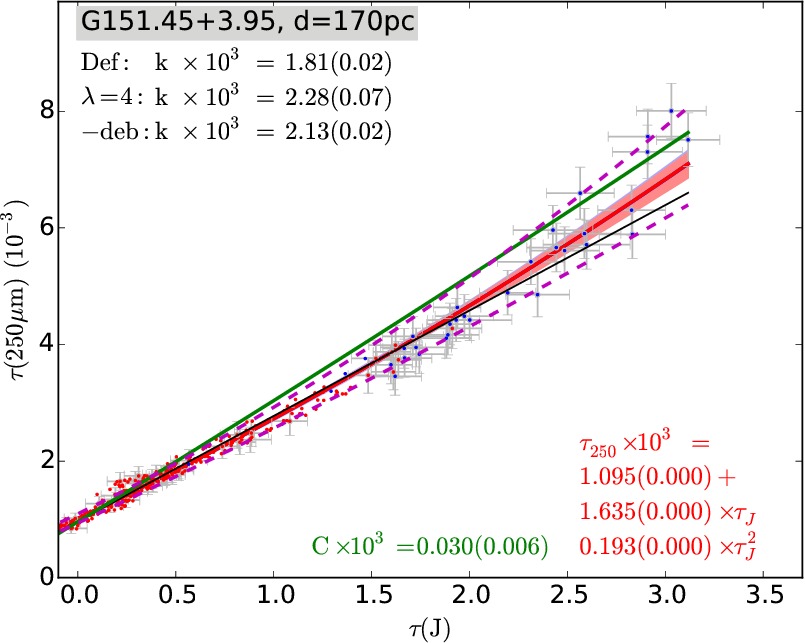}
\includegraphics[width=6.0cm]{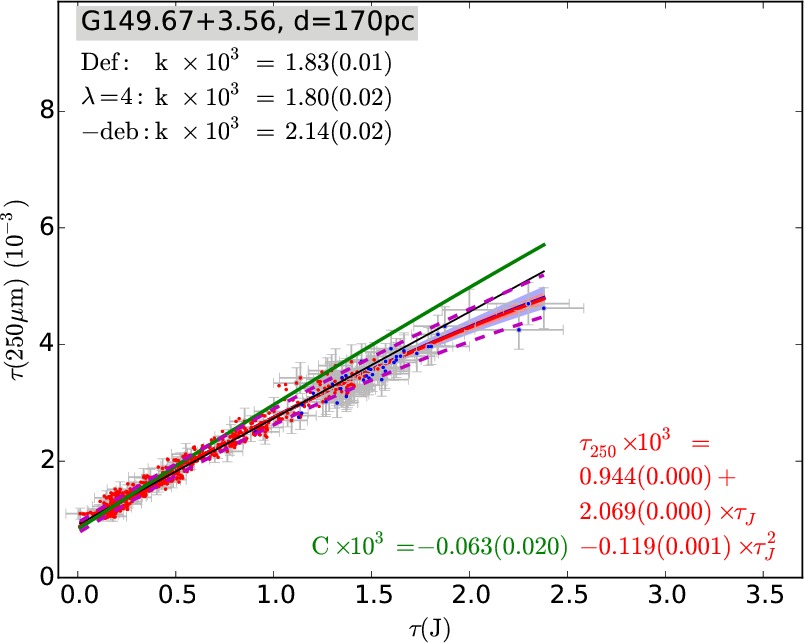}
\includegraphics[width=6.0cm]{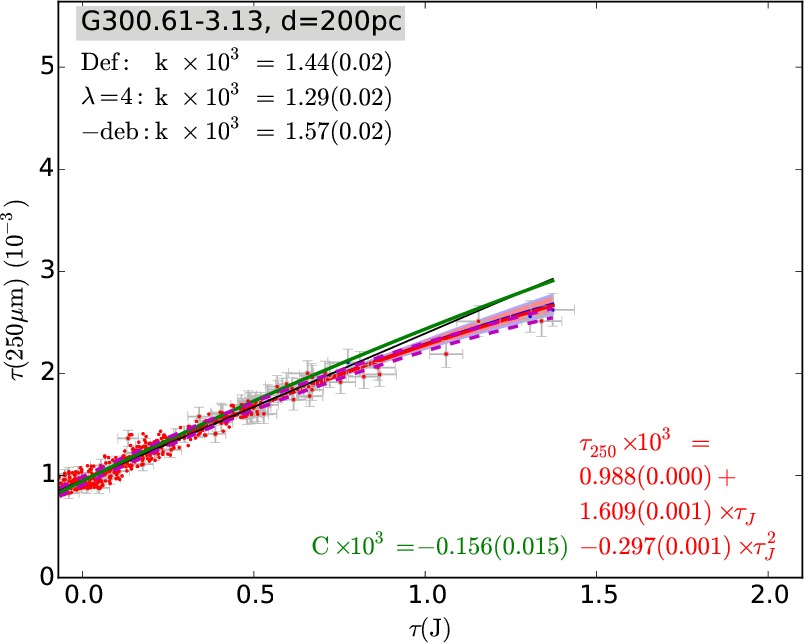}
\end{center}
\caption{
Fits of $\tau(250\mu {\rm m})$ vs. $\tau(J)$ in selected fields ordered by increasing
distance. The red and blue points (dust temperature above and below 14\,K) with error
bars are the bias-corrected data points, where $\tau (250\mu {\rm m})$ is based on SPIRE
data. 
The slopes of linear fits are listed in the upper left corner for (1) the default data set
based on SPIRE data alone (``Def.''), (2) 160--500\,$\mu$m data (``$\lambda$=4''), (3)
SPIRE data but without bias corrections (``-deb'').  The linear fit of the default case
is shown with a black line.
The non-linear fits are shown with solid blue lines (MCMC) and solid red lines
(bootstrapping) with associated shaded 68\,\% confidence regions. The dashed magenta
lines correspond to different bias correction of $\tau(J)$ using distances $d-\delta
d$ and $d+ \delta d$.
The parameters from bootstrapping are given in the lower right corner. The non-linear
fit to data without bias corrections is plotted with a solid green curve (without error
region) with the parameter $C$ given at the bottom of the figure.
The zero points of the ${\tau(J)}$ axes are not absolute.
}
\label{fig:indi_1}
\end{figure*}
\end{center}

\begin{center}
\begin{figure*}
\begin{center}
\includegraphics[width=6.0cm]{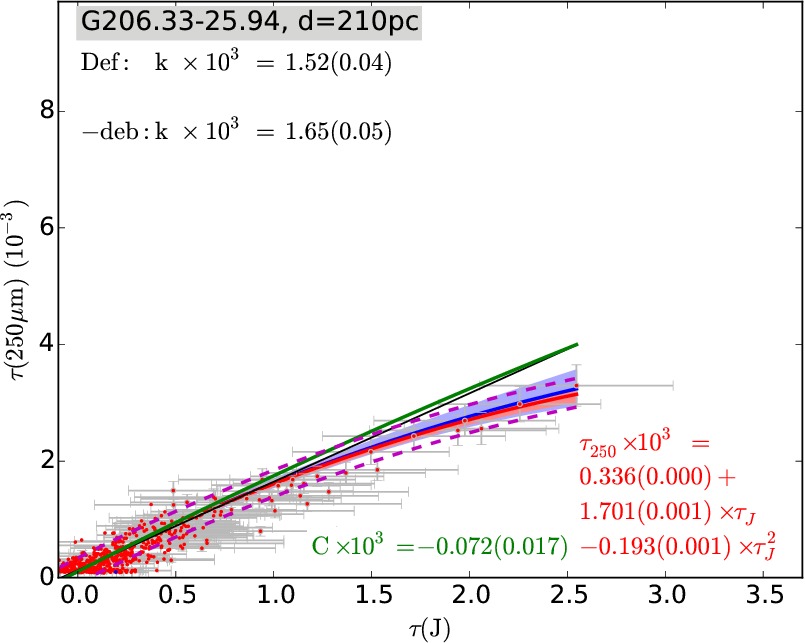}
\includegraphics[width=6.0cm]{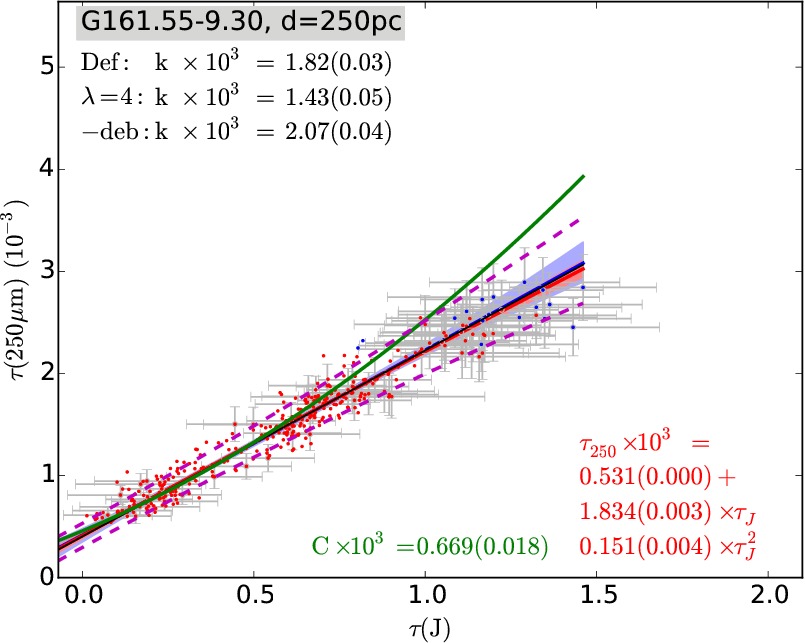}
\includegraphics[width=6.0cm]{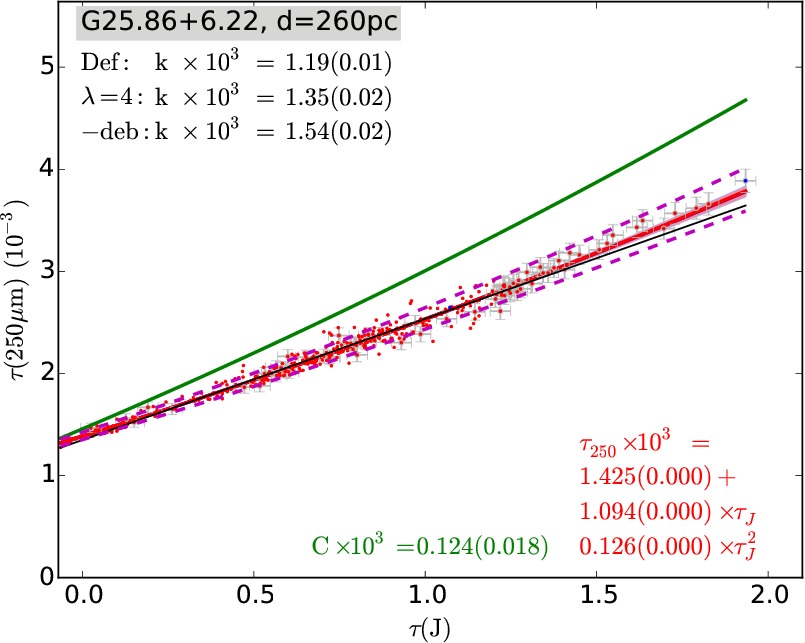}
\includegraphics[width=6.0cm]{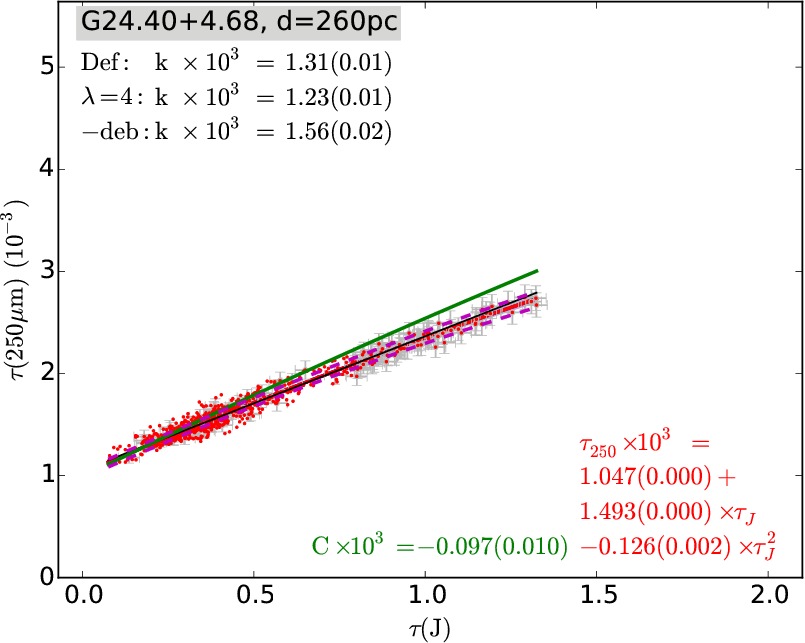} 
\includegraphics[width=6.0cm]{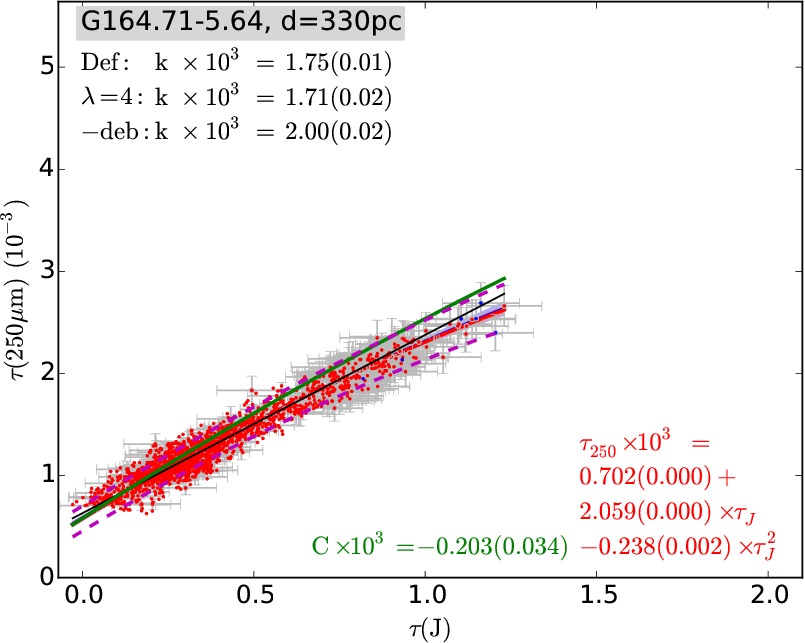} 
\includegraphics[width=6.0cm]{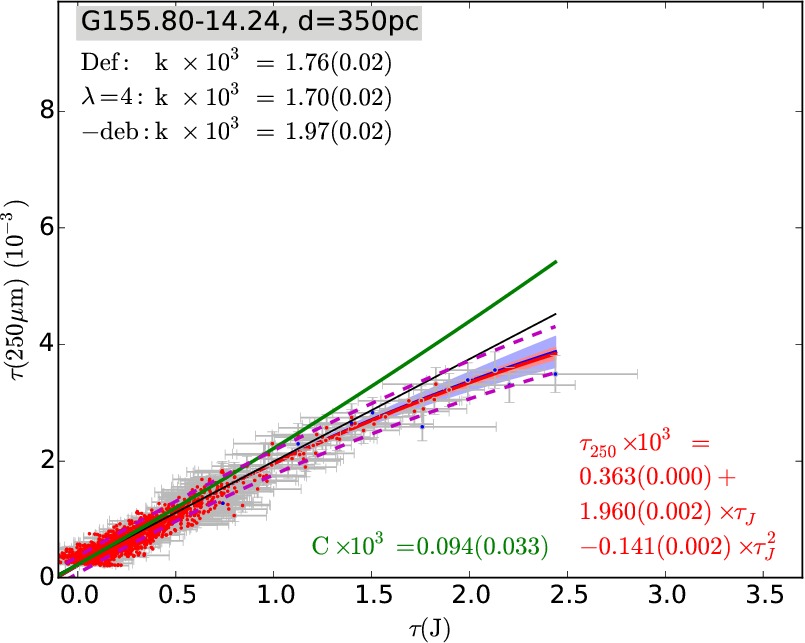} 
\includegraphics[width=6.0cm]{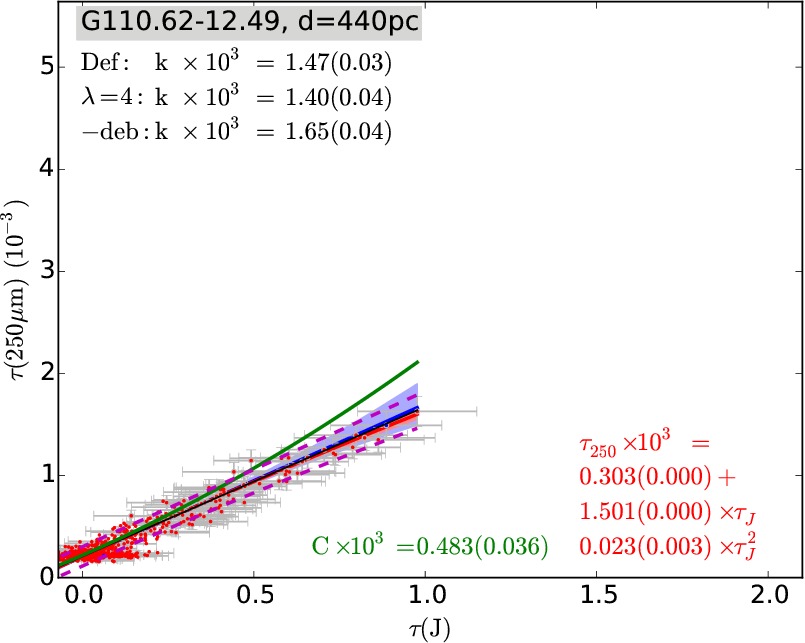} 
\includegraphics[width=6.0cm]{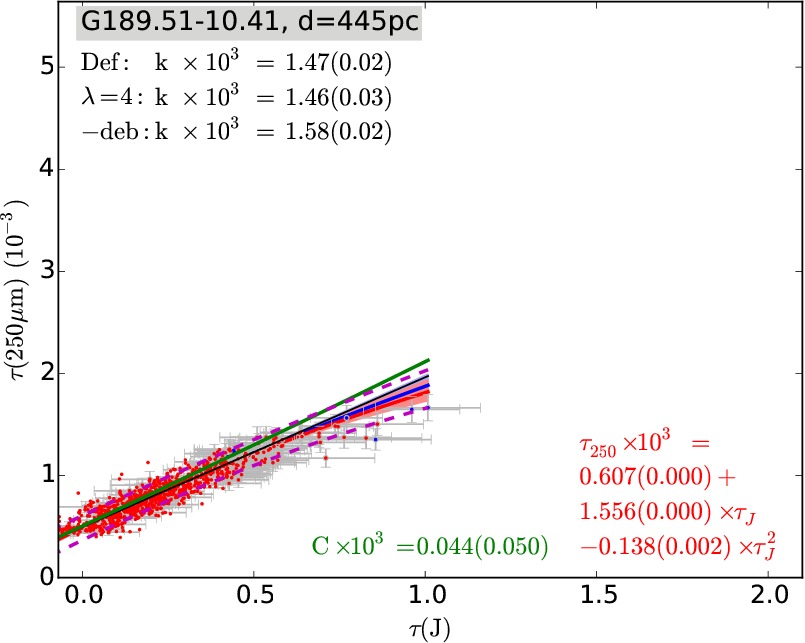} 
\includegraphics[width=6.0cm]{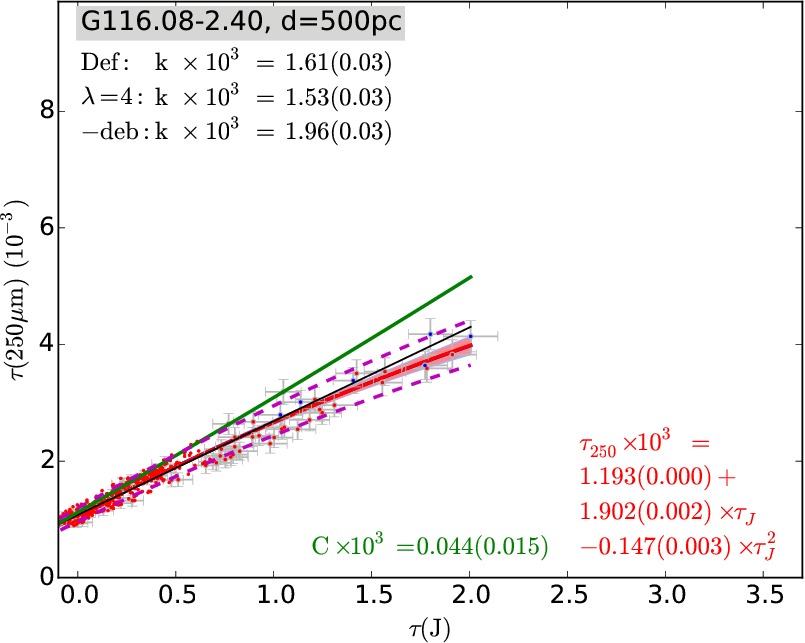} 
\end{center}
\caption{
Continued{\ldots}
}
\label{fig:indi_2}
\end{figure*}
\end{center}

\subsection{Correlations in selected fields} \label{sect:individual}

Figure~\ref{fig:MCMC_TLS} showed hints of an increase of the ${\tau(250\mu {\rm m})/\tau(J)}$
values as the column density increases, but the global statistics may be confused by the
mix of different fields. Furthermore, the sign of the $C$ parameter is determined by the
highest ${\tau(J)}$ points that originate in a small number of individual fields. Each
field may be affected by different systematic effects related to the surface brightness
zero points, distance uncertainty (via bias correction), and differences in the local
radiation field. Several diffuse fields are even entirely below $\tau(J)\sim 1.0$.
Therefore, we also need to examine the fields individually.

Three criteria were used to select a subset of fields. We required that (1) the
uncertainty of the fitted parameter $C$ is below ${\rm 0.3\times 10^{-4}}$, (2) there are
at least ten data points (selected from the maps at 90$\arcsec$ steps) with $\tau(J)$
above 0.6, and (3) the bias corrections change the slope of the
linear fit  of ${\rm \tau(250\mu
m)}$ vs. $\tau(J)$ by less than 30\%. The first two criteria ensure that there are enough
data points at large $\tau(J)$ with a small scatter to gain some
insight about
the column density dependence of the $\Delta \tau(250\mu {\rm m})/\Delta \tau(J)$. The
third criterion excludes distant fields for which the uncertainty of the bias correction
of $\tau(J)$ renders the results uncertain, even for the apparently well-defined
relation between ${\tau(J)}$ and ${\rm \tau(250\mu m)}$.
The selection leaves 23 fields with distances mostly in the range 100-500\,pc. There are
two exceptions, G216.76-2.58 and G111.41-2.95, for which the estimated distances are
2.4\,kpc and 3.0\,kpc. We kept the two fields in the sample even
though the results are known to be unreliable because of the
large distance.

To avoid underestimating the fit errors, we scaled the error estimates of $\tau(J)$
and $\tau(250\mu {\rm m}) $ up to correspond to the actual scatter of points (see
Sect.~\ref{sect:global}). All observations, sampled at 90$\arcsec$ steps, are fitted
with a linear model ${\tau (250\mu {\rm m})=b + k \times \tau(J)}$ using total least-squares. We continued to use as the default data set one with $\tau(250\mu {\rm m})$
derived from SPIRE bands alone, including bias corrections in $\tau(250\mu {\rm
m})$ and $\tau(J)$. However, for comparison, we also examined results obtained with four
{\em Herschel} bands (160--500\,$\mu$m; including the bias corrections) and, finally,
with three {\em Herschel} bands but without any bias corrections. 

The non-linear fits were made using MCMC (with $2\times 10^5$ samples per field) and bootstrap sampling (2000 realisations per field). Both fits used total least-squares\footnote{Distance between a ($\tau(J)$, $\tau(250\mu {\rm m})$) point and the
model curve is measured in a coordinate system where the error region of the point is
circular. We used the smallest distance to the curve, ignoring the marginal effect
resulting from the curvature of the model curve.} and the error estimates of the
individual points. The parameter values and the uncertainties derived with these two
methods should be similar, except for rare cases in which the result depends on a small
number of influential points, which are always present in the MCMC calculation, but not in
all bootstrap samples.

Figures~\ref{fig:indi_1}-\ref{fig:indi_2} show the results of these fits. Each frame
shows the values of the linear slopes for the three cases discussed above. The
parameters of the non-linear fits are shown together with the error estimates
calculated with the bootstrap method. In addition to the fit to the default data set
(solid red curve), the dashed magenta lines show the effect of the distance
uncertainty. Using the distance uncertainties $\delta d$ listed in
Table~\ref{table:fields}, we also calculated the bias corrections for $\tau(J)$  for
distances $d-\delta d$ and $d+\delta d$. Thus, the upper dashed line corresponds to
distance $d-\delta d$ and a smaller bias correction.

\begin{center}
\begin{figure}
\includegraphics[width=8.5cm]{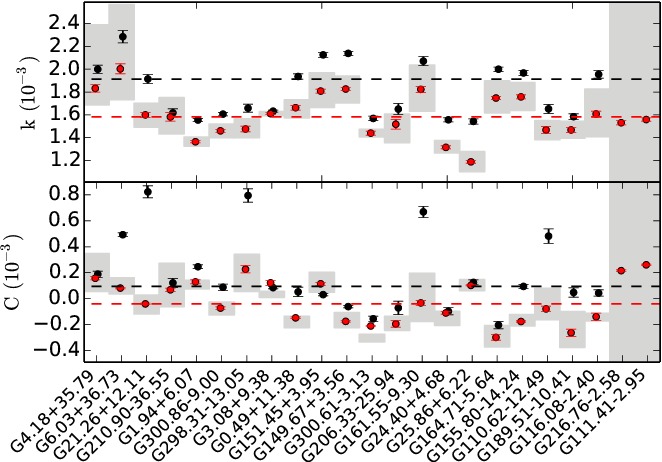}
\caption{
Linear slope $k$ (upper frame) and the parameters $C$ (lower frame) for the 23 selected fields. The values obtained without bias corrections are shown with black symbols. The
values obtained with corrected $\tau(J)$ and ${\rm \tau(250\mu m)}$ data are shown with
red symbols, the shaded area corresponding to the uncertainty of the bias correction that
is due to the uncertainty of the distance estimates. The dashed lines show the median
values corresponding to the black and red symbols. 
The fields are arranged in order of increasing distance, and the ${\rm \tau(250\mu m)}$
values are based on SPIRE data alone.
}
\label{fig:indipara}
\end{figure}
\end{center}

Figure~\ref{fig:indipara} shows the linear slopes $k$ and the values of parameter $C$ for
this sample of fields. The fits were performed to all data, without a $\tau(J)$
threshold. If the data below $\tau(J)=0.6$ were removed, the median slope $\tau(250\mu
{\rm m})/\tau(J)=1.6\times 10^{-3}$ did not change appreciably (by less than 0.1$\times
10^{-3}$). The median value of $C$ increased to $2.0\times 10^{-4}$, which in that
case is still lower than the scatter. The differences between the fields are larger than the estimated formal uncertainties (including the statistical errors of
$\tau(J)$ given by NICER and $\tau(250\mu{\rm m})$ derived from the uncertainty of the
surface brightness measurements). The error bars only reflect the statistical errors of
the fits and do not include the uncertainty of the bias correction, for example.
Figure~\ref{fig:indi_1} demonstrates that the uncertainty of the distances can be a
significant source of error. In Fig.~\ref{fig:indipara}, the shaded areas show the
difference between the values obtained with distances $d - \delta d$ and $d + \delta d$,
as listed in Table~\ref{table:fields}. These were estimated directly by repeating the
analysis using these two distance values. A smaller distance corresponds to smaller bias
correction in $\tau(J)$ and, thus, to a steeper slope $k$ and typically a higher value of
$C$. 
In some cases, the value of $C$ obtained with the default distance $d$ is
outside the shaded region, showing that the effect is not always this simple. The
distance uncertainty is not yet enough to explain all the scatter in $k$ and especially
in $C$. 
A change in the distance estimate results at first approximation in a nearly linear
scaling of $\tau(J)$ values (see Fig.~\ref{fig:indi_1}, comparison of the dashed
magenta lines).
In reality, the situation may be more complex. In particular, if a field contains cloud
structures at different distances, this might result in large errors in both $k$ and
$C$. Figure~\ref{fig:indipara} also shows that in spite of the small formal errors of
the least-squares fits, we cannot constrain the opacity values in the last two fields
with distances exceeding 2\,kpc.

In the sample of Fig.~\ref{fig:indipara}, the median value of $C$ is close to zero with
a number of fields with negative values. The positive values of $C$ in
Fig.~\ref{fig:MCMC_TLS} are due to a small number of fields, and the increase of
$\tau(250\mu {\rm m})/\tau(J)$ values was only visible above $\tau(J)\sim 5$. There are
only 20 fields with any data points above $\tau(J)=5$. Only six fields have ten or more
data points above this limit:
G6.03+36.73, G70.10-1.69, G82.65-2.00 G92.04+3.93 G107.20+5.52, and G202.02+2.85. Of
these, only G6.03+36.73 is included in the sample of Fig.~\ref{fig:indipara}. All the
others were excluded because the bias correction changed the slope $k$ by more than
30\%. Thus, a clear steepening of the relation $\tau(250\mu {\rm m}$ vs. $\tau(J)$ is
seen exclusively in fields with the highest column densities, which for the same
reason also have the largest uncertainty regarding the bias corrections.

\subsection{Maps of $\tau(250\mu {\rm m})/\tau(J)$ ratio}

\begin{center}
\begin{figure}
\includegraphics[width=8.8cm]{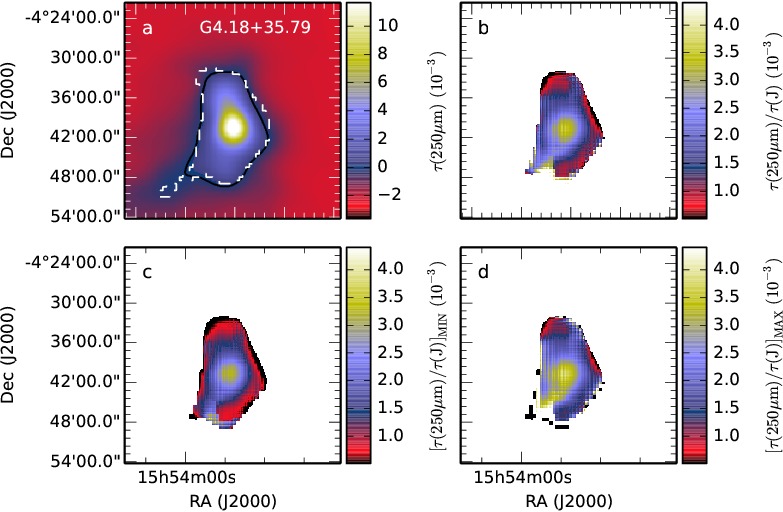}
\caption{
Field G4.18+35.79 (LDN~134). The upper frames show $\tau(250\mu {\rm m})$ (frame a) and
the ratio $\tau(250\mu {\rm m}) / \tau(J)$ (frame b). The lower frames show the lower
(frame c) and upper (frame d) limits of $\tau(250\mu {\rm m})/\tau(J)$ calculated as
$(\tau(250\mu {\rm m})+\delta \tau(250\mu {\rm m}))/(\tau(J)-\delta \tau(J))$ and
$(\tau(250\mu {\rm m})-\delta \tau(250\mu {\rm m}))/(\tau(J)+\delta \tau(J))$. The areas
not covered by {\it Herschel} observations and regions with a
SN below 0.5 have been
masked. In frame $a$, the solid black contour and the dashed white contour correspond to
$\tau(250\mu{\rm m}= \delta \tau(250\mu{\rm m})$ and $\tau(J)=\delta \tau(J)$. The maps have a resolution of 180\,$\arcsec$ and $\tau(J)$ is derived
using 2MASS data.
}
\label{fig:ratio_map}
\end{figure}
\end{center}

We also examined the ratios $\tau(250\mu {\rm m})/\tau(J)$  in the form of maps. This is
useful if $k$ changes in small regions that have little effect when all data of a field
are fitted. Unlike in Fig.~\ref{fig:indi_1}, where the offset between $\tau(250\mu {\rm
m})$ and $\tau(J)$ is a free parameter, the appearance of the ratio maps depends on the
consistency of the $\tau(250\mu {\rm m})$ and $\tau(J)$ zero points. Because we do not
have an absolute zero point for $\tau(J)$, we used the reference areas listed in
Table~\ref{table:fields} and subtracted from $\tau(250\mu{\rm m})$ and $\tau(J)$
the average value found in the reference area. This limits the region where a reliable
ratio can be calculated, excluding regions of low column density. The details of the
calculations are given in Appendix~\ref{sect:ratiomaps},  where we also show the figures of
selected fields. As an example, Fig.~\ref{fig:ratio_map} shows the field
G4.18+35.79 (LDN~134), where the ratio $\tau(250\mu {\rm m})/\tau(J)$ is strongly
correlated with column density.
In Fig.~\ref{fig:ratio_map}, the increase of $k$ remains clear even in maps of
$(\tau(250\mu {\rm m}) \pm \delta \tau(250\mu {\rm m})/(\tau(J) \pm \delta \tau(J))$.
The error estimates $\delta \tau(250\mu {\rm m})$ include the statistical errors due to
{\it Herschel} photometry and uncertainty of the surface brightness zero point (see
Sects.~\ref{sect:tau250} and~\ref{sect:zeropoint}).
The parameter $\delta \tau(J)$ corresponds to the uncertainty of the $\tau(J)$ zero
point (see Appendix~\ref{sect:ratiomaps}).  
For $\tau({\rm J})$ the formal error estimates calculated with NICER are below
10\%.
For high-opacity sources like G4.18+35.79 the bias correction of $\tau(J)$ is very
important. If no background stars are visible through some part of the core, the
values of $k$ naturally remain more uncertain (see, however,
Sect.~\ref{sect:biasbias}).
Conversely, the zero-point uncertainty only becomes important in diffuse regions but
might even reverse the correlation with column density.
The ratio maps are also affected by the assumption of a constant value of $\beta$ and
of potential errors in the bias corrections. However, we argue in
Sect.~\ref{sect:biasbias} that these are mainly multiplicative errors (that do not affect
the morphology of the maps) and/or tend to decrease the variations seen in the ratio
maps. Therefore, we are confident that the increase of submillimetre opacity that is
seen in some of the maps is real.

Based on the maps, the submillimetre opacity is correlated with column density in
the fields G4.18+35.79, G6.03+36.73, G111.41-2.95, G161.55-9.30, G151.45+3.95, and
G300.86-9.00 (see Appendix~\ref{sect:ratiomaps}). In G4.18+35.79 and G6.03+36.73 the
values rise to close to $4 \times 10^{-3}$. In some fields the background subtraction
reduces the available map area to such an extent that no conclusions can be drawn.

\begin{center}
\begin{figure*}
\sidecaption
\includegraphics[width=12cm]{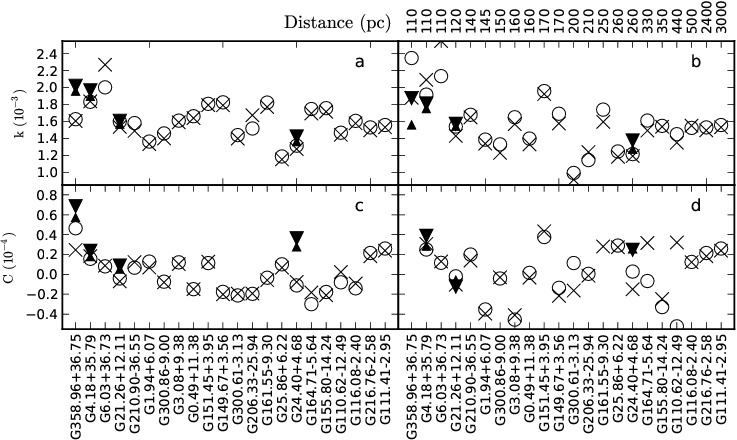}
\caption{
Comparison of parameters $k$ (upper frames) and $C$ (lower frames) obtained with data at resolutions of
180$\arcsec$ and 120$\arcsec$ . Frames $a$ and $c$ show fits to all data,
frames $b$ and $d$ fits to values ${\tau(J)}>0.6$ alone. The open circles and crosses show
the estimates obtained with 2MASS data at resolutions of 180$\arcsec$ and 120$\arcsec$ . The filled triangles show the results for VISTA observations, the larger
triangles corresponding to a resolution of 180$\arcsec$ , the smaller to a resolution of 120$\arcsec$
. The fields are the same as in Fig.~\ref{fig:indipara}, with the addition of
G358.96+36.75, which has fewer data points above ${\tau_J>0.6}$  and
for which parameter $C$ could not be fitted with extinction maps
with a resolution of 120$\arcsec$ .
}
\label{fig:kC_tables}
\end{figure*}
\end{center}

So far, all NIR extinction maps were calculated at 180$\arcsec$ resolution.  Depending
on the number of background stars, extinction map could be derived at a higher
resolution and possibly with smaller bias. This especially applies to the four fields
for which VISTA observations are available.
We recalculated the extinction maps at 120$\arcsec$ resolution, repeating Monte Carlo
simulations to estimate the bias of ${\tau(J)}$. The smaller beam increases the noise
per resolution element, but does not yet cause holes in the extinction maps. The
analysis was repeated for the four fields with VISTA data with resolutions of 180$\arcsec$ and
120$\arcsec$ .
The results are summarised in Fig.~\ref{fig:kC_tables}. The resolution has no strong
systematic effect on the parameters. The largest differences appear when parameter $C$ is
estimated excluding low column density points. However, even in that case the
difference between the results with a resolution of 120$\arcsec$ and 180$\arcsec$  is smaller than the
effect of excluding low $\tau(J)$ data from the fits. When VISTA data are available,
the results are close to those obtained with 2MASS.
Because we used the same {\it Herschel} data and bias corrections derived in the
same way, the results are not independent. However, because the uncertainty of
$\tau(J)$ is expected to be one of the most significant sources of error, this gives
us some confidence that the observed differences between the fields are real.

\begin{center}
\begin{figure}
\includegraphics[width=8.8cm]{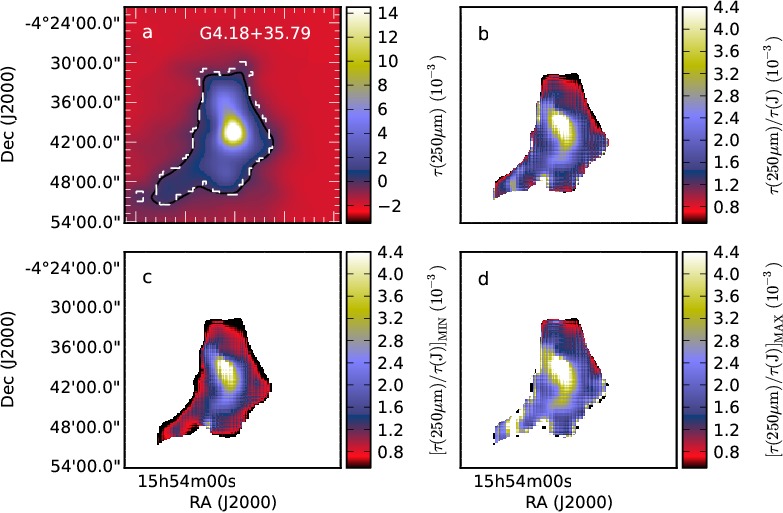}
\caption{
Ratio ${\tau(250\mu {\rm m})/\tau(J)}$ in field G4.18+35.79 at a resolution of
120$\arcsec$, based on VISTA NIR data. Frame $a$ shows a map of $\tau(250\mu {\rm m}),$
frame $b$ a map of the ratio $\tau(250\mu {\rm m}) / \tau(J)$. Frames $c$ and $d$
are estimated lower and upper limits of $\tau(250\mu {\rm m})/\tau(J)$ (cf
Fig.~\ref{fig:ratio_map}).  
}
\label{fig:VHS_map}
\end{figure}
\end{center}

The ${\tau(250\mu {\rm m})/\tau(J)}$ ratio in the field G4.18+35.79 was shown in
Fig.~\ref{fig:ratio_map}. Figure~\ref{fig:VHS_map} shows the corresponding figure
obtained with VISTA NIR data and a spatial resolution of 120$\arcsec$. The highest
value of ${\tau(250\mu {\rm m})/\tau(J)}$ has increased from $3.6 \times 10^{-3}$ in
Fig.~\ref{fig:ratio_map} to $6.7 \times 10^{-3}$. This is mainly attributed to the
increased spatial resolution, although also at the 180$\arcsec$ resolution the peak
value is $\sim$25\% higher than in Fig.~\ref{fig:ratio_map}. Even in VISTA data there
are only $\sim$10 stars within the $2\arcmin \times 2\arcmin$ area centred on the
$\tau(250\mu{\rm m}$ maximum, and therefore the peak value of ${\tau(250\mu {\rm
m})/\tau(J)}$ is subject to some uncertainty.

\subsection{Potential systematic errors} \label{sect:biasbias}


Because of the significance of the bias corrections (Sects.~\ref{sect:bias_tauJ} and~\ref{sect:bias_tau250}), we tried to characterise the effects that systematic
errors in these corrections could have on the $\tau(250\mu {\rm m})/\tau(J)$ ratios.
Furthermore, the assumption of a constant dust emissivity spectral index may be
incorrect. Below we examine the possible systematic effects caused by these factors.

\begin{center}
\begin{figure}
\includegraphics[width=8.5cm]{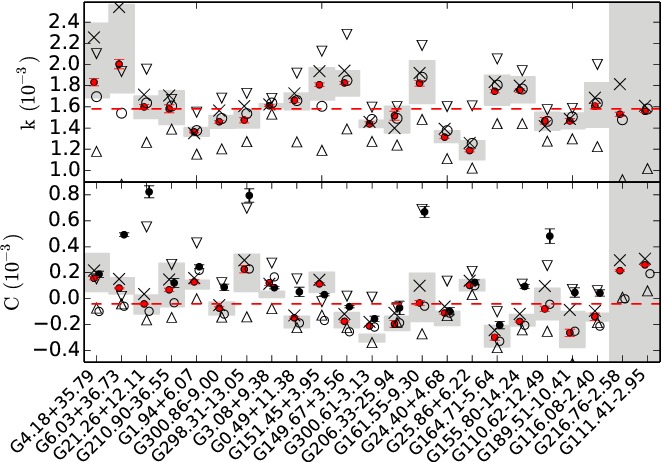}
\caption{
As Fig.~\ref{fig:indipara}, but comparing our default $\tau(250\mu {\rm m})/\tau(J)$
estimates (red solid circles) and results from alternative analyses: $\tau(250\mu {\rm
m}$ estimates derived assuming a spatially varying spectral index (crosses), $\tau(250\mu
{\rm m}$ bias estimated with \citet{Ossenkopf1994} dust model (open circles), $\tau(J)$
bias estimated using $k=1.0\times 10^{-3}$ (triangles pointing upwards), and $\tau(J)$
bias estimated using $k=3.0\times 10^{-3}$ (triangles pointing downwards).
}
\label{fig:MOD_BIAS}
\end{figure}
\end{center}

The bias correction made to the ${\rm \tau(250\mu m)}$ values is in itself small (see
Fig.~\ref{fig:hist_bias}) and, consequently, the errors made in that correction are expected to be small. The correction was derived from radiative transfer models with
dust properties corresponding to $R_{\rm V}$=5.5 \citep{Draine2003}. If the submillimetre
dust emissivity is in fact higher, the dust column density will be overestimated and
models will also overestimate the cloud opacity at visual and NIR wavelengths, thus
exhibiting stronger temperature variations than the real clouds. Because the ${\rm
\tau(250\mu m)}$ bias is related to the line-of-sight temperature variations, we could in
this case systematically overestimate the bias in ${\rm \tau(250\mu m)}$. To check the
magnitude of the effect, we repeated the modelling using a dust model with higher long-wavelength emissivity. We used a dust model from \citet{Ossenkopf1994} with coagulated
grains with thin ice mantles accreted in 10$^5$ years at a density of 10$^6$\,cm$^{-3}$.
Compared to NIR (the wavelengths contributing to most of the heating deep inside a
cloud), this dust model has an emissivity higher by $\sim$50\% at SPIRE wavelengths. The
results are shown in Fig.~\ref{fig:MOD_BIAS}, the open circles corresponding to this
alternative modelling. Because of the smaller estimated bias, these should be below the
$k$ values of our previous analysis (red symbols). The median value of $k$ is 1.58$\times
10^{-3}$ and thus practically unchanged. The strongest change is seen for
G6.03+36.73 (LDN~183), where the estimate has been reduced by more than 20\%. This is a
source with very high column density and thus, in its central parts, a very uncertain
estimate of ${\rm \tau(250\mu m)}$. However, the uncertainty of ${\rm \tau(250\mu m)}$
bias must also be considered in connection with the uncertainty of the ${\tau(J)}$
correction, which could compensate for some of the change (see below).

The ${\tau(J)}$ bias corrections are more significant than the ${\rm \tau(250\mu m)}$
bias corrections. The estimation of the ${\tau(J)}$ bias is, in principle, more reliable because it only
depends on the assumption of the ${\tau(J)}$ structure of the clouds. In our
calculations (see Sect.~\ref{sect:bias_tauJ}), $\tau(J)$ was derived from {\it Herschel}
observations, dividing ${\rm \tau(250\mu m)}$ by the constant factor of $k=$1.5$\times
10^{-3}$ to obtain a template map of $\tau(J)$. There are two possible sources of error.
First, if the targets contain much structure below the 18$\arcsec$ resolution of the
${\rm \tau(250\mu m)}$ maps, we will underestimate the bias and our $k$ estimates will be
too high. We cannot directly estimate this error, but it is expected to be a small
fraction of the total bias estimate. This is because $18\arcsec \ll 3\arcmin$ and, at
least for the closest fields, {\it Herschel} already resolves most of the
cloud structure. 
The second potential source of error is again connected with dust emissivity. If the
local ratio between ${\rm \tau(250\mu m)}$ and ${\tau(J)}$ is higher than 1.5$\times
10^{-3}$ (strongly increased submillimetre opacity), we have
overestimated the cloud opacity at NIR wavelengths in the modelling, the bias correction of ${\tau(J)}$ is
too large, and we underestimate the true value of $k$. Thus, a change in the value of $k$
that is used to estimate the ${\tau(J)}$ bias will change the recovered value of $k$ in
the same direction. To examine this potential problem more quantitatively, we repeated
the bias estimation using $k$ values of 1.0$\times 10^{-3}$ and 3.0$\times 10^{-3}$. The
resulting values of $k$ and $C$ parameters are shown in Fig.~\ref{fig:MOD_BIAS} as
triangles. The initial assumption of $k=$1.0$\times 10^{-3}$ leads to a recovered median
value of $k=$1.26$\times 10^{-3}$. The initial assumption of $k=$3.0$\times 10^{-3}$
leads to a recovered median value of $k=$1.92$\times 10^{-3}$. In both cases the input
and output values are inconsistent, unlike in our previous analysis,
where an assumption of 1.5$\times 10^{-3}$ led to a recovered value of 1.6$\times
10^{-3}$. Furthermore, an error in the assumed value of $k$ leads to a systematic error
in the recovered value that is about half of the original error, even lower if the true value
of $k$ was initially overestimated. 

The previous test shows that the estimates of $k$ will be biased towards the selected
value of 1.5$\times 10^{-3}$. Our previously recovered median value of 1.6$\times
10^{-3}$ is thus not significantly affected (bias lower than 0.05), but the effect can be
stronger for individual fields. For example, in G4.18+35.79 the estimate was 1.8$\times
10^{-3}$ , but the true value is probably higher by $\sim$10\%. The calculations could
be iterated, field by field, to carry out the bias correction self-consistently with the
final $k$ estimate. However, the errors are typically below 10\%, and rough estimates of
their magnitude can be seen in Fig.~\ref{fig:MOD_BIAS}.

There is a specific consequence of the way the $\tau(250\mu {\rm m})$ and ${\tau(J)}$
bias corrections are implemented. If a cloud included regions of such a high opacity that
no 2MASS stars were visible through the cloud, the ratio of $\tau(250\mu {\rm m}$ and
$\tau(J)$ would normally be overestimated. The resulting apparent increase of
submillimetre opacity could thus be an artefact resulting from errors in extinction
values. However, in our analysis we also correct ${\tau(J)}$ in this case based on
the assumed opacity derived using the $\tau(250\mu {\rm m})$ input map and the extinction
calculated with simulated 2MASS stars. If the gap in the distribution of background stars
is increased, the ratio $\tau(250\mu {\rm m})/\tau(J)$ does not continue to increase, but
instead tends towards the assumed ratio, 1.5$\times 10^{-3}$. Higher values should thus
not be the result of gaps in extinction data.

To investigate the potential effects of a spatially varying spectral index, we repeated
the analysis using an ad hoc $\beta(T)$ law to introduce $\beta$ variations in all of our
maps. We took the temperatures calculated with $\beta=2.0$ and fixed new $\beta$ values
pixel by pixel using a functional dependence $\beta=2.0 \times (T/15.0)^{-0.24}$. We then
repeated the full analysis, starting with the zero point and colour corrections and
continuing with the calculation of colour temperatures and submillimetre opacity. We did
not solve the ($T$, $\beta$) values (which are very susceptible to noise effects), but
simply assumed that $\beta$ could vary in a systematic way so that the
values are higher when the dust temperature is lower. The parameters of the $\beta(T)$
formula were selected so that $\beta$ changes from $\sim 1.8$ in warm regions with $T\sim
23$\,K to $\sim 2.2$ in the coldest spots $T\sim 10$\,K \citep[cf][]{Dupac2003,
Desert2008, planck2011-7.7b, Paradis2010, Veneziani2010, Juvela2011}. The average $\beta$
is still close to the original $\beta=2.0,$ and we mainly examined the effects of
correlated changes of $\beta$ rather than the effects of absolute $\beta$ values that can
be estimated more directly. 
The crosses in Fig.~\ref{fig:MOD_BIAS} show the slopes $k=\Delta \tau(250\mu {\rm m}/
\Delta\tau(J)$ and the parameters $C$ obtained with the spatially varying $\beta$. The
values are typically within $\sim$10\% of the values obtained with $\beta=2.0$. The
strongest changes of $k$ are seen in the two closest fields, G4.18+35.79 and G6.03+3673,
which have well-resolved clumps with very low temperatures. The increase of the slope
values is $\sim$20\%. Note that this is mostly consistent with the general dependence
between $\tau(250\mu {\rm m})$ and $\beta$ and not necessarily an effect of the spatial
variation of $\beta$. In this respect, it is very interesting to note that the effect on
the parameter $C$ is weak. In other words, if the variations of $\beta$ are as assumed
above, this will be reflected in the slope $\tau(250\mu {\rm m})/\tau(J),$ but without a
noticeable non-linearity in the $\tau(250\mu {\rm m})$ vs. $\tau(J)$ relation.

All the above suggests an uncertainty of 10--20\% in $k$ and $\sim$0.1 units in $C$. In
particular, if the increase of dust opacity is associated with values $\beta>2.0$, our
highest $k$ estimates could still systematically underestimate the true values of $k$
because of the lower assumed value of $\beta$ and because the $\tau(J)$ correction biases
the $k$ values towards 1.5$\times 10^{-3}$.

\section{Discussion}  \label{sect:discussion}

We have examined the submillimetre opacity by correlating the 250\,$\mu$m optical depth
${\rm \tau(250\mu m)}$ with the near-infrared optical depth, $\tau(J)$, assuming the
latter to be an independent tracer of the total dust column density. 
Because the comparison was made at a resolution of 2$\arcmin$ , it is not sensitive to
dense cores ($A_{\rm V}>>10$\,mag), at which both the ${\rm \tau(250\mu m)}$ and $\tau(J)$
estimates would become very uncertain. Nevertheless, corrections for systematic bias in
${\rm \tau(250\mu m)}$ and especially in $\tau(J)$ are important.

The sample consists of the heterogeneous set of 116 Galactic fields that were mapped
with {\it Herschel} as part of the {\em Galactic Cold Cores} project. The main
objectives were to estimate the typical ratio of ${\rm \tau(250\mu m)/\tau(J)}$ and to
search for variations of this quantity, between the fields and as function of column
density. Such variations were then related to differences in the properties of
interstellar dust grains. For the present sample, high column densities also imply low
dust temperatures and thus conditions where submillimetre dust opacity is expected to be
enhanced by grain aggregation. 
The limited resolution means that we did not probe the full range of opacity
variations, if these are partly limited inside compact cores.

\subsection{Main results and their reliability} \label{sect:summary}

By restricting the analysis to $\sim$20 fields for which the results appeared most reliable, we
derived a median value of $k={\rm \tau(250\mu m)/\tau(J) = 1.6 \times 10^{-3}}$ (see
Sect.~\ref{sect:individual} and Fig.~\ref{fig:indipara}). In Fig.~\ref{fig:indipara},
50\% of the most nearby fields had positive values of $C$, the multiplier of the second-order term, indicating some degree of positive correlation between column density and
submillimetre emissivity measured by $\tau(250\mu {\rm m})$. In the maps of the ratio
$k$, the same tendency was very clear in only six cases. The low percentage is partly
caused by the noise in ${\tau(J)}$, whose magnitude is strongly correlated with
the cloud distances. For two of the best examples, G4.18+35.79 and G6.03+36.73, the ratio
${\rm \tau(250\mu m)/\tau(J)}$ increases to $\sim 4\times 10^{-3}$, almost a factor of
three higher than the median value. The peak values are uncertain because of the large bias
corrections, but because of the low spatial resolution used, the strongest effect can be
even greater.

No dependence on either Galactocentric distance or on Galactic height was observed
(Fig.~\ref{fig:vs_distance}). The reliability of the estimates decreases with distance,
but nevertheless, our sample extends over more than $\sim$4\,kpc in Galactocentric
distance.  No trends are seen; if they were present at 10\% level, they should still be visible
over the scatter of individual data points. The result might be affected by systematic
errors in the bias correction. However, these probably depend either on the distance or
on the morphology of the field and at first approximation are
not expected to be different in the
inner and in the outer Galaxy.

In Fig.~\ref{fig:indipara}, the only clear trend is the decrease of $k$ as a function
of distance (also visible as larger scatter around the Galactocentric distance of
8.5\,kpc). As mentioned in Sects.~\ref{sect:bias_tauJ} and~\ref{sect:bias_tau250},
there are two possible explanations. First, the bias correction of $\tau(J)$ might be
overestimated so that as the distance increases, the error increases and $k$ becomes
underestimated.  This is probably not the main reason because the bias correction
depends as much on column density values and column density gradients as on distance.
The trend probably is a combination of selection and resolution effects. At
1\,kpc the 180$\arcsec$ resolution corresponds to almost 1\,pc in linear scale.
Therefore, we measure the mean cloud properties for the most distant fields.
In the nearby fields individual clumps are resolved and the slopes $k$ reflect more
the contrast between diffuse regions and compact, core-sized objects. Thus, the trend
would be compatible with the hypothesis that $\tau(250\mu {\rm m})/\tau(J)$ increases
in the densest and coldest regions of interstellar clouds.

The global non-linear fits of Fig.~\ref{fig:MCMC_TLS} also indicated an increase of
the ratio $k=\tau(250\mu {\rm m})/\tau(J)$  as the function of column density. The plots
show clear deviations from a linear dependence beyond $\tau(J)\sim 4$. For $\tau(J)\sim
10$, $k$ is already twice as high at low column densities. The results are dependent
on a few fields with the highest column densities for which the bias
corrections reach about ten per cent. However, if the distance
dependence of
Fig.~\ref{fig:vs_distance} means that the bias correction of ${\tau(J)}$ is probably overestimated and not underestimated, the true values of $k$ might be even
higher.

Figure~\ref{fig:indipara} concentrated on selected fields for
which linear and non-linear
fits were more reliable than on average. Within this subset, no clear distance
dependence was visible in $k$ values, but the scatter is larger than the estimated
uncertainties. The linear slopes are sensitive to the highest column densities within a
field. The parameter $C$ of the non-linear fits is expected to
be even more sensitive to these
data points, which probe the variations of $\tau(250\mu m)/\tau(J)$ inside the fields. For
the whole sample, the median value of $C$ was very close to zero.
Nevertheless, it may be significant that when the two fields at kiloparsec
distances are excluded, the value of $C$ appears to decrease somewhat systematically with the
distance of the field. There is some preference for the most nearby fields to have positive values; here, the
densest parts of the clouds are resolved. 

Two of the first fields with high $k$ values are G6.03+36.73 and G4.18+35.79, better
known as LDN~183 and LDN~134. At the {\it Herschel} scale, the clouds have a simple
morphology, each consisting of a single column density maximum that is partially resolved
by the 180$\arcsec$ beam. This is illustrated by Figs.~\ref{fig:ratio_map} and
\ref{fig:ratiomaps}. The ratio $k=\tau(250\mu {\rm m})/\tau(J)$ closely follows the
morphology of the column density distribution, making these the best examples of clumps
with increased submillimetre opacity. The highest values are $\sim 4\times 10^{-3}$,
almost three times the average value over all fields.

In our results, some of the main sources of uncertainty are the corrections made for
the expected bias in $\tau(J)$ and, to lesser extent, in ${\rm \tau(250\mu m)}$.  
According to Fig.~\ref{fig:indipara}, the net effect of all bias corrections is a
$\sim$20\% decrease in the value of $k$. This number applies to nearby fields but is
more dramatic if all fields are taken into account. Figure~\ref{fig:vs_distance} shows
that for fields at $\sim$1\,kpc distance the correction is almost a factor of two. The
corrections have been remarkably successful in decreasing the scatter of $\tau(250\mu
m)/\tau(J)$ values, which strongly suggests that they are approximately of the
correct magnitude. 

In the statistical sense, estimating the ${\tau(J)}$ bias is straightforward,
using a model of the Galactic stellar distribution and the higher resolution {\it
Herschel} data as a template for the column density structure of the field (see
Appendix~\ref{sect:NIR_simu} and Sect.~\ref{sect:biasbias}).
The correction is large, but {\em \textup{on average}}, the correction itself probably does not suffer
from major systematic errors. Errors could arise either from incorrect distance estimates
or from the use of an incorrect model of the NIR opacity distribution. The distances
are uncertain, and through the ${\rm \tau(J)}$ bias, their effect on the main parameters is
shown in Fig.~\ref{fig:indipara} (the grey bands) and in Fig.~\ref{fig:indi_1}. For a
sample of fields, this is mainly a statistical and not a systematic error. 

The model of $\tau(J)$ distribution in each field was derived from {\it Herschel} data
at 18$\arcsec$ resolution. If there is still significant structure below this scale,
our correction of $\tau(J)$ values will be too low. The difference of the
2-3$\arcmin$ scale and the 18$\arcsec$ scale is so large, however, that most of the
effects of column density variations are already included. Thus, after the distance,
the other main error in the $\tau(J)$ bias correction arises from scaling the {\it
Herschel} estimates of ${\rm \tau(250\mu m)}$ into a template of the J-band opacity. We
showed in Sect.~\ref{sect:biasbias} that this amounts to an uncertainty
of $\sim$10\% in the
final listed values of $k=\Delta \tau(250\mu {\rm m})/ \Delta \tau(J)$. For the sample
in Fig.~\ref{fig:indipara}, the value of $k$ assumed during the bias correction is
consistent with the recovered value (to be precise, the assumed value of $k=1.5\times 10^{-3}$
, the recovered value of $k=1.6\times 10^{-3}$). For individual fields, the values are
biased towards the initial assumed value. Figure~\ref{fig:MOD_BIAS} showed that if bias
correction assumed a value of $k=1.0\times 10^{-3}$, the recovered median value was
still higher than $k=1.2\times 10^{-3}$. This shows that $k$ cannot be significantly
overestimated because of an erroneous bias correction in $\tau(J)$. Thus, the result
that the average ratio $k=\Delta \tau(250\mu m)/ \Delta \tau(J)$ is clearly higher than the
normal value found in diffuse medium remains robust.

Because of the bias in the $\tau(J)$ correction, our estimates are probably conservative
for fields where values $k>1.5\times 10^{-3}$ were obtained. The dependence on the
assumed value of $k$ decreases as the true value of $k$ increases. This is because
higher $k$ means a lower NIR opacity and lower overall bias in $\tau(J)$. Nevertheless,
for fields like G4.18+35.79, we might systematically underestimate $k$ by
$\sim$10\% because of this error in the $\tau(J)$ bias correction. At the highest column
densities {\it Herschel} emission data themselves will underestimate the cloud opacity, which leads to the corrections discussed in\
Sect. \ref{sect:bias_tau250}. For the optical
depth ratio, the effect of ${\rm \tau(250\mu m)}$ bias is weaker than that of
$\tau(J)$. Nevertheless, the errors made in the corrections of ${\rm \tau(250\mu m)}$ and $\tau(J)$
 (for example, those associated with a change of submillimetre opacity) would
partly cancel each other out.

In the analysis we assumed that apart from the problems associated with the
sampling provided by the background stars, NIR reddening is an independent and reliable
measure of column density. Unlike in the optical range, the NIR extinction curve is
often assumed to be constant for a wide range of column densities \citep{Cardelli1989,
MartinWhittet1990, Roy2013}. Nevertheless, some cloud-to-cloud variations are observed,
the ratio $E_{J-H}/E_{H-K}$ ranging from values lower than 1.5 to higher than 2.0 \citep{Racca2002,
Draine2003ARAA}. Clear changes take place at high optical depths, above $A_{\rm V} \sim
20$\,mag, but only in the form of the flattening of the MIR extinction curve, at
wavelengths above 3\,$\mu$m \citep{Indebetouw2005, Cambresy2011, Ascenso2013}. Recently,
\citet{Whittet2013} studied cloud LDN~183 (which is also included in our sample).
Comparison with the 9.7\,$\mu$m silicate absorption feature suggested that the NIR
colour excess might not be a perfect tracer of the total dust column. The ratio between
$E_{J-K}$ and the 9.7\,$\mu$m feature was observed to increase as the column density
exceeded $A_{\rm V} \sim$20\,mag. The observation might partly
arise because the
9.7\,$\mu$m feature is dampened by the formation of ice mantles \citep[see
also][]{Chiar2007}. However, if we assume that NIR extinction indeed \textup{{\em
\textup{overestimates}}} the total dust column in this object, the ratio of ${\rm \tau(250\mu
m)}$ relative to column density would increase by $\sim$20\% \citep[see][Fig.
12]{Whittet2013}. This is still a weaker effect than the increase by more than a
factor of two that was observed in $\tau(250\mu {\rm m})/\tau(J)$. We recall
that at high column densities, above $A_{\rm V}\sim 20$\,mag, the values of ${\rm
\tau(250\mu m)}$ are also uncertain and can be underestimated by a significant fraction
(\citet{Pagani2004} discussed a similar limitation at the nearby wavelength of
200\,$\mu$m).

Finally, we recall that $\tau(250\mu{\rm m})$ values were derived using a fixed value of
the spectral index, $\beta=2.0$. By assuming $\beta=1.8$ instead, the $\tau(250\mu{\rm
m})$ and $k$ values would decrease by up to 30\%. Conversely, if the value of $\beta$
increased towards dense and cold regions \citep{Dupac2003, Desert2008, planck2011-7.7b,
Paradis2010, Veneziani2010, Juvela2011}, we underestimate the
dust opacity changes if we use a constant value of $\beta$ . In this sense (and regarding possible bias in
${\tau(J)}$ corrections), 
Figs.~\ref{fig:MCMC_TLS}--8 give conservative estimates of the
possible increase of submillimetre dust opacity. Clearly, the assumed
values of $\beta$ must also be taken into account when comparing our
results with other studies (see Sect. \ref{sect:comparison}). On the
other hand, our tests indicate that spatial variations of $\beta$
probably do not have a strong additional effect on $k$ (see
Sect.~\ref{sect:biasbias}).

\subsection{Comparison with other studies} \label{sect:comparison}

The submillimetre dust opacity has previously been studied in relation to both dust
extinction \citep{Terebey2009, Flagey2009, Juvela2011, Martin2012, Roy2013,
Malinen2013, Malinen2014} and to H{\sc I} \citep{Boulanger1996, Lagache1999,
planck2011-7.12, planck2011-7.13, Martin2012}. \citet{Boulanger1996} compared COBE
observations of FIR dust emission with the hydrogen 21\,cm observations. At high
latitudes, where H{\sc I} is a good tracer of the full gas column density, the derived
value was $\tau/N_H = 1.0 \times 10^{-25} \, (\lambda/250\mu{\rm
m})$\,cm$^{2}$\,H$^{-1}$. Similar values
were obtained in the first studies using {\it Planck} data \citep{planck2011-7.12}.
In the most recent {\it Planck} papers, somewhat lower values were reported, corresponding to
$\tau(250\mu {\rm m})/N_{\rm H} \sim 0.55 \times 10^{-25}$\,cm$^{2}$\,H$^{-1}$
\citep{planck2013-p06b, planck2013-XVII}. The change is associated with the revised
calibration of the highest frequency channels and, correspondingly, a lower value of
$\beta$. In \citet{planck2013-XVII} the spectral index above 353\,GHz was found to be
$\beta \sim 1.65$, lower than the value of 1.8 assumed in \citet{planck2011-7.12} or
the value of 2.0 used in \citet{Boulanger1996}. The higher dust opacity value of
\citet{Boulanger1996} is thus largely explained by the higher value of $\beta$.

The {\it Planck} result for the diffuse medium can be compared to our result most directly
by converting their $N_{\rm H}$ to $\tau(J)$. We can use the conversion factor $N({\rm
H}_2)/A_{\rm V}= 9.4 \times 10^{20}$\,cm$^{-2}$\,mag$^{-1}$ derived by \citet{BSD} at low
extinction, $E(B-V)$<0.5 \citep[see also][]{Nozawa2013}. Some studies \citep{Rachford2009,
planck2013-p06b, Liszt2014a} have found $N_{\rm H}/E(B-V)$ values that are 10--30\% higher
than in \citet{BSD}, these results apply partly to even more diffuse lines of sight
($E(B-V)\la 0.1$). On the other hand, \citet{Gudennavar2012} examined a sample of lines of
sight with $E(B-V)$ extending to values higher than one. The result, $N({\rm
H})/E(B-V)=(6.1\pm 0.2)\times 10^{21}$\,H\,cm$^{-2}$\,mag$^{-1}$, was close to that of
\citet{BSD}. With the \citet{BSD} relation and $R_{\rm V}=3.1$ extinction curve
\citep{Cardelli1989}, the {\it Planck} result $\tau(250\mu {\rm m})/N_{\rm H} \sim 0.55
\times 10^{-25}$\,cm$^{2}$\,H$^{-1}$ \citet{planck2013-p06b} corresponds to ${\tau(250\mu {\rm
m})/\tau(J)}=0.41\times 10^{-3}$. Our median value of ${\tau(250\mu {\rm
m})/\tau(J)}=1.6\times 10^{-3}$ is thus 3.9 times higher, and our highest local values,
4.5$\times 10^{-3}$ in G4.18+35.79, are higher by one order of magnitude. In
\citet{planck2013-p06b} the estimated value of $N({\rm H}/E(B-V))$ was $\sim$20\% higher
than in \citet{BSD}. With this value, our median value of ${\tau(250\mu {\rm m})/\tau(J)}$
would be $\sim 3.3$ times higher than the {\it Planck} estimate.

We can alternatively convert our result into $\tau(250\mu {\rm m})/N_{\rm H}$ , but in
dense regions the shape of the extinction curve (dependence on $R_{\rm V}$) and
the ratio between visual extinction and total column density are more uncertain. Using
the $R_{\rm V}=3.1$ and $N({\rm H}_2)/A_{\rm V}$ from \citet{BSD}, our median value
corresponds to $\tau(250\mu {\rm m})/N_{\rm H} = 2.16 \times
10^{-25}$\,cm$^{2}$\,H$^{-1}$ (again, of course, 3.9 times the {\em Planck} value).
However, $R_{\rm V}$ is expected to be higher than 3.1 in dense clouds. Using $R_{\rm
V}=5.5$ instead of $R_{\rm V}=3.1$ in converting $\tau(J)$ into visual extinction, the
values of $\tau(250\mu {\rm m})/N_{\rm H}$ would decrease by $\sim 15$\% (change in
$A_{\rm V}/E(J-K)$). However, the scaling we used above (corresponding to the $R_{\rm
V}$=3.1 extinction curve and the ratio of $N({\rm H}_2)/A_{\rm V}$ taken from
\citet{BSD}) corresponds to $N({\rm H})/E(J-K)=11.0\times
10^{21}$\,cm$^{-2}$\,mag$^{-1}$, which is very close to the value of $11.5\times
10^{21}$\,cm$^{-2}$\,mag$^{-1}$ that \citet{Martin2012} derived in Vela molecular
cloud using 2MASS NIR data up to $E(J-K)\sim0.55$\,mag ($E(B-V)\sim 1.1$\,mag).

When $R_{\rm V}$ is modified, the NIR extinction curve remains practically unchanged
and the differences take place between the optical and NIR wavelengths. Compared to shorter (optical) and longer (MIR) wavelengths, the extinction curve is
considered relatively constant in the NIR regime \citep[e.g.][]{Indebetouw2005,
Lombardi2006_Pipe, RomanZuniga2007, Ascenso2013, Wang2014} and the power-law slope of the
NIR extinction curve typically only varies at a level of $\sim$5\%
\citep{Stead2009, Fritz2011}. In very dense cores the formation of ice mantles and the
grain growth could have an additional impact. \citet{RomanZuniga2007} examined the
cloud Barnard 59 up to $A_{\rm V}=59$\,mag, but found no significant changes in the NIR
extinction law (compatible with $R_{\rm V}=5.5$). Similarly, \citet{Lombardi2006_Pipe}
found no changes in the Pipe nebula in the reddening NIR law up to $E(H-K)\sim
1.5$\,mag ($A_{\rm V}\sim 9$\,mag). Therefore, it is not likely that our results are significantly biased because of the assumed NIR extinction curve. At the resolution
of our extinction maps, practically all our data points are lower
than$A_{\rm V} \sim
10$\,mag and are thus not affected by extreme optical depths. However,
\citet{Whittet2013} observed a decrease in L183 in the ratio of $E(J-K)$ and the
9.7\,$\mu$m silicate absorption feature. Above $E(J-K)\sim 1.0$ ($A_{\rm J} \sim
1.7$\,mag, $A_{\rm V} \sim 5.0$\,mag for $R_{\rm V}=5.0$), the ratio deviated from
diffuse medium value by $\sim 20$\%. Stronger deviations were
only seen beyond
$E(J-K)\sim 3$ ($A_{\rm J} \sim 5$\,mag, $A_{\rm V} \sim 15$\,mag for $R_{\rm
V}=5.0$). This is above the range probed by our measurements. It is not clear that
changes in this ratio would only be caused by the NIR extinction curve. However, if,
in some sources, the extinction curve does flatten above $A(J)\sim 2$, this might
contribute to the observed increase of $\tau(250\mu{\rm m})/\tau(J)$ (possibly at the
20\% level).

The effect of the bias corrections on $\tau(250\mu{\rm m})/\tau(J)$ is typically
$\sim$20\% or weaker, and their uncertainty is smaller. The largest uncertainties in
$\tau(250\mu {\rm m})/N_{\rm H}$ are caused by $\beta$ and possibly by $R_{\rm V}$.
With $\beta=1.8$ and $R_{\rm V}$=5.5, the $\tau(250\mu{\rm m})/\tau(J)$ values would
be 40\% lower than with $\beta=2.0$ and $R_{\rm V}$=3.1. At this lower limit, our
median value of $\tau(250\mu{\rm m})/\tau(J)$ would not be 3.9 times, but $\sim$2.3
times the value found in the diffuse high-latitude sky. The 40\% uncertainty may be a
realistic 1 $\sigma$ lower limit for the linear fits that include all pixels in the
maps. However, it is a conservative estimate for the clumps where the average value of
$\beta$ is expected to be clearly higher than 1.8. In fact, preliminary results
indicate that the average value of $\beta$ in our fields (including the more diffuse
regions) is close to $\beta=1.9,$ and that with $\beta=2.0$ we underestimate the
$\tau(250\mu {\rm m})$ values of many clumps \citep{Juvela2014_TB}.

Increased far-infrared and submillimetre opacity has been reported by many authors
\citep[][etc.]{Kramer2003, Lehtinen2004, delBurgo2005, Ridderstad2010, Bernard2010,
Suutarinen2013}. One famous example is the Taurus filament L1506, for which the models of
\citet{Stepnik2003} suggested an increase by more than a factor of three. With the recent
detailed modelling of the dust properties and the structure of this filament,
\citet{Ysard2013} estimated the increase of the 250\,$\mu$m opacity to be $\sim$2.
\citet{Martin2012} studied the relation in the Vela cloud, comparing BLAST and IRAS data with
the reddening of 2MASS stars. The properties of the examined areas corresponded to the
average properties of our fields, with column densities extending to $10^{22}$\,cm$^{-2}$
and with typical dust temperatures of $\sim 15\,K$. They found a very similar range of
dust opacities, $\tau(250\mu m)/N_{\rm H} = (2-4)\times 10^{-25}$\,cm$^2$\,H$^{-1}$
{(assuming $\beta=1.8$)}. 
In the Orion A cloud, a comparison of {\it Herschel} and 2MASS data led to the detection of
a dependence of $N^{0.28}$ on the 250\,$\mu$m opacity \citep{Roy2013}. The range of column
densities in Orion A was similar to our fields, and the derived dust opacities were mainly in
the range of $\tau(250\mu m)/N_{\rm H} = (1-3)\times 10^{-25}$\,cm$^2$\,H$^{-1}$. These
estimates were derived with $\beta=1.8$ and would become $\sim 20$\% higher (depending on
the details of the fitting) if a value of $\beta=2.0$ were used. 
More recently, \citet{Lombardi2014} derived ratios $A({\rm K})/\tau(850\mu m)$ of
2640\,mag and 3460\,mag for the Orion A and B molecular clouds, respectively. The 850\,$\mu$m
optical depth was derived from {\em Herschel} observations rescaled using a comparison
with {\em Planck} measurements, and the NIR extinction was calculated with the method NICEST
\citep{Lombardi2009}. The modified blackbody fits used $\beta$ values that were estimated
in \citet{planck2013-p06b} at a resolution of 30$\arcmin$. With the reported average value
of $\beta=1.8$ and adopting a ratio 0.40 between $A({\rm K})$ and $A({\rm J})$, the
results correspond to ${\rm \tau(250\mu m)/\tau(J)}$ values of $1.56\times 10^{-3}$ and
$1.19\times 10^{-3}$ for Orion A and B, respectively. These fits were made for data ${\rm
\tau(850\mu m)}< 2\times 10^{-4}$ (${\rm \tau(250\mu m)} \la 1.8\times 10^{-3}$).  The
optical depth range is similar to many of our fields, and the result for Orion A is close
to our median value for the fits concerning entire fields. By adopting $\beta=1.8$, our
median value would fall between the Orion A and Orion B estimates of
\citet{Lombardi2014}.

\subsection{Implications for dust evolution}

As shown by simulations, an observed increase of the dust far-IR/submm opacity towards
dense regions cannot be due to radiative transfer effects, but must originate in
intrinsic variations of dust properties \citep{Malinen2011, JuvelaYsard2012,
Ysard2012}. Theoretical studies have shown that such an increase, coupled with a
decrease in dust temperature, can be explained by the formation of large aggregate
particles \citep{Ossenkopf1994, Stognienko1995, Ormel2011, Kohler2011, Kohler2012}.
\citet{Kohler2012} showed that the coagulation of just four big grains of the diffuse
ISM type, coated by smaller carbon grains, already leads to an increase in the opacity
at 250\,$\mu$m of a factor 2.6. These authors also showed that these aggregates can
form within a typical cloud lifetime of 10 million years \citep{Walmsley1991}.
Consequently, we interpret the observed increase in ${\rm \tau(250\mu m)}$ in our cold
core sample as grain growth in dense molecular regions. 

In our sample, the clouds LDN~183 and LDN~134 (G6.03+36.73 and G4.18+35.79) were the
most convincing examples of increased submillimetre dust opacity. LDN~183 has been
studied thoroughly in both continuum and line emission \citep[e.g.][]{Juvela2002,
Pagani2005, Pagani2007}. A slight increase of 200\,$\mu$m opacity was already reported
based on ISOPHOT observations \citep{Juvela2002}, but ISOPHOT data $\lambda \le 200\mu$m
and, to some extent even {\it Herschel} observations, are not sufficient to fully probe
the inner parts of the cloud where the high visual extinction is approaching 100\,mag
and the dust temperature drops well below 10\,K \citep{Pagani2004, Pagani2014}. LDN~183
was the first object where enhanced mid-infrared (MIR) light scattering, the so-called
coreshine phenomenon, was detected in Spitzer data \citep{Steinacker2010, Pagani2010}.
The effect was also seen in \citet{GCC-III}, where WISE MIR observations were analysed
and both LDN~183 and LDN~134 were found to be sources of strong MIR emission. If the
signal is interpreted as scattering of the interstellar radiation field, it seems to
imply the presence of very large, micrometre-sized dust particles \citep{Steinacker2010,
Steinacker2014, Lefevre2014}. Thus, our detection of increased submillimetre opacity in
these clouds agrees with the evidence provided by MIR wavelengths.

\section{Conclusions}  \label{sect:conclusions}

We have examined dust optical depths by comparing measurements of submillimetre dust
emission and the reddening of the light of background stars in the near-infrared. The
goal was to measure the value of dust submillimetre opacity and to search for variations
that might be correlated with the physical state and the environment of the cloud. The
study led to the following conclusions:
\begin{itemize}
\item For a subsample of 23 fields with well-defined correlation between the 
two variables, we obtained a median ratio of $\tau(250\mu {\rm m})/\tau(J) = (1.6 \pm 0.2)
\times 10^{-3}$. This is more than three times the value that was derived from {\it
Planck} data for the diffuse medium at high Galactic latitudes. Assuming $\beta=1.8$
instead of $\beta=2.0$, the value decreases by 30\%, but is still more than twice the
diffuse value.
\item The conversion to $\tau(250\mu {\rm m})/N_{\rm H}$ involves more assumptions.
Using the $R_{\rm V}=3.1$ extinction curve and the $N({\rm H}_2)/A_{\rm V}$ ratio of
\citet{BSD}, our median estimate corresponds to $\tau(250\mu {\rm m})/N_{\rm H} = 2.16
\times 10^{-25}$\,cm$^{2}$\,H$^{-1}$.
\item 
The fit to all data above $\tau(J)=1$ gives a relation $\tau(250\mu {\rm m}) \sim
1.25 \times 10^{-3} \, \tau(J) + 0.11 \times 10^{-3} \, \tau(J)^2$. The positive
second-order coefficient $C=0.11 \times 10^{-3}$ is determined by a small number of
fields that, because of the high column density, are subject to large uncertainty in
the bias corrections.
\item
For the same sample, the scatter in the coefficients of the second-order terms $C$ is
$\sim 2 \times 10^{-4}$ and the median value is consistent with zero. Spatial
variations of the ratio $\tau(250\mu {\rm m}) / \tau(J)$ are only seen in a few
fields.
\item
From the maps of $\tau(250\mu{\rm m})/\tau(J)$, we identified six fields
where the ratio appears to increase further at the location of the main column
density peaks. The highest values in the fields G4.18+35.79 and G6.03+36.73 are $\sim
4 \times 10^{-3}$ at the resolution of 180$\arcsec$. Thus, although the densest
clumps are associated with the largest uncertainties, we consider this increase of
submillimetre opacity to be real. 
\end{itemize}

\begin{acknowledgements}
This publication makes use of data products from the Two Micron All Sky Survey, which
is a joint project of the University of Massachusetts and the Infrared Processing and
Analysis Center/California Institute of Technology, funded by the National Aeronautics
and Space Administration and the National Science Foundation.
MJ, JMo, VMP, and JMa acknowledge the support of the Academy of Finland Grant No.
250741. 
\end{acknowledgements}

\bibliography{biblio_with_Planck}

\clearpage

\onecolumn

\begin{landscape}

\begin{longtable}{lcccccccc}
\caption{
Coordinates and sizes of the {\it Herschel} fields. The last columns
give the position and radius of reference regions used in case of
local background subtraction.}\label{table:fields} \\
\hline \hline
Field &  \multicolumn{2}{c}{Centre coordinates} &  Area           &  Distance &  \multicolumn{3}{c}{Reference region} 
&   Aliases \\
&   RA(2000.0) & DEC(2000.0)             &  $(\arcmin)^2$ &  (kpc)     &  RA(2000) & DEC(2000) & $r$ ($\arcmin$) 
&    \\
\hline 
\endfirsthead
\caption{continued.} \\ 
\hline \hline
Field &  \multicolumn{2}{c}{Centre coordinates} &  Area           &  Distance &  \multicolumn{3}{c}{Reference region}  &
Aliases \\
&   RA(2000.0) & DEC(2000.0)             &  $(\arcmin)^2$ &  (kpc)     &  RA(2000) & DEC(2000) & $r$ ($\arcmin$)  &         \\
\hline      
\endhead      
\hline 
\endfoot

G0.02+18.02 &    16 40 56.7 & -18 34 59.9 & 1332   & 0.16(0.05)   &   16 40 27.6 & -18 32 16.8   & 2.5 &   - \\
G0.49+11.38 &    17 04 41.9 & -22 13 51.1 & 1775   & 0.16(+0.20/-0.16)   &   17 04 21.8 & -22 02 20.4   & 2.8 &   LDN15 \\ 
G1.94+6.07 &    17 28 08.2 & -24 01 53.9 & 2276   & 0.14(0.05)   &   17 28 18.0 & -23 44 56.4   & 3.9 &   LDN69, B77 \\
G2.83+21.91 &    16 34 28.8 & -14 11 27.8 & 2279   & 0.30(0.30)   &   16 34 27.1 & -13 58 58.8   & 4.7 &   LDN83, MBM135 \\
G3.08+9.38 &    17 17 51.1 & -21 26 16.4 & 3424   & 0.16(0.05)   &   17 17 59.3 & -21 44 13.2   & 4.2 &    \\ 
G3.72+21.02 &    16 39 45.9 & -14 02 09.3 & 1776   & 0.16(0.05)   &   16 39 45.4 & -13 51 28.8   & 3.5 &   LDN121 \\
G4.18+35.79 &    15 53 31.4 & -04 37 55.5 & 1331   & 0.11(0.01)   &   15 52 57.4 & -04 32 27.6   & 2.7 &   LDN134, MBM36, MLB40 \\
G6.03+36.73 &    15 54 13.9 & -02 51 40.1 & 1332   & 0.11(0.01)   &   15 54 04.6 & -02 42 28.8   & 2.5 &   LDN183, MBM37 \\
G9.45+18.85 &    17 00 22.2 & -10 51 07.0 & 1777   & 0.28(0.10)   &   17 00 49.7 & -10 44 31.2   & 2.9 &   -  \\
G10.20+2.39 &    17 59 20.6 & -18 56 22.5 & 1331   & 0.83(0.40)   &   17 59 07.9 & -18 49 37.2   & 2.3 &   -  \\
G20.72+7.07 &    18 03 38.9 & -07 30 24.5 & 1331   & 0.26(0.26)   &   18 03 01.2 & -07 33 21.6   & 2.8 &   -  \\
G21.26+12.11 &    17 46 55.7 & -04 36 50.6 & 1331   & 0.12(0.12)   &   17 46 48.2 & -04 26 34.8   & 2.9 &   LDN425, LDN428, LM240 \\ 
G24.40+4.68 &    18 19 21.5 & -05 32 44.7 & 2284   & 0.26(0.05)   &   18 18 17.3 & -05 36 25.2   & 3.6 &   LDN475, LDN477, LDN470 \\
G25.86+6.22 &    18 16 20.7 & -03 24 48.9 & 1331   & 0.26(0.05)   &   18 15 39.4 & -03 26 27.6   & 2.5 &   LDN500 \\
G26.34+8.65 &    18 08 37.9 & -01 51 26.2 & 1331   & 0.40(0.40)   &   18 08 40.3 & -01 42 32.4   & 2.6 &   LDN502, CB112, P61 \\
G37.49+3.03 &    18 48 57.0 & +05 26 08.7 & 1333   & 0.80(0.60)   &   18 48 31.2 & +05 29 13.2   & 2.4 &   BDN31.48+3.02 \\
G37.91+2.18 &    18 52 45.4 & +05 26 01.0 & 1331   & 1.06(0.79)   &   18 52 26.4 & +05 31 51.6   & 2.3 &   - \\
G39.65+1.75 &    18 57 00.1 & +06 47 00.2 & 1774   & 0.99(0.48)   &   18 56 50.6 & +06 41 13.2   & 3.5 &   - \\
G62.16-2.92 &    19 59 47.3 & +24 16 15.0 & 1329   & 1.02(0.35)   &   19 59 29.3 & +24 22 40.8   & 2.7 &   - \\
G69.57-1.74 &    20 13 24.7 & +31 16 01.4 & 1772   & 1.78(0.81)   &   20 13 46.1 & +31 08 09.6   & 2.6 &   - \\ 
G70.10-1.69 &    20 13 58.5 & +31 49 25.7 & 3420   & 2.09(0.83)   &   20 13 39.1 & +32 09 18.0   & 4.3 &   - \\
G71.27-11.32 &    20 53 14.0 & +26 50 56.4 & 1332   &    -         &   20 53 17.0 & +26 58 04.8   & 3.0 &   - \\
G82.65-2.00 &    20 53 11.3 & +41 35 16.7 & 4744   & 1.00(+1.00/-0.60)   &   20 52 35.0 & +41 14 27.6   & 5.5 &   LDN914 \\
G86.97-4.06 &    21 17 20.9 & +43 24 21.6 & 1325   & 0.70(0.10)   &   21 16 53.8 & +43 32 13.2   & 2.3 &   LDN943, LDN944 \\
G89.65-7.02 &    21 38 23.7 & +43 16 23.6 & 3422   & 0.70(+1.00/-0.70)   &   21 37 30.2 & +43 02 24.0   & 4.5 &   B159, LDN977 \\
G91.09-39.46 &    23 10 32.6 & +17 06 32.7 & 1330   & 0.52(0.20)   &   23 10 15.1 & +17 12 28.8   & 2.7 &   -  \\
G92.04+3.93 &    21 03 00.6 & +52 31 46.3 & 2278   & 0.80(0.80)   &   21 01 09.8 & +52 33 50.4   & 3.5 &   LDN1003, ArchG092.03+03.93 \\
G92.63-10.43 &    22 03 05.7 & +42 14 34.9 & 1330   & 0.90(0.90)   &   22 02 48.7 & +42 07 58.8   & 2.3 &   -  \\
G93.21+9.55 &    20 37 00.2 & +56 54 46.7 & 1332   & 0.44(+0.46/-0.14)   &   20 37 35.3 & +56 46 12.0   & 2.8 &   LDN1033 \\
G94.15+6.50 &    20 59 22.8 & +55 35 54.0 & 2279   & 0.80(0.80)   &   20 57 55.4 & +55 37 22.8   & 3.9 &   B357 \\
G95.76+8.17 &    20 57 24.1 & +58 10 19.9 & 2829   & 0.80(0.80)   &   20 55 23.8 & +58 17 06.0   & 3.5 &   LDN1071, B354  \\
G98.00+8.75 &    21 03 55.9 & +59 59 36.1 & 1777   & 1.10(0.20)   &   21 03 31.7 & +60 10 22.8   & 4.1 &   ArchG097.82+08.67 \\
G105.57+10.39 &    21 41 35.6 & +66 33 11.6 & 2829   & 0.90(0.30)   &   21 43 16.8 & +66 25 12.0   & 3.0 &   ArchG105.55+10.45 \\
G107.20+5.52 &    22 21 14.4 & +63 41 46.0 & 3733   & 0.90(0.30)   &   22 24 24.7 & +63 44 09.6   & 6.3 &   PCC249, LDN1204, S140,  S9 \\
G108.28+16.68 &    21 09 45.7 & +72 52 58.3 & 1332   & 0.30(+0.20/-0.15)   &   21 08 25.7 & +72 47 13.2   & 2.5 &   -  \\
G109.18-37.59 &    00 03 50.3 & +24 00 10.8 & 1320   &    -         &   00 04 25.9 & +24 06 32.4   & 4.1 &   -  \\
G109.80+2.70 &    22 53 30.7 & +62 30 57.7 & 1331   & 0.70(0.10)   &   22 53 20.6 & +62 38 24.0   & 2.8 &   PCC288, S8 \\
G110.62-12.49 &    23 37 27.7 & +48 29 40.8 & 2280   & 0.44(0.10)   &   23 37 26.2 & +48 43 19.2   & 5.0 &   -  \\
G110.89-2.78 &    23 18 15.5 & +58 04 01.7 & 2265   & 3.00(1.00)   &   23 17 28.6 & +57 53 52.8   & 4.0 &   -  \\
G110.80+14.16 &    21 59 02.6 & +72 48 56.3 & 2281   & 0.40(0.10)   &   22 02 28.3 & +72 42 39.6   & 3.8 &   -  \\
G111.41-2.95 &    23 22 15.4 & +57 50 00.0 & 2264   & 3.00(1.00)   &   23 20 43.4 & +57 55 15.6   & 3.6 &   ArchG111.11-03.01 \\
G115.93+9.47 &    23 24 05.5 & +71 09 29.0 & 1331   & 1.00(0.50)   &   23 25 43.4 & +71 07 58.8   & 2.8 &   -  \\
G116.08-2.40 &    23 57 06.7 & +59 42 27.3 & 1332   & 0.50(0.50)   &   23 58 29.0 & +59 41 02.4   & 2.9 &   LDN1257, LDN1257B  \\
G126.24-5.52 &    01 15 46.6 & +57 12 37.5 & 1327   & 1.00(0.10)   &   01 16 59.3 & +57 08 52.8   & 2.9 &   -  \\
G126.63+24.55 &    04 23 46.3 & +85 47 06.5 & 1350   & 0.12(0.02)   &   04 28 02.9 & +85 41 20.4   & 3.0 &   S1, LDN1320 \\
G127.79+2.66 &    01 37 49.1 & +65 05 26.6 & 1344   & 0.80(0.10)   &   01 38 32.9 & +65 12 50.4   & 3.2 &   ArchG127.69+02.65 \\
G128.78-69.46 &    00 59 22.1 & -06 44 41.2 & 2819   &    -         &   00 59 50.6 & -06 57 57.6   & 4.7 &   -  \\
G130.37+11.26 &    02 32 07.2 & +72 39 15.1 & 1325   & 0.60(0.10)   &   02 30 49.7 & +72 30 18.0   & 2.4 &   LDN1340 \\
G130.42-47.07 &    01 12 33.7 & +15 28 59.0 & 1328   &    -         &   01 13 10.1 & +15 34 44.4   & 3.2 &   -  \\
G131.65+9.75 &    02 39 25.4 & +70 47 11.4 & 1328   & 1.07(+0.52/-0.49)   &   02 41 39.6 & +70 46 01.2   & 2.3 &   S3  \\
G132.12+8.95 &    02 39 50.5 & +69 48 54.9 & 1762   & 0.85(0.10)   &   02 37 25.0 & +69 52 51.6   & 3.7 &   -  \\ 
G139.60-3.06 &    02 52 31.6 & +55 39 35.5 & 4074   & 0.85(0.10)   &   02 50 13.4 & +55 24 10.8   & 5.4 &   -  \\
G141.25+34.37 &    08 48 17.8 & +72 43 16.1 & 1329   & 0.11(0.01)   &   08 46 18.7 & +72 47 38.4   & 3.3 &   MBM27  \\
G149.67+3.56 &    04 18 07.7 & +55 15 04.8 & 2275   & 0.17(0.05)   &   04 17 25.4 & +55 30 39.6   & 3.6 &   LDN1400, LDN1394 \\
&               &             &        &              &              &               &     &   B8, B9, MBL71 \\
G150.47+3.93 &    04 24 37.9 & +55 02 21.6 & 1776   & 0.17(0.05)   &   04 24 30.2 & +55 14 02.4   & 3.5 &   LDN1399, MLB72, MLB74, \\
&               &             &        &              &              &               &     &   LM17, ArchG150.41+03.91 \\  
G151.45+3.95 &    04 29 53.9 & +54 14 53.0 & 1331   & 0.17(0.05)   &   04 30 15.6 & +54 06 36.0   & 2.6 &   B12, LDN1407, LDN1400F, \\
&               &             &        &              &              &               &     &   MLB77, LM25 \\  
G154.08+5.23 &    04 47 34.4 & +53 05 01.9 & 1329   & 0.17(0.05)   &   04 48 04.8 & +53 13 26.4   & 2.6 &   LDN1426, LM56 \\
G155.80-14.24 &    03 37 09.4 & +37 42 31.8 & 4765   & 0.35(0.10)   &   03 37 35.0 & +38 03 36.0   & 6.2 &   LDN1434  \\
G157.08-8.68 &    04 01 39.8 & +41 14 20.4 & 1806   & 0.15(0.15)   &   04 02 37.7 & +41 17 27.6   & 2.9 &   (LDN1443) \\
G157.92-2.28 &    04 28 51.8 & +45 15 06.3 & 2278   & 2.50(1.00)   &   04 30 20.2 & +45 22 26.4   & 3.9 &   -  \\
G159.23-34.51 &    02 56 02.5 & +19 39 10.2 & 6527   & 0.33(0.05)   &   02 57 25.2 & +19 11 06.0   & 8.0 &   LDN1457, MBM12, ArchG159.17-34.44  \\
G159.12-14.30 &    03 50 36.0 & +35 41 58.0 & 1331   & 0.80(0.80)   &   03 50 21.8 & +35 49 44.4   & 3.1 &   -  \\
G159.34+11.21 &    05 41 17.5 & +52 11 38.8 & 2279   & 0.70(0.70)   &   05 41 30.2 & +51 55 55.2   & 3.6 &   -  \\
G161.55-9.30 &    04 16 11.2 & +37 46 17.5 & 1358   & 0.25(+0.28/-0.25)   &   04 15 50.4 & +37 39 39.6   & 2.4 &   S7  \\
G163.82-8.44 &    04 27 13.7 & +36 46 47.2 & 8385   & 0.42(0.10)   &   04 23 21.8 & +36 27 43.2   & 8.6 &   -  \\
G164.71-5.64 &    04 40 42.8 & +38 09 06.4 & 5002   & 0.33(0.20)   &   04 42 46.1 & +38 20 34.8   & 5.3 &   LDN1481 \\
G167.20-8.69 &    04 36 34.9 & +34 16 50.4 & 2829   & 0.16(0.16)   &   04 35 08.9 & +34 10 44.4   & 4.2 &   -  \\
G168.85-10.19 &    04 37 04.3 & +31 49 17.9 & 1330   & 1.30(1.30)   &   04 36 12.7 & +31 49 12.0   & 2.7 &   -  \\
G171.35-38.28 &    03 18 04.2 & +10 27 01.2 & 1330   &    -         &   03 17 23.3 & +10 25 40.8   & 2.8 &   MBM16  \\
G173.43-5.44 &    05 08 42.0 & +31 22 38.5 & 3424   & 0.15(0.15)   &   05 08 37.0 & +31 02 13.2   & 4.5 &   -  \\
G174.22+2.58 &    05 41 42.4 & +35 11 58.2 & 1331   & 1.80(0.20)   &   05 42 26.2 & +35 14 24.0   & 2.3 &   -  \\
G176.27-2.09 &    05 28 14.3 & +30 57 27.0 & 1356   & 1.57(0.26)   &   05 28 04.3 & +31 06 18.0   & 3.0 &   S6 \\
G181.84-18.46 &    04 43 56.1 & +16 57 22.9 & 1356   & 0.50(0.50)   &   04 43 53.8 & +16 50 09.6   & 2.8 &   -  \\
G188.24-12.97 &    05 17 05.1 & +14 54 32.7 & 3422   & 0.45(0.05)   &   05 15 37.4 & +14 55 48.0   & 4.6 &   -  \\ 
G189.51-10.41 &    05 29 55.3 & +15 27 03.2 & 2831   & 0.45(0.05)   &   05 28 38.6 & +15 25 30.0   & 4.0 &   -  \\
G195.74-2.29 &    06 10 58.1 & +14 09 56.7 & 1330   & 0.60(0.60)   &   06 11 01.2 & +14 02 02.4   & 2.7 &   ArchG195.73-02.39 \\
G198.58-9.10 &    05 52 28.8 & +08 19 33.8 & 2279   & 0.45(0.45)   &   05 51 54.7 & +08 16 58.8   & 3.5 &   LDN1598 \\
G202.23-3.38 &    06 19 33.2 & +08 02 56.7 & 1331   & 3.80(1.00)   &   06 18 49.2 & +08 03 00.0   & 2.7 &   -  \\
G202.02+2.85 &    06 41 07.4 & +10 47 22.8 & 4079   & 0.76(0.10)   &   06 42 42.7 & +10 42 39.6   & 4.3 &   -  \\
G203.42-8.29 &    06 04 35.6 & +04 22 31.0 & 2278   & 0.39(0.10)   &   06 03 56.4 & +04 15 57.6   & 3.8 &   -  \\
G205.06-6.04 &    06 16 27.5 & +04 09 44.0 & 2832   & 0.40(0.10)   &   06 16 40.8 & +04 28 15.6   & 4.6 &   -  \\
G206.33-25.94 &    05 06 49.1 & -06 14 56.6 & 2276   & 0.21(0.03)   &   05 06 17.0 & -06 00 00.0   & 5.6 &   IC~2118, Witch Head Nebula \\
G210.90-36.55 &    04 35 07.0 & -14 14 35.3 & 4763   & 0.14(+0.020/-0.028)   &   04 34 50.2 & -14 26 45.6   & 4.3 &   LDN1642, MBM20, IREC305 \\
G212.07-15.21 &    05 55 57.7 & -06 09 26.4 & 1332   & 0.23(0.10)   &   05 56 21.8 & -06 02 56.4   & 2.8 &   -  \\
G215.37-3.04 &    06 45 03.3 & -03 32 35.3 & 1329   & 2.40(0.50)   &   06 44 30.5 & -03 30 18.0   & 2.2 &   -  \\
G215.44-16.38 &    05 57 02.8 & -09 33 26.2 & 1353   & 1.45(1.45)   &   05 57 26.4 & -09 31 15.6   & 2.6 &   S4  \\
G216.76-2.58 &    06 48 59.8 & -04 36 09.6 & 1330   & 2.40(0.50)   &   06 49 06.2 & -04 24 39.6   & 2.6 &   -  \\
G218.06+2.12 &    07 08 22.0 & -03 34 59.3 & 1329   &    -         &   07 07 59.3 & -03 38 31.2   & 3.4 &   -  \\
G219.36-9.71 &    06 28 01.3 & -09 57 53.7 & 2279   & 0.91(0.03)   &   06 28 14.6 & -09 42 54.0   & 4.6 &   LDN1652  \\
G219.29-9.25 &    06 29 43.0 & -09 48 37.5 & 1331   & 0.91(0.03)   &   06 29 17.0 & -09 41 49.2   & 2.8 &   IREC317  \\
G227.95-2.98 &    07 07 50.7 & -14 46 23.6 & 2826   & 2.00(0.50)   &   07 07 30.0 & -15 03 57.6   & 4.4 &   -  \\
G247.55-12.27 &    07 09 26.1 & -36 16 39.4 & 2824   & 0.17(0.17)   &   07 08 39.1 & -36 35 52.8   & 4.0 &   IREC365, DCld 247.5-12.3 \\
G253.71+1.93 &    08 25 04.9 & -34 26 33.1 & 1318   & 2.26(0.10)   &   08 25 42.2 & -34 23 09.6   & 3.1 &   -  \\
G255.33-4.88 &    08 01 23.8 & -39 38 07.5 & 1764   & 0.80(0.40)   &   08 00 33.6 & -39 45 36.0   & 2.8 &   DCld 255.4-04.9  \\
G258.90-4.10 &    08 14 35.7 & -42 08 01.8 & 1323   & 1.04(0.44)   &   08 14 20.9 & -41 59 56.4   & 2.7 &   HMSTG259.1-4.0C  \\
G265.04+6.08 &    09 17 48.9 & -40 36 11.7 & 1755   & 0.92(0.26)   &   09 18 40.8 & -40 43 26.4   & 2.8 &   -  \\
G265.60-5.82 &    08 27 59.9 & -48 36 50.8 & 1319   & 2.40(+0.71/-0.85)   &   08 29 07.9 & -48 38 16.8   & 3.2 &   -  \\
G268.21+2.02 &    09 13 10.0 & -45 36 45.4 & 1756   & 1.87(+0.86/-1.39)   &   09 14 11.3 & -45 33 03.6   & 2.7 &   SDN114, HMSTG268.0+1.8  \\
G271.06+4.84 &    09 36 23.9 & -45 39 13.0 & 2260   & 1.32(0.50)   &   09 37 00.7 & -45 27 36.0   & 3.9 &   SDN118, HMSTG271.4+4.8  \\
G271.51+5.14 &    09 39 27.5 & -45 49 41.0 & 3400   & 1.32(0.50)   &   09 41 13.2 & -45 59 20.4   & 4.6 &   HMSTGG271.4+4.8  \\
G276.78+1.75 &    09 50 16.3 & -51 39 46.6 & 5064   & 2.00(2.00)   &   09 52 58.8 & -51 25 33.6   & 5.3 &   S5, FeSt2-72, DCld 276.9+01.7, \\
&               &             &        &              &              &               &     &   DCld 276.8+0.1.9 \\
G298.31-13.05 &    11 38 50.7 & -75 17 27.2 & 1330   & 0.15(0.10)   &   11 40 48.0 & -75 23 24.0   & 3.4 &   SDN138, FeSt2-129, \\
&               &             &        &              &              &               &     &   HMSTG298.3-13.1, FeSt1-188, \\
&               &             &        &              &              &               &     &   DCld 298.3-13.1 \\
G299.57+5.61 &    12 26 45.9 & -57 05 33.6 & 1331   &    -         &   12 25 33.6 & -57 04 44.4   & 3.0 &   HMSTG299.6+5.6   \\
G300.61-3.13 &    12 28 54.8 & -65 47 40.3 & 1761   & 0.20(0.05)   &   12 27 04.6 & -65 50 34.8   & 2.8 &   HMSTG300.6-3.0 \\
G300.86-9.00 &    12 25 17.3 & -71 46 05.6 & 1357   & 0.15(0.03)   &   12 26 07.4 & -71 50 38.4   & 3.0 &   PCC550, S10, SDN143, VMF32, \\
&               &             &        &              &              &               &     &   Musca DN Complex \\
G315.88-21.44 &    17 19 40.0 & -76 55 16.8 & 1331   & 0.25(0.01)   &   17 16 27.8 & -76 49 22.8   & 2.5 &    -   \\
G320.84+5.09 &    14 55 10.5 & -53 24 46.6 & 1328   & 1.00(1.00)   &   14 55 15.4 & -53 18 28.8   & 2.7 &   HMSTG320.8+5.1   \\
G325.54+5.82 &    15 18 44.6 & -50 22 09.3 & 2276   & 0.64(0.44)   &   15 19 04.1 & -50 06 18.0   & 3.7 &   SDN175, HMSTG325.5+5.8, DCld 325.5+05.8 \\
G332.70+6.77 &    15 49 43.7 & -45 28 32.1 & 1775   & 0.65(0.20)   &   15 50 11.0 & -45 37 19.2   & 2.8 &   -  \\
G334.65+2.67 &    16 14 40.4 & -47 11 48.8 & 1778   & 1.50(0.50)   &   16 13 45.8 & -47 08 24.0   & 3.7 &   -   \\
G339.22-6.02 &    17 12 09.7 & -49 34 31.0 & 1334   & 2.09(1.00)   &   17 13 08.9 & -49 37 58.8   & 3.2 &   -   \\
G341.18+6.51 &    16 25 05.1 & -39 59 10.1 & 1332   & 0.14(0.02)   &   16 25 14.6 & -39 49 40.8   & 3.3 &   SLDN16, HMSTG341.2+6.5, LM170,  \\
&               &             &        &              &              &               &     &   DCld 341.2+06.5, FeSt1-357 \\
G343.64-2.31 &    17 10 29.2 & -43 46 28.7 & 2278   & 1.00(0.50)   &   17 10 52.1 & -43 57 43.2   & 3.8 &   HMSTG343.7-2.3, (FeSt1-373) \\
G344.77+7.58 &    16 33 30.3 & -36 38 59.7 & 1332   & 0.24(0.24)   &   16 33 25.0 & -36 31 48.0   & 2.5 &   -  \\
G345.39-3.97 &    17 22 50.0 & -43 28 24.7 & 1776   & 0.23(+0.20/-0.10)   &   17 23 42.5 & -43 34 01.2   & 3.8 &   -  \\
G358.96+36.75 &    15 39 37.9 & -07 13 09.2 & 1355   & 0.11(0.01)   &   15 38 52.6 & -07 10 58.8   & 3.0 &   LDN1780, LDN1778, MBM33 \\
\end{longtable}
\tablefoot{The aliases refer to the following catalogues:
Arch = \citet{Desert2008},
B    = \citet{Barnard1919,Barnard1927}, 
BDN  = \citet{Bernes1977},              
CB   = \citet{Clemens1988},             
FeSr = \citet{Feitzinger1984}
HMST = \citet{Hartley1986},             
LDN  = \citet{Lynds1962},               
MBM  = \citet{MBM},                     
MLB  = \citet{MLB1983},                 
SLDN = \citet{Sandqvist1976}.           
Most entries are looked up from \citet{Dutra2002}. Names starting with
PCC and the entries S1-S10 refer to names used in \citet{Juvela2010}
and \citet{planck2011-7.7a} for some of the {\it Herschel} fields.}
\end{landscape}

\clearpage

\twocolumn
\appendix

\section{Alternative data analysis} \label{sect:alternative}

To investigate the robustness of the results to details of the data reduction, we
calculated a set of alternative $\tau(250\mu{\rm m})$ maps. In addition to the analysis
of three or four {\it Herschel} bands (see Sect.~\ref{sect:Herschel_data}), we considered
the use of local background subtraction and the possibility of making a correction for
residual mapping artefacts with the help of other all-sky surveys. The resulting
$\tau(250\mu {\rm m})/\tau(J)$ ratios were compared to the values found in
Sect.~\ref{sect:apparent}. The comparison was carried out without bias corrections,
comparing the results with the 160-500\,$\mu{\rm m}$ fits of
Sect.~\ref{sect:correlations}.

\subsection{Local background subtraction} \label{sect:subtraction}

Our default analysis is based on {\it Herschel} data for which the absolute zero points
were derived from a comparison with Planck and IRAS maps. As an alternative, we used {\it
Herschel} surface brightness maps from which the diffuse background was subtracted using the
reference regions listed in Table~\ref{table:fields}. Thus, the average surface
brightness of the reference region was subtracted from each {\it Herschel} surface
brightness map separately, before calculating the colour temperatures and the dust
optical depths.

The local background subtraction might be a more reliable way to ensure a consistent zero
level for the compared quantities. However, it also means that colour temperature and
column density can only be estimated in the part of the map in
which the surface brightness
values are significantly higher than those of the reference area. We masked the
area in which the signal is lower than twice the estimated statistical uncertainty of the
surface brightness in each band. The final mask is a combination of these masks and the original mask
that eliminated the map boundaries for which the convolution to the resolution of the
${\tau(J)}$ data is only poorly defined.

\subsection{Checks for mapping artefacts} \label{sect:grad}

Although {\it Herschel} data are usually of very good quality, there can still be some
small artefacts that affect some parts of the maps. Errors might result from data reduction
or from instrumental effects such as striping or general gain changes
\citep{SPIRE_mapmaking, Paladini2013}. If processing includes high-pass filtering, the
large-scale surface brightness gradients may be affected and the contrast between faint
and bright regions may be decreased. Our maps often contain significant emission up to
the map boundary. Without a flat border region with very low emission, it is difficult to
estimate whether the baseline assumed for the scans is correct. Such effects could be
more important and more difficult to detect for small maps. Thus, this could mostly
affect PACS maps, for which the signal-to-noise ratio is also typically lower than in
the SPIRE data \citep{Juvela2010}.

These effects were investigated with the help of independent FIR and submillimetre data.
At 100\,$\mu$m, 350\,$\mu$m, and 500\,$\mu$m we can compare {\it Herschel} data almost
directly with the corresponding IRIS and {\it Planck} bands. The {\it Planck} 545\,GHz
data were corrected to 500\,$\mu$m using a modified black body with the {\it Herschel}
colour temperature map and $\beta$ equal to 2.0. Keeping the other parameter constant, an
error of 2\,K in temperature or an error of 0.2 in $\beta$ would both correspond to only
a $\sim$2\% error in the extrapolated value. At 160\,$\mu$m and 250\,$\mu$m we used
values interpolated from IRIS 100\,$\mu$m, AKARI 140\,$\mu$m \citep{Murakami2007}, and
{\it Planck} 350\,$\mu$m channels. Here $\Delta \beta \sim 0.2$ translates into a change of less than
1\% at 160\,$\mu$m. For a direct extrapolation from the 350\,$\mu$m to 160\,$\mu$m,
an error of $\Delta \beta$=0.1 would result in an error in the interpolated value that is
still lower than 8\%. Typically interpolation errors should thus be below the statistical
errors. The same considerations apply to the zero-point corrections of
Sect~\ref{sect:zeropoint}, with the difference that they are not affected by any
multiplicative errors. 

The reference data were compared with the original {\it Herschel} maps at 6$\arcmin$
resolution to derive an additive correction that leaves the median value of the maps
unchanged and only affects scales larger than $\sim 6\arcmin$. We also checked similar
multiplicative corrections, assuming that the zero points of the different surveys are
compatible, and even calculating corrections where linear fits between {\it Herschel} and
reference data were first used to estimate the differences in zero-point and gain
calibration. The last alternative would avoid the assumptions of consistent calibration
and zero points between the data sets. In most cases, there are no significant
differences between the three choices. A typical map has no clear artefacts,
and the proposed correction will probably decrease the data quality. However, when the
local artefacts are clear (e.g., excessive surface brightness some corner of a {\it
Herschel} map), the corrected map should give a better description of the true surface
brightness. Thus, we do not believe that the corrected maps represent a clear improvement,
but the difference between the corrected and uncorrected data should give some idea of
the potential effect that artefacts of that magnitude could have.

\subsection{Comparison of the data sets} \label{sect:altcomp}

We compared the $\tau(250\mu {\rm m})/\tau(J)$ values (without bias corrections) among
six data sets: (1) three bands at 250-500\,$\mu$m (our default data set), (2) four bands
at 160--500\,$\mu$m, (3) three bands with local background subtraction, (4) three bands
with the corrections of Sect.~\ref{sect:grad}, and (5) four bands with the corrections of
Sect.~\ref{sect:grad}. Table~\ref{table:comp} shows the results, comparing the dispersion
of the obtained $\tau(250\mu {\rm m})/\tau(J)$ values between fields and the change in
the values compared to the default case where three bands and the absolute zero points
were used.

\begin{table}{}
\caption{
Comparison of the mean values of $\tau(250\mu {\rm m})/\tau(J)$ obtained with different
versions of {\em Herschel} data, without bias corrections. The columns are (1) data
version, (2) number of fields with $<10\%$ error in $k$, (3) mean and standard deviation
for that sample of fields, (4) difference and standard deviation when compared to the
values in the default case (3-bands), for a common set of 81 fields).
}
\label{table:comp}
\centering
\begin{tabular}{cccc}
\hline
\hline
Data set   &   Fields & $\tau(250\mu {\rm m})/\tau(J)$
&   $\Delta(\tau(250\mu {\rm m})/\tau(J))$ \\
&         &   ($10^{-4}$) & ($10^{-4}$) \\
\hline
3-band              & 105 &  22.4$\pm$9.6    &  --             \\
4-band              &  83 &  22.2$\pm$12.5   &  -0.04 $\pm$  2.80 \\
Bg-subtracted       & 102 &  20.1$\pm$9.5    &  -0.96 $\pm$  1.88 \\
3-band, corr.$^1$   & 105 &  20.7$\pm$8.6    &  -0.91 $\pm$  1.51 \\
4-band, corr.$^1$   &  82 &  21.3$\pm$11.8   &  -0.54 $\pm$  2.43 \\
\hline
\multicolumn{4}{l}{$^1$ Correction at large scales using ancillary data.}
\end{tabular}
\end{table}

The second column of the table lists the number of fields where the formal error of the
slope $\tau(250\mu {\rm m})/\tau(J)$ is lower than 10\%. This number is smaller when PACS
data are included, 82--83 fields compared to the 102-105 fields with SPIRE data alone. The
numbers do not include the Witch Head Nebula, for which we have no PACS data and which
therefore was excluded from this comparison. The background subtraction and the
Sect.~\ref{sect:grad} corrections both decrease the mean value, but not significantly.
Note that on physical grounds one could have expected the values to increase with
background subtraction (if dense material has higher $\tau(250\mu {\rm m})/\tau(J)$) and
to decrease with the inclusion of 160\,$\mu$m channel (if the inclusion of shorter
wavelengths increases estimated colour temperatures). The last column shows the
difference relative to our default case. In this column we only
list the common set of 81
fields where the error estimates are lower than 10\% for all the five cases. On average, the
changes in the $\tau(250\mu {\rm m})/\tau(J)$ values are less than 1.0 units (lower than
5\%). The dispersion between different SPIRE analyses is lower
than 2.0 units and somewhat
higher when analyses of four and three bands are compared.

\section{Simulated NIR extinction maps} \label{sect:NIR_simu}

In addition to photometric errors, the reliability of NIR optical depth estimates is mainly affected by the sampling provided by the background stars and the possible contamination by
foreground stars and galaxies.

We used 2MASS catalogue flags to eliminate most of the obvious galaxies. In addition to
requiring a photometric quality corresponding to {\em ph\_qual} in classes A--C, we
excluded all point sources that were extended (flag {\em ext\_key} is set) or were
flagged with {\em gal\_contam}. These only remove part of the galaxies. The increased
dispersion of intrinsic colours caused by galaxies is taken into account in the error
estimates provided by the method NICER. Because our simulations use actual 2MASS data
near the target fields, this extragalactic contamination is also automatically present in
the simulations described below.

The simulations are based on dust 250\,$\mu$m optical depth maps derived from {\it
Herschel} observations. Using only SPIRE data, we derive column densities at
25$\arcsec$ resolution as a combination of
\begin{equation}
\tau  =  \tau(500) + [ \tau(350) - \tau(350 \rightarrow 500)] ,
\end{equation}
where $\tau(500)$ is calculated using 250, 350, and 500\,$\mu$m maps at the lowest common
resolution, $\tau(350)$ is calculated similarly from 250\,$\mu$m and 350\,$\mu$m maps,
and $\tau(350 \rightarrow 500)$ is the latter convolved to the resolution of the
500\,$\mu$m observations \citep[see][]{Palmeirim2013}. Thus, the expression in square
brackets describes structures that are seen at the resolution of 350\,$\mu$m data
(25$\arcsec$), but not at the resolution of 500\,$\mu$m data (36$\arcsec$). The
calculations assume a fixed value $\beta$=2.0. The differences to the $\tau(250\mu$m)
maps used in Sect.~\ref{sect:results} are small and, furthermore, we  only
consider differences between these input maps and the values recovered by NICER. On the
other hand, we wish to retain the highest resolution possible (18$\arcsec$ instead of
36$\arcsec$) because the bias in NICER estimates is probably linked to the amount of
small-scale structure.

The Besancon model \citep{Robin2003} was used to create a simulated catalogue of stars
over a $0.5\degr \times 0.5\degr$ area centred on each target field. The catalogue
includes stellar distances and H-band magnitudes, and together with the distance
estimates listed in Table~\ref{table:fields}, this was converted into the probability that
a star of given magnitude resides between the cloud and the observer. In the simulation,
the corresponding fraction of stars was assumed to be located in front of the cloud.

To simulate NIR observations, we used the same 2MASS data that were used to calculate the
actual $\tau(J)$ maps of the fields. The stars in the reference area (see
Table~\ref{table:nicer}) were used to determine an empirical probability distribution of H-band magnitudes and the dependence between the J, H, and Ks magnitudes and their
uncertainties, as given in the 2MASS catalogue. This reference area may be affected by
small amounts of absorption by diffuse dust, but gives a good approximation of the
brightness distribution of stars that are unextincted by the main cloud. 

We simulated a uniform distribution of stars over the {\it Herschel} field, generating the
magnitudes from the empirical H-band magnitude distribution and matching the average
stellar density of the reference region. The J and Ks magnitudes of each star were
generated using the J-H and H-Ks colours of a random star selected from the reference
region. This ensures that the distribution of intrinsic colours is realistic and that the
simulations reproduce proper correlations (also in errors) between the bands. 

Based on the Besancon model, a fraction of stars was assumed to reside in front of the
cloud and to be unaffected by extinction. For the remaining stars, the line-of-sight
optical depth in J band was calculated using the input column density map and a fixed
ratio of $\tau(250\mu {\rm m})/ \tau(J)=1.5 \times 10^{-3}$. The optical depths in H and
Ks bands then follow from the \citet{Cardelli1989} extinction curve. The magnitudes were
adjusted according to the line-of-sight optical depths. The magnitudes and their
uncertainties in the reference area were used to calculate the typical photometric
uncertainties as a function of magnitude, and they were used in the NICER calculation.
Because the intrinsic colours of the stars were generated based on observed stars, no
additional scatter needed to be added for intrinsic colours. However, because the extinction
makes the stars fainter, the typical photometric errors increase as well. This was taken into
account by adding normal distributed noise to
the magnitudes, which corresponds to the
difference in the typical uncertainties between the original and the extincted
magnitudes.

\begin{figure*}
\sidecaption
\includegraphics[width=12.0cm]{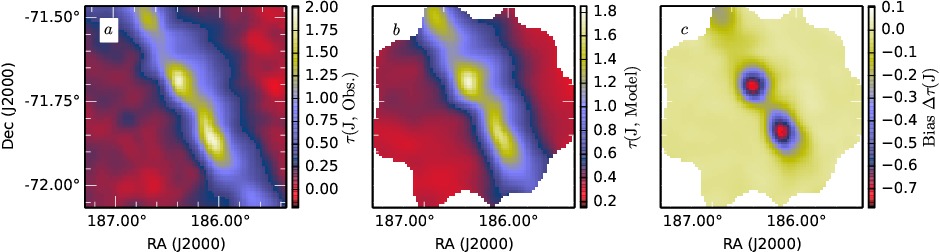}
\caption{
Field G300.86-9.00 as an example of $\tau(J)$ bias maps estimated with simulated NICER
observations. The frames show the optical depth derived from actual observations (frame a),
average recovered extinction map in simulations (frame b), and the bias as the difference
between the output and input maps in the simulation (frame c).
}
\label{fig:plot_extinction}
\end{figure*}

The simulated stellar catalogues were fed to the NICER algorithm to derive extinction
maps with the same parameters as in Sect.~\ref{sect:NIR_data}. For each target field, one
hundred realisations of the ${\tau(J)}$ maps were calculated to obtain maps for the
standard deviation and the bias of the estimated ${\tau(J)}$ values.
Figure~\ref{fig:plot_extinction} shows one example.

\section{Simulated {\it Herschel} observations} \label{sect:Herschel_simu}

The estimation of the ${\rm \tau(250\mu m)}$ bias is more uncertain than the estimation
of $\tau(J)$ bias. Because of line-of-sight temperature variations, the colour temperature
overestimates the mass-averaged dust temperature, which translates into too low estimates
of ${\rm \tau(250\mu m)}$. The magnitude of the effect cannot be estimated precisely
because the line-of-sight temperature distribution is unknown. Order of magnitude
estimates can be obtained with radiative transfer modelling by making assumptions of the
radiation field, dust properties, and the cloud structure. 

We carried out radiative transfer calculations individually for all the 116 fields.
We assumed constant dust properties corresponding to Milky Way dust with
a selective extinction of $R_{\rm V}$=5.5 \citep{Draine2003}.
The initial radiation field was assumed to correspond to the \citet{Mathis1983} model of
solar neighbourhood. The spectrum of the illuminating radiation has an impact on the
temperature contrasts. We have no way to independently determine the shape of the
spectrum of the radiation field. However, this typically remains a second-order effect, and
the main effect, the level of the radiation field intensity, is part of the modelling.
One exception are the possible stars that heat clouds from the inside and may locally
have a strong effect. These are considered later in the analysis, but are not part of the
simulations.

For each field, we built a model that attempts to reproduce the 250--500\,$\mu$m
observations of that field. From the background-subtracted surface brightness maps we
selected the central $30\arcmin \times 30\arcmin$ area. The model cloud had the same
angular dimensions and was discretised onto a $181^3$ cell grid. Each cell of the model
therefore corresponds to an angular size of 10$\arcsec$ , but the linear scale depends on
the distance of the cloud. In the line-of-sight direction we assumed a Gaussian density
distribution with a FWHM equal to 25\% of the field size. The linear size again depends on
the cloud distance because in more distant fields we are also  probably concerned with
larger structures. In the line-of-sight direction the density peak is always in the
central plane of the model cube. This increases mutual shadowing of dense regions, which
in reality can reside at different distances and increases the temperature contrasts in
the model. On the other hand, for a field at 200\,pc distance, the selected line-of-sight
FWHM extent is $\sim$0.4\,pc, which is larger than the size of typical cores.
Therefore, it is difficult to say whether the selection of this particular line-of-sight
density profile leads to an over- or underestimation of the final ${\rm \tau(250\mu
m)}$ bias. These are usually second-order effects \citep{Juvela2013_colden} except for
very dense clumps that can remain practically invisible in $\tau(J)$ maps as well.

We carried out radiative transfer calculations to produce synthetic surface brightness maps
at {\it Herschel} wavelengths that were then convolved to the resolution of the
observations. The ratio of observed and modelled 350\,$\mu$m maps was used to adjust the
column densities, applying the same multiplicative factor to all cells along the same
line-of-sight. The intensity of the external radiation field was scaled based on the
ratios between the observed and modelled 250\,$\mu$m and 500\,$\mu$m surface brightness.
The aim is to also reproduce the average shape of the spectra. The full procedure was
iterated until the model matched the observed 350\,$\mu$m map at $\sim$1\% accuracy and
the average ratio 250\,$\mu$m/500\,$\mu$m were correct within the same tolerance.

The final model takes into account the range of column density, the morphology, and the
radiation field intensity of a field. It is not necessarily a perfect match to all
surface brightness maps, but is a good facsimile of the pixel-by-pixel column density
structure. We analysed the synthetic surface brightness maps as in
Sect.~\ref{sect:tau250}. The comparison of the obtained ${\rm \tau(250\mu m)}$ estimates
and the true values known for the model cloud gives a $30\arcmin \times 30\arcmin$ map of
the ${\rm \tau(250\mu m)}$ bias (resolution 40$\arcsec$). The remaining border areas
usually have a low column density and therefore low bias. However, to extend bias estimates
over the whole map area, we assigned values calculated using the
average bias vs. column density relation estimated from central $30\arcmin \times
30\arcmin$ area to the remaining pixels.

\begin{figure*}
\includegraphics[width=18.0cm]{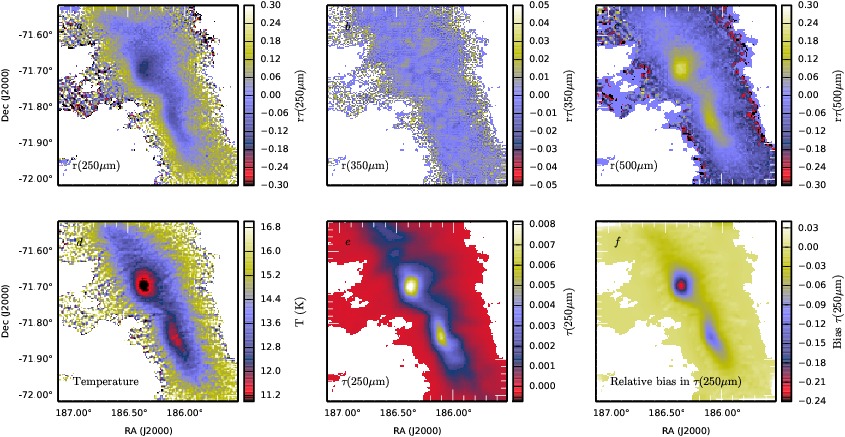} 
\caption{
Field G300.86-9.00 as an example of the ${\rm \tau(250\mu m)}$ bias estimation with radiative
transfer modelling. The upper row shows the relative error between the model predicted
surface brightness and the observations at 250\,$\mu$m, 350\,$\mu$m, and 500\,$\mu$m. The
lower frames show the colour temperature and ${\rm \tau(250\mu m)}$ maps calculated from
the synthetic observations and the relative bias in ${\rm \tau(250\mu m)}$ obtained by
comparison with the actual ${\rm \tau(250\mu m)}$ values in the model.
}
\label{fig:emission_model}
\end{figure*}

Figure~\ref{fig:emission_model} shows one example, the surface brightness maps produced
by the model and the bias in the ${\rm \tau(250\mu m)}$ values estimated from these
synthetic observations.

\section{Additional checks of correlations between ${\rm \tau(250\mu m)}$ and
$\tau(J)$}
\label{sect:additional}

In addition to the factors examined in Sect.~\ref{sect:correlations}, we checked the
importance of two additional factors, the technical implementation of the least-squares
fits, and the importance of internally heated regions.

The total least-squares fits of Sect.~\ref{sect:correlations} used the formal error
estimates of $\tau(J)$ and ${\rm \tau(250\mu m)}$. The former were obtained from NICER
routine, the latter were estimated with MCMC calculations starting with the assumption of
7\% (SPIRE) or 15\% (PACS) relative errors in the surface brightness data. If the
correlation is poor, the estimated slope becomes sensitive to the error estimates. For
example, if the true errors of ${\rm \tau(250\mu m)}$ were much larger, for example
because of some artefacts in map making, the use of too low error estimates would
increase the slope estimates. We checked this by repeating the analysis using twice the
original ${\rm \tau(250\mu m)}$ error estimates. In an extreme case, we can ignore the
error estimates altogether and perform unweighted least-squares fits. Based on the error
estimates used, the true relative uncertainty is significantly larger in $\tau(J)$ than
in ${\rm \tau(250\mu m)}$. Therefore, the unweighted least-squares fit probably
underestimates the true slope.

The third test concerns the internally heated regions in which because of strong
temperature variations combined with compact, high column density clumps, both optical
depth estimates are particularly uncertain. Furthermore, the estimates of ${\rm
\tau(250\mu m)}$ bias are probably incorrect in the same areas. This is caused by two
factors. First, without the internal heating source, the models are unable to produce
sufficient surface brightness values and result in very high column densities and thus
high estimates of the bias \citep{Juvela2013_colden}. Second, in the real clump the
internal heating may help to {\rm decrease} the actual bias if the clump centre remains
warm instead of being too cold to be registered in {\it Herschel} bands
\citep{Malinen2011}. We repeated the analysis of Sect.~\ref{sect:correlations} by masking
warm regions. We first masked all pixels for which the dust colour temperature was higher than 20\,K.
The mask was then extended to cover areas in which after convolution to 180$\arcsec$
resolution, the influence of the $T>20$\,K region was more than 10\% of the convolved
value.

\begin{figure}
\includegraphics[width=8.5cm]{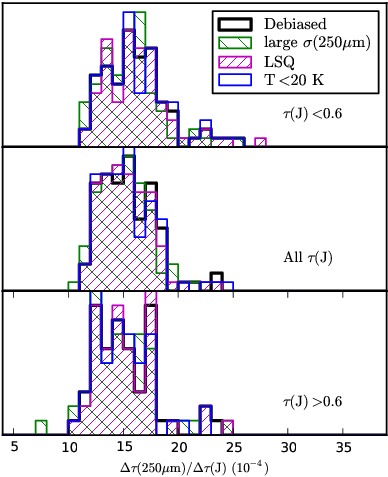} 
\caption{
Comparison of $\Delta \tau(250\mu {\rm m})$/$\Delta \tau(J)$ bias-corrected
distributions. In addition to the default case, derived distributions are shown for tests
with larger ${\rm \tau(250\mu m)}$ error estimates, normal unweighted least squares, and
fits excluding data affected by regions with dust temperatures exceeding 20\,K.
}
\label{fig:MOD}
\end{figure}

Figure~\ref{fig:MOD} compares the result with the bias-corrected results already shown
in Fig.~\ref{fig:hist_bias}. It shows that the shape of the distribution is not
sensitive to the fitting procedure, nor is it significantly affected by the warm
regions.

\section{Maps of $\tau(250\mu {\rm m})/\tau(J)$ ratio} \label{sect:ratiomaps}

In the least-squares fits of Sect.~\ref{sect:global}, the interesting parameters,
$k$ and $C$ were independent of additive errors in the correlated quantities. The
highest $\tau(J)$ points were particularly important, both visually and regarding the
fitted parameters. As mentioned in Sect.~\ref{sect:individual}, we also calculated maps
of the $\tau(250\mu {\rm m})/\tau(J)$ ratio. Because the column densities are typically
low over most of the mapped area, the visual appearance of the maps is dominated by
regions with low $\tau(250\mu {\rm m})$ and $\tau(J)$ values where, by definition, the
results become sensitive to any zero-point mismatch.

The maps were calculated by first correcting the $\tau(250\mu {\rm m})$ and $\tau(J)$
maps for the bias that was estimated with modelling (see Sect.~\ref{sect:NIR_simu} and
Sect.~\ref{sect:Herschel_simu}). The maps of $\tau(250\mu {\rm m})$ were then convolved
to the resolution of the $\tau(J)$ map. A 2$\arcmin$ wide region near the {\it Herschel}
map borders was masked because the convolved values would be affected by data outside our
map coverage. Before calculating the ratio $\tau(250\mu {\rm m})/\tau(J)$, we subtracted
from both quantities the values in the reference regions that are listed in
Table~\ref{table:fields}. A typical diameter of these reference areas is $\sim 6\arcmin$.
For $\tau(250\mu{\rm m})$ the statistical error of the reference value is very small. The
true uncertainty of the remaining $\tau(250\mu{\rm m})$ is completely dominated by
systematic errors. Because the main purpose of the ratio maps is to determine variations
in the $\tau(250\mu {\rm m})/\tau(J)$ ratio, we are not very concerned with
multiplicative errors. 

We expect the statistical errors to be more significant in $\tau(J)$ and, because the
variable is in the denominator, we need to mask areas with a
low SN of $\tau(J)$ . NICER
has provided error maps for $\tau(J),$ but here we estimated the uncertainty using the
following procedure: We took the data plotted in Fig.~\ref{fig:indi_1}, selected 20\% of
the lowest $\tau(J)$ points, and subtracted from them the prediction of the non-linear fit
(red line in Fig.~\ref{fig:indi_1}). The uncertainty of the reference value, $\Delta
\tau(J)$, was calculated as the standard deviation of the residuals, scaled by the ratio
of (90$\arcsec$)$^2$ and the area of the reference region. This should be a very
conservative estimate because it assumes that in Fig.~\ref{fig:indi_1} the scatter would
be due to $\tau(J)$ errors alone.

\begin{center}
\begin{figure*}
\begin{center}
\includegraphics[width=8.7cm]{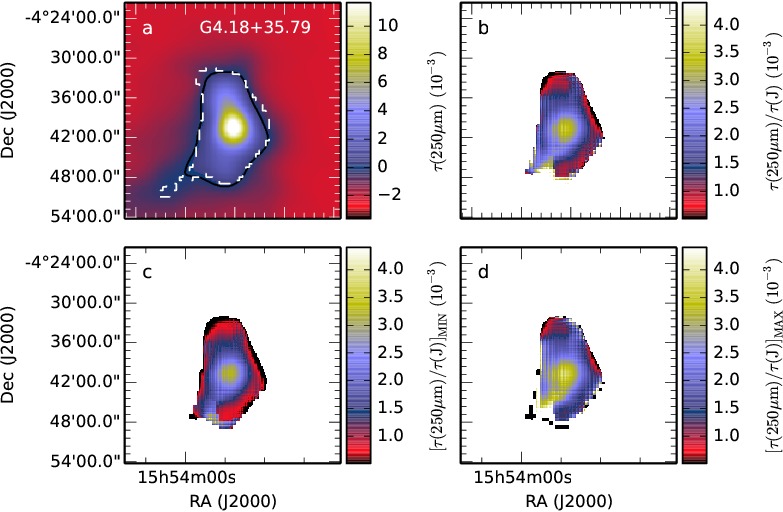}
\includegraphics[width=8.7cm]{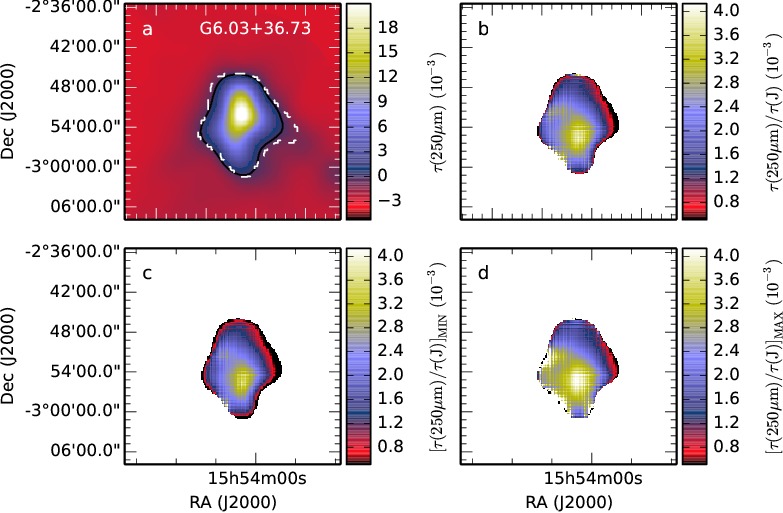}
\includegraphics[width=8.7cm]{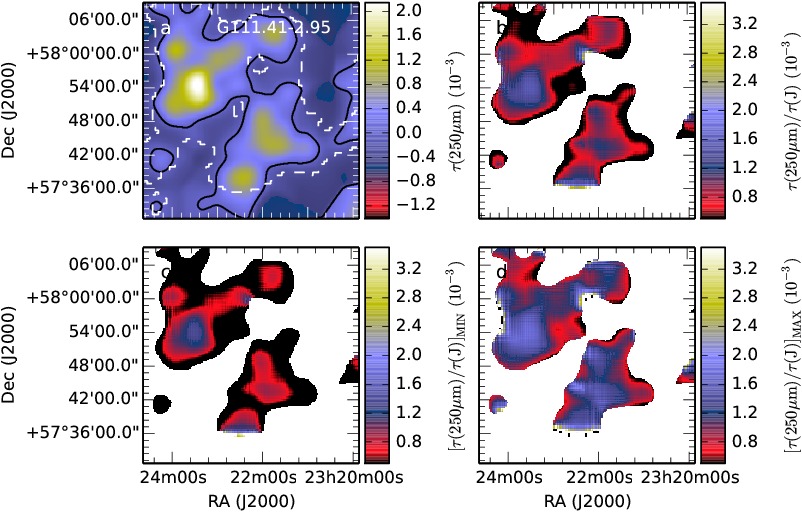}
\includegraphics[width=8.7cm]{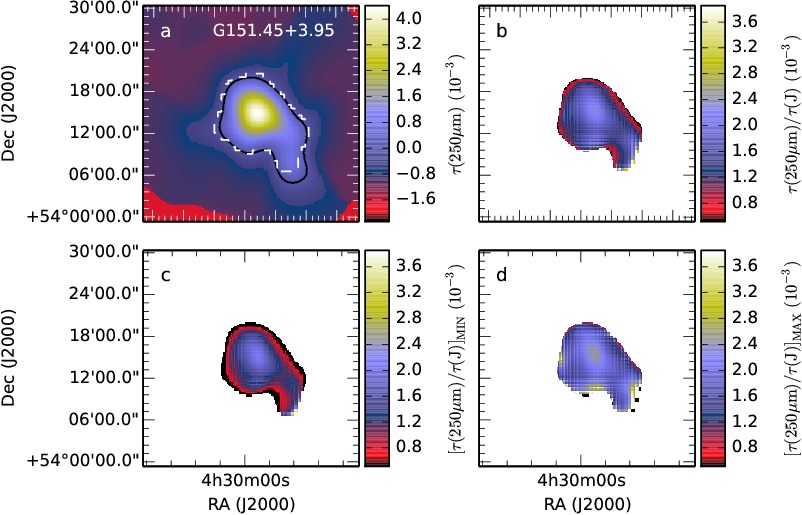}
\includegraphics[width=8.7cm]{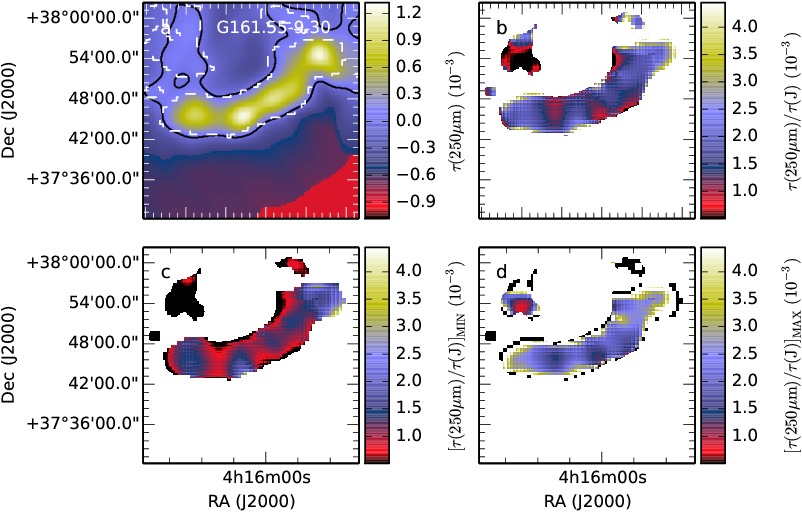}
\includegraphics[width=8.7cm]{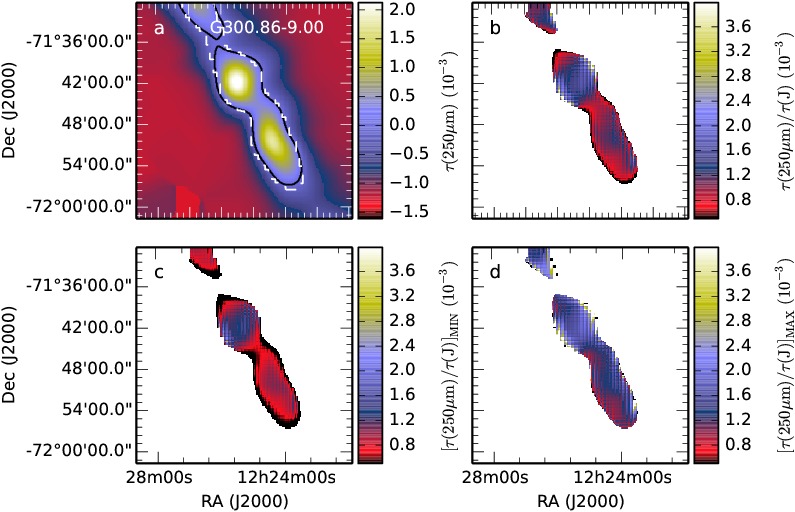}
\end{center}
\caption{
Maps of $\tau(250\mu{\rm m})/\tau(J)$ for the selected fields. The upper frames show 
$\tau(250\mu {\rm m})$ (frame a) and the ratio $\tau(250\mu {\rm m}) / \tau(J)$. The
lower frames show the lower (frame c) and upper (frame d) limits of $\tau(250\mu {\rm
m})/\tau(J)$ calculated as $(\tau(250\mu {\rm m})+\delta \tau(250\mu {\rm
m}))/(\tau(J)-\Delta \tau(J))$ and $(\tau(250\mu {\rm m})-\delta \tau(250\mu {\rm
m}))/(\tau(J)+\Delta \tau(J))$ where $\delta \tau(250\mu{\rm m})$ is the error map of
$\tau(250\mu{\rm m})$ and $\Delta \tau(J)$ is the estimated uncertainty of the $\tau(J)$
zero point. The areas in which the SN of either variable drops below 0.5 have been masked. In
the first frame, the black contour corresponds to $\tau(250\mu{\rm m})=\delta
\tau(250\mu{\rm m})$ and the dashed white contour to $\tau(J)=\Delta \tau(J)$.
}
\label{fig:ratiomaps}
\end{figure*}
\end{center}

The results are shown in Fig.~\ref{fig:ratiomaps}. The first frames show the $\tau(250\mu
{\rm m})$ maps, the contours indicating the region with a SN of each parameter higher than
one. The second frames show the calculated maps of $\tau(250\mu {\rm m})/\tau(J)$. The
regions where either parameter falls below SN=0.5 were masked. The remaining frames
show the extreme cases corresponding to $\tau(J) \pm \Delta \tau(J)$ and $\tau(250\mu{\rm
m}) \pm \delta \tau(250\mu {\rm m})$, where $\delta \tau(250\mu{\rm m})$ is the estimated
error map (see Sect.~\ref{sect:tau250}).

\clearpage
\onecolumn
\renewcommand{\footnoterule}{}

\begin{longtable}{lcccl}
\caption{
Coordinates and radii of the reference regions used to 
calculate the near-infrared optical depth maps. 
The last column lists the id numbers of the {\it Herschel} fields themselves.}
\label{table:nicer} \\
\hline \hline
Field &  \multicolumn{2}{c}{Reference area} &  Radius & {\it Herschel} Observation IDs\\
&    RA(2000) & DEC(2000) &  ($\arcmin$)  &          \\
\hline 
\endfirsthead
\caption{continued.} \\ 
\hline \hline
Field &  \multicolumn{2}{c}{Reference area} &  Radius & {\it Herschel} Observations IDs\\
&    RA(2000) & DEC(2000) &  ($\arcmin$)  &          \\
\hline      
\endhead      
\hline 
\endfoot

G0.02+18.02 &    16 43 59.5 & -18 29 20.4 & 16.8 &  1342227737, 1342216067, 1342216066 \\
G0.49+11.38 &    17 01 01.2 & -21 23 27.6 & 10.1 &  1342227645, 1342216500, 1342216499 \\
G1.94+6.07 &    17 26 46.8 & -23 17 09.6 & 13.7 &  1342216944, 1342216588, 1342216587 \\
G2.83+21.91 &    16 30 59.3 & -13 27 46.8 &  9.6 &  1342227646, 1342227040, 1342227039 \\
G3.08+9.38 &    17 18 49.9 & -20 40 01.2 & 14.8 &  1342252016, 1342252015, 1342251955 \\
G3.72+21.02 &    16 43 08.9 & -13 19 37.2 &  7.7 &  1342227647, 1342216502, 1342216501 \\
G4.18+35.79 &    15 50 46.1 & -05 04 37.2 & 18.8 &  1342215385, 1342215384, 1342213473 \\
G6.03+36.73 &    15 56 12.2 & -02 02 45.6 & 10.0 &  1342204167, 1342204166, 1342203075 \\
G9.45+18.85 &    17 03 07.7 & -10 09 54.0 & 18.0 &  1342227648, 1342216504, 1342216503 \\
G10.20+2.39 &    17 56 04.1 & -18 09 07.2 & 12.1 &  1342217762, 1342217761, 1342216945 \\
G20.72+7.07 &    18 02 13.7 & -07 34 51.6 & 10.0 &  1342216948, 1342216592, 1342216591 \\
G21.26+12.11 &    17 43 37.9 & -03 47 27.6 & 11.3 &  1342216949, 1342216590, 1342216589 \\
G24.40+4.68 &    18 16 02.4 & -04 43 19.2 & 10.9 &  1342217409, 1342217408, 1342216950 \\
G25.86+6.22 &    18 18 31.7 & -03 17 56.4 & 11.3 &  1342217772, 1342217771, 1342216951 \\
G26.34+8.65 &    18 05 08.9 & -01 01 22.8 &  8.4 &  1342216952, 1342216594, 1342216593 \\
G37.49+3.03 &    18 46 05.3 & +05 53 42.0 & 17.7 &  1342219075, 1342219074, 1342216953 \\
G37.91+2.18 &    18 49 47.3 & +06 09 46.8 & 16.0 &  1342229180, 1342219077, 1342219076 \\
G39.65+1.75 &    18 53 57.8 & +07 30 46.8 & 15.1 &  1342219633, 1342219079, 1342219078 \\
G62.16-2.92 &    20 02 27.4 & +23 30 36.0 & 14.9 &  1342219986, 1342219043, 1342219042 \\
G69.57-1.74 &    20 15 55.7 & +30 40 19.2 & 14.2 &  1342220090, 1342220089, 1342219988 \\
G70.10-1.69 &    20 17 18.0 & +31 06 46.8 & 17.2 &  1342220092, 1342220091, 1342219987 \\
G71.27-11.32 &    20 56 03.6 & +27 36 25.2 & 15.0 &  1342244225, 1342244224, 1342244143 \\
G82.65-2.00 &    20 57 19.2 & +40 46 30.0 & 11.2 &  1342222137, 1342221287, 1342221286, 1342219611 \\
G86.97-4.06 &    21 17 01.4 & +42 51 00.0 & 14.7 &  1342222116, 1342220280, 1342220279, 1342219610 \\
G89.65-7.02 &    21 42 49.9 & +44 06 32.4 &  9.1 &  1342220078, 1342220077, 1342213455 \\
G91.09-39.46 &    23 14 05.0 & +17 15 18.0 & 10.9 &  1342221461, 1342221285, 1342221284 \\
G92.04+3.93 &    20 57 44.9 & +53 18 28.8 & 12.8 &  1342220862, 1342219031, 1342219030 \\
G92.63-10.43 &    22 03 29.5 & +41 39 25.2 &  7.4 &  1342220869, 1342220080, 1342220079 \\
G93.21+9.55 &    20 31 49.7 & +57 36 39.6 & 18.2 &  1342220821, 1342220820, 1342213457 \\
G94.15+6.50 &    20 53 42.5 & +56 08 31.2 & 12.9 &  1342220553, 1342220552, 1342213456 \\
G95.76+8.17 &    20 50 34.3 & +59 01 55.2 &  8.3 &  1342244196, 1342243763, 1342243762 \\
G98.00+8.75 &    21 07 10.8 & +59 13 26.4 & 15.1 &  1342220551, 1342220550, 1342220525 \\
G105.57+10.39 &    21 33 44.9 & +67 18 28.8 & 13.7 &  1342219970, 1342220819, 1342220818 \\
G107.20+5.52 &    22 13 59.8 & +63 38 09.6 & 11.7 &  1342187332, 1342187331 \\
G108.28+16.68 &    21 07 11.8 & +72 31 48.0 & 13.5 &  1342216072, 1342216071, 1342213452 \\
G109.18-37.59 &    00 07 23.0 & +24 49 22.8 &  9.0 &  1342213511, 1342213510, 1342213494 \\
G109.80+2.70 &    22 49 03.8 & +63 14 56.4 & 16.5 &  1342187655, 1342187009, 1342187008, 1342187007 \\
G110.62-12.49 &    23 41 20.9 & +47 50 16.8 & 19.0 &  1342222136, 1342222110, 1342222109 \\
G110.89-2.78 &    23 17 37.7 & +57 17 45.6 & 13.0 &  1342223343, 1342223342, 1342213453 \\
G110.80+14.16 &    21 49 39.1 & +73 23 02.4 & 12.4 &  1342220622, 1342220817, 1342220816 \\
G111.41-2.95 &    23 16 59.8 & +57 02 52.8 & 13.0 &  1342223341, 1342223340, 1342213454 \\
G115.93+9.47 &    23 20 49.2 & +72 01 12.0 &  8.8 &  1342222598, 1342220666, 1342220665 \\
G116.08-2.40 &    00 01 23.8 & +58 57 28.8 & 16.6 &  1342222599, 1342222158, 1342222157 \\
G126.24-5.52 &    01 21 49.9 & +57 04 19.2 & 10.9 &  1342224972, 1342216074, 1342216073 \\
G126.63+24.55 &    04 04 11.0 & +84 55 55.2 &  8.6 &  1342243735, 1342243734, 1342219426, 1342219425, \\
&               &             &      &  1342219424, 1342199364 \\
G127.79+2.66 &    01 44 44.9 & +65 20 38.4 & 17.2 &  1342216512, 1342216511, 1342203610 \\
G128.78-69.46 &    00 56 41.0 & -06 16 12.0 & 18.2 &  1342246680, 1342246679, 1342246579 \\
G130.37+11.26 &    02 33 16.1 & +71 50 20.4 &  9.8 &  1342218719, 1342218718, 1342216918 \\
G130.42-47.07 &    01 12 09.8 & +14 48 43.2 & 19.2 &  1342213525, 1342213524, 1342213486 \\
G131.65+9.75 &    02 36 24.2 & +71 39 21.6 &  9.5 &  1342243761, 1342243760, 1342223882, 1342218640, \\
&               &             &      &  1342203609 \\
G132.12+8.95 &    02 43 03.4 & +70 37 44.4 & 11.4 &  1342223881, 1342223880, 1342216919 \\
G139.60-3.06 &    02 47 26.6 & +54 55 48.0 & 16.5 &  1342226625, 1342216414, 1342216413 \\
G141.25+34.37 &    08 37 44.9 & +73 19 22.8 & 13.3 &  1342244197, 1342243739, 1342243738 \\
G149.67+3.56 &    04 15 01.4 & +55 59 38.4 & 14.1 &  1342217533, 1342217532, 1342216920 \\
G150.47+3.93 &    04 22 01.2 & +55 45 36.0 & 17.1 &  1342217531, 1342217530, 1342214702 \\
G151.45+3.95 &    04 31 30.5 & +53 51 28.8 &  9.9 &  1342205043, 1342205042, 1342203607 \\
G154.08+5.23 &    04 52 23.5 & +53 47 24.0 & 18.1 &  1342217529, 1342217528, 1342216921 \\
G155.80-14.24 &    03 40 03.4 & +38 29 06.0 & 13.0 &  1342226626, 1342224781, 1342224780 \\
G157.08-8.68 &    04 02 21.6 & +42 01 15.6 & 13.1 &  1342216416, 1342216415, 1342203615 \\
G157.92-2.28 &    04 28 53.0 & +44 30 25.2 & 13.7 &  1342217527, 1342217526, 1342216923 \\
G159.23-34.51 &    02 59 58.6 & +19 12 03.6 & 13.8 &  1342239264, 1342239263 \\
G159.12-14.30 &    03 46 43.0 & +34 50 09.6 &  7.8 &  1342226627, 1342216419, 1342216418 \\
G159.34+11.21 &    05 45 49.0 & +51 29 31.2 & 18.6 &  1342218721, 1342218720, 1342216922 \\
G161.55-9.30 &    04 12 05.5 & +37 11 20.4 & 11.5 &  1342205045, 1342205044, 1342203616 \\
G163.82-8.44 &    04 23 32.6 & +36 22 48.0 & 16.7 &  1342205048, 1342205047 \\
G164.71-5.64 &    04 43 22.8 & +38 52 48.0 & 14.5 &  1342217525, 1342217524, 1342216925 \\
G167.20-8.69 &    04 34 25.9 & +34 10 44.4 &  6.6 &  1342217392, 1342217391, 1342216926 \\
G168.85-10.19 &    04 33 29.5 & +31 49 19.2 & 14.1 &  1342217523, 1342217522, 1342214701 \\
G171.35-38.28 &    03 19 13.7 & +09 39 25.2 & 12.8 &  1342224970, 1342224215, 1342224214 \\
G173.43-5.44 &    05 07 51.4 & +30 37 55.2 & 17.3 &  1342217507, 1342217506, 1342216927 \\
G174.22+2.58 &    05 42 31.9 & +34 22 48.0 &  9.9 &  1342244179, 1342243759, 1342243758 \\
G176.27-2.09 &    05 24 21.8 & +30 08 06.0 & 10.3 &  1342205197, 1342205196, 1342203617 \\
G181.84-18.46 &    04 47 25.7 & +16 18 21.6 & 10.3 &  1342217469, 1342217468, 1342203625 \\
G188.24-12.97 &    05 14 20.9 & +14 06 54.0 & 11.9 &  1342217455, 1342217454, 1342214700 \\
G189.51-10.41 &    05 28 18.5 & +15 00 07.2 & 10.9 &  1342217457, 1342217456, 1342216930 \\
G195.74-2.29 &    06 13 50.2 & +14 48 46.8 & 17.9 &  1342219409, 1342219408, 1342216931 \\
G198.58-9.10 &    05 55 25.0 & +08 15 57.6 & 17.9 &  1342218702, 1342218701, 1342216932 \\
G202.23-3.38 &    06 18 57.4 & +07 14 02.4 & 10.2 &  1342218773, 1342218772, 1342216933 \\
G202.02+2.85 &    06 42 51.8 & +11 30 43.2 & 12.2 &  1342228372, 1342228371, 1342228342 \\
G203.42-8.29 &    06 07 25.2 & +05 00 18.0 & 12.6 &  1342218777, 1342218776, 1342216935 \\
G205.06-6.04 &    06 19 59.0 & +04 49 15.6 &  7.6 &  1342218775, 1342218774, 1342216934 \\
G206.33-25.94 &    05 04 09.1 & -05 37 58.8 &  9.3 &  1342249237 \\
G210.90-36.55 &    04 31 38.4 & -13 23 34.8 &  9.7 &  1342225213, 1342225212, 1342216940 \\
G212.07-15.21 &    05 59 01.9 & -05 58 30.0 & 14.0 &  1342218783, 1342218782, 1342216936 \\
G215.37-3.04 &    06 44 21.4 & -04 16 37.2 & 15.7 &  1342219957, 1342219405, 1342219404 \\
G215.44-16.38 &    05 59 46.6 & -08 52 40.8 & 19.9 &  1342204306, 1342204305, 1342203631 \\
G216.76-2.58 &    06 45 47.8 & -05 23 56.4 & 12.0 &  1342219956, 1342219407, 1342219406 \\
G218.06+2.12 &    07 11 45.8 & -02 43 48.0 &  8.5 &  1342219958, 1342220785, 1342220784 \\
G219.36-9.71 &    06 28 11.3 & -09 30 07.2 & 15.6 &  1342227708, 1342219403, 1342219402 \\
G219.29-9.25 &    06 28 20.9 & -09 27 43.2 & 16.5 &  1342227707, 1342219401, 1342219400 \\
G227.95-2.98 &    07 04 29.5 & -15 27 39.6 & 10.9 &  1342219982, 1342220901, 1342220900 \\
G247.55-12.27 &    07 13 20.6 & -36 48 00.0 & 12.7 &  1342222831, 1342220781, 1342220780 \\
G253.71+1.93 &    08 25 07.0 & -33 44 45.6 & 10.8 &  1342222828, 1342220783, 1342220782 \\
G255.33-4.88 &    07 58 19.9 & -39 40 51.6 & 16.7 &  1342222830, 1342220779, 1342220778 \\
G258.90-4.10 &    08 10 17.0 & -42 19 15.6 & 10.6 &  1342222896, 1342220777, 1342220776 \\
G265.04+6.08 &    09 21 18.7 & -40 14 38.4 &  7.8 &  1342222845, 1342220309, 1342220308 \\
G265.60-5.82 &    08 32 48.5 & -47 47 06.0 & 12.0 &  1342222829, 1342220775, 1342220774 \\
G268.21+2.02 &    09 17 09.8 & -44 53 49.2 & 17.7 &  1342222827, 1342221275, 1342221274 \\
G271.06+4.84 &    09 41 12.0 & -45 28 51.6 &  9.9 &  1342222895, 1342221270, 1342221269 \\
G271.51+5.14 &    09 44 13.9 & -45 54 39.6 & 10.2 &  1342222894, 1342221268, 1342221267 \\
G276.78+1.75 &    09 54 27.1 & -51 06 21.6 &  9.8 &  1342198593, 1342198592 \\
G298.31-13.05 &    11 26 36.0 & -75 02 52.8 & 11.7 &  1342223601, 1342223600, 1342216941 \\
G299.57+5.61 &    12 22 23.3 & -56 28 33.6 & 23.5 &  1342223605, 1342223604, 1342223259 \\
G300.61-3.13 &    12 27 23.0 & -66 32 20.4 & 15.1 &  1342223603, 1342223602, 1342213480 \\
G300.86-9.00 &    12 14 18.7 & -72 08 02.4 & 11.4 &  1342188162, 1342188101, 1342188100, 1342188099 \\
G315.88-21.44 &    17 06 03.4 & -77 16 37.2 & 14.6 &  1342218738, 1342218737, 1342216942 \\
G320.84+5.09 &    14 55 25.2 & -52 38 56.4 & 13.2 &  1342227725, 1342215598, 1342215597 \\
G325.54+5.82 &    15 16 29.0 & -49 30 54.0 &  9.8 &  1342227693, 1342225000, 1342224999 \\
G332.70+6.77 &    15 47 46.1 & -44 43 04.8 & 15.5 &  1342227692, 1342226700, 1342226699 \\
G334.65+2.67 &    16 10 07.0 & -46 25 26.4 & 13.5 &  1342216490, 1342216489, 1342214758 \\
G339.22-6.02 &    17 15 55.9 & -50 25 01.2 &  9.3 &  1342216582, 1342216581, 1342214757 \\
G341.18+6.51 &    16 21 30.2 & -39 15 07.2 &  9.4 &  1342227691, 1342216492, 1342216491 \\
G343.64-2.31 &    17 14 32.4 & -44 28 40.8 & 17.3 &  1342216943, 1342216584, 1342216583 \\
G344.77+7.58 &    16 29 49.0 & -37 18 21.6 & 13.7 &  1342227690, 1342216496, 1342216495 \\
G345.39-3.97 &    17 26 34.1 & -43 44 13.2 & 20.6 &  1342252027, 1342251963 \\
G358.96+36.75 &    15 36 34.1 & -07 12 54.0 & 12.2 &  1342204169, 1342204168, 1342202205 \\

\hline 
\end{longtable}

\clearpage

\end{document}